\documentclass[traditabstract]{aa}

\usepackage{natbib}
\usepackage{mathrsfs}
\usepackage{amsmath}
\usepackage{txfonts}
\usepackage{array}
\usepackage{color}
\usepackage{graphicx}

\def\msun{M_\mathrm{\odot}}
\def\lsimeq
{\hbox{\raise0.5ex\hbox{$<\lower1.06ex\hbox{$\kern-1.07em{\sim}$}$}}}
\begin{document}

\title{Constraining $\gamma$-ray pulsar gap models with a simulated pulsar population}

\author{M. Pierbattista\inst{1,2} 
\and I. A. Grenier\inst{1,3} 
\and A. K. Harding\inst{4} 
\and P. L. Gonthier\inst{5}
}

\institute{Laboratoire AIM, Universit\'e Paris Diderot/CEA-IRFU/CNRS, Service d'Astrophysique, CEA Saclay, 91191 Gif sur Yvette, France\\
\email{marco.pierbattista@cea.fr}
\and Fran\c{c}ois Arago Centre, APC, Universit\'e Paris Diderot, CNRS/IN2P3, CEA/Irfu, Observatoire de Paris, Sorbonne Paris Cit\'e, 10 rue A. Domon et L. Duquet, 75205 Paris Cedex 13, France 
\and Institut Universitaire de France
\and Astrophysics Science Division, NASA Goddard Space Flight Center, Greenbelt, MD 20771, U.S.A. 
\and Hope College, Department of Physics, Holland MI, U.S.A.
 }

\date{}

  \abstract
{With the large sample of young $\gamma$-ray pulsars discovered by the \emph{Fermi} 
Large Area Telescope (LAT), population synthesis has become a powerful tool for 
comparing their collective properties with model predictions. We synthesised a pulsar 
population based on a radio emission model and four $\gamma$-ray gap models (Polar 
Cap, Slot Gap, Outer Gap, and One Pole Caustic). Applying  
$\gamma$-ray and radio visibility criteria, we normalise the simulation to the number of detected radio pulsars 
by a select group of ten radio surveys. 

The luminosity and the wide beams from the outer gaps can easily account for the number of 
\emph{Fermi} detections in 2 years of observations.  The wide slot-gap beam requires an increase 
by a factor of $\sim 10$ of the predicted luminosity to produce a reasonable number of $\gamma$-ray pulsars.  Such large increases in
the luminosity may be accommodated by implementing  offset polar 
caps. The narrow polar-cap beams contribute at most only a handful of LAT pulsars. 
Using standard distributions in birth location and pulsar spin-down power ($\dot{E}$), we 
skew the initial magnetic field and period distributions in a an attempt to account for the high $\dot E$ \emph{Fermi} pulsars.  
While we compromise the agreement between simulated and detected distributions of radio pulsars, the simulations fail to reproduce the 
 LAT findings:
all models under-predict  the number of LAT pulsars with high $\dot{E}$,
and they cannot explain the high probability of detecting both the radio and $\gamma$-ray beams at high $\dot{E}$. 
The beaming factor remains close to 1.0 over 4 decades in $\dot{E}$ evolution for the slot gap whereas 
 it significantly decreases with increasing age for the outer gaps. The 
 evolution of the enhanced slot-gap luminosity with $\dot{E}$ is compatible with the large dispersion of $\gamma$-ray luminosity seen in the LAT data. 
 The stronger evolution predicted for the outer gap, which is linked to the polar cap heating by the return current, 
 is apparently not supported by the LAT data.
 The LAT sample of  $\gamma$-ray pulsars therefore provides a fresh perspective on the early evolution  of the luminosity and beam width
 of the $\gamma$-ray emission from young pulsars, calling for thin and more luminous gaps.}  

\authorrunning{Pierbattista et al. 2012}
\titlerunning{Gamma-ray pulsar population}

\maketitle

\section{Introduction}

After the  radio detection of the first pulsar signal in 1967 \citep{hbp+68}, 
a pulsar magnetosphere model was formulated by
\cite{gj69}.  A direct consequence of the Goldreich \& Julian model is the establishment 
of a magnetospheric charge density that creates a force-free pulsar magnetosphere. 
However, such a magnetosphere has no electric field along the magnetic field 
to accelerate charges and produce $\gamma$-rays.

The detection, a few years later, of pulsed emission at $\gamma$-ray energies from the Crab \citep{mbc+73} and Vela 
\citep{tfko75} pulsars, and the detection of four more $\gamma$-ray pulsars by \cite{tab+94}  
established that pulsars accelerate particles to energies of at least a few TeV  suggesting that there are magnetospheric 
regions where the charge density departs from that of Goldreich \& Julian, locally violating the force-free condition  
and allowing particle acceleration. These regions were identified in two magnetospheric zones. In the inner magnetosphere, acceleration 
can take place both above the polar cap and in the \emph{slot gap}, which extends to high-altitude
along the last open magnetic field lines. In the outer magnetosphere, the \emph{outer gap} extends from the null charge surface to the light cylinder.
These gap regions correspond to three models: the low-altitude slot-gap model, hereafter Polar Cap (PC, \cite{mh03}), the Slot Gap model (SG,
\cite{mh04a}), and the Outer Gap model (OG, \cite{crz00}).

In the \emph{polar-cap model} the emission comes from a region close to the
neutron star (NS) surface and well confined above the magnetic polar cap. 
Charged particles from the neutron star are initially accelerated in the strong
electrostatic field generated by a departure from the Goldreich-Julian charge density \citep{as79}.
Aided by inertial frame dragging \citep{mt92}, pulsars
emit high energy photons by curvature radiation (CR) and inverse Compton scattering (ICS). 
The most energetic of these photons reach threshold for electron-positron 
pair production in the strong magnetic field at a Pair Formation Front (PFF), above which the 
secondary pairs can screen the electric field in a short distance.  
The pairs, produced in excited Landau states,  emit synchrotron photons
which trigger a pair cascade with high multiplicity. A small fraction of the pairs is actually 
accelerated. The pair plasma likely establishes force-free conditions
along the magnetic field lines above the PFF, as well as radiate 
$\gamma$-rays. Over most of the polar cap, the PFF and 
$\gamma$-ray emission occurs well within a few stellar radii of the NS surface.
The main contribution to the $\gamma$-ray emission comes from CR
from the pairs moving upward. 
Since the CR intensity scales with the magnetic field lines curvature, it decreases from the polar cap edge
toward the magnetic axis, conferring to the emission beam the structure of an hollow cone.

The \emph{slot-gap} emission is generated from the same polar cap electromagnetic pair cascade 
near the boundary of the closed magnetic field lines region where the parallel electric field
$E_\mathrm{\parallel} \rightarrow 0$ and the PFF rises to higher altitude.  Here electrons are accelerated over longer 
distances to produce the pair cascade. A narrow gap, \emph{the slot gap}, is formed along the closed magnetic field surface where 
the PFF is never established, and electrons continue to be accelerated and radiating $\gamma$-rays by self-limited curvature radiation  into the outer magnetosphere.
The resulting hollow beam is much broader and less collimated near the magnetic axis than the lower-altitude PC emission
(see Section \ref{Phase-plot calculation and normalisation}).

The outer gaps are vacuum regions characterised by a strong electric field along the magnetic field lines
\citep{hol73,crs76} above the null charge surface. Two outer gap regions  \citep{crs76,ry95,crz00,h06} can exist in the 
\emph{angular velocity-magnetic momentum} plane, one for each pole. In the physical OG model, in the case of a 
non-aligned rotator,  the gap region closer to the pulsar surface is more active than the other  gap further away from the surface
due to the pair production screening operating more efficiently at lower altitude.
In the OG model a charge-deficient region forms in the outer magnetosphere above the null charge surface where
 a charge-separated flow is formed. The induced electric field accelerates pairs radiating 
$\gamma$-rays in a direction tangent to the ${\bf B}$ lines. The $\gamma$-ray 
photons interact with thermal X-rays from the NS surface to produce pairs on field lines interior to the last open field line. 
The pair formation surface screening the electric field defines the interior surface of the gap.

More than 2000 pulsars are listed in the ATNF database \citep{mhth05}, most of which were first observed at radio wavelength. 
We employ the following ten selected pulsar radio surveys in this study:
Molonglo2  \citep{mlt+78}, 
Green Bank 2 \& 3   \citep{dtws85,stwd85}, 
Parkes 2 (70 cm)  \citep{lml+98}, 
Arecibo 2 \& 3  \citep{sstd86,nft95}, 
Parkes 1  \citep{jlm+92}, 
Jodrell Bank 2 \citep{cl86}, 
Parkes Multi-beam  \citep{mlc+01} 
and the extended Swinburne surveys  \citep{ebsb01,jbo+09}.
For these, the survey parameters are known with a high accuracy and they cover the largest possible sky surface while minimising 
the overlapping regions. 

The advent of the LAT telescope on the \emph{Fermi} satellite \citep{aaa+09a} led to a drastic increase in the 
number of $\gamma$-ray pulsars. After three years of observations the LAT detected about 106  pulsars, 
more than doubling the number of detections listed in the first pulsar catalog \citep{aaa+10} leading to the discovery of two
well defined $\gamma$-ray pulsar populations consisting of 31 millisecond pulsars, and 75 young or middle aged isolated, normal pulsars. To study and compare the collective properties of the LAT normal isolated pulsars and investigate
the emission mechanisms that best explain the observed emission, we synthesised a pulsar population incorporating four important high-energy radiation gap models.
The  simulation takes into account the axisymmetric structure of our Galaxy and is designed to match the known characteristics 
of the group of older radio pulsar population than the younger group of pulsars sampled in $\gamma$-rays. Four $\gamma$-ray emission gap models 
have been assumed: the previously described Polar Cap (PC), Slot Gap (SG), and Outer Gap (OG), and a variation of 
the OG, hereafter the One Pole Caustic (OPC)  \citep{rw10,wrwj09} that differs from the OG in the energetics. 
We model the radio emission at two different frequencies, 1400 MHz and 400 MHz  \citep{gvh04,hgg04},
comparing simulated radio fluxes with the flux thresholds of existing surveys.

The outline of this paper is as follows.
In Sections \ref{NSchoice} and  \ref{birthAndEvolution}, we describe the neutron star characteristics and evolution.
In Sections \ref{RadioM}, \ref{SG}, and \ref{OG}, we give a brief overview of the radio luminosity computation, $\gamma$-ray gap 
widths, and $\gamma$-ray luminosities computations. Sections \ref{Phase-plot calculation and normalisation} and \ref{Flux calculations}
describe the pulsar light-curve and flux computation. Section \ref{Radio pulsar visibility} reviews the radio and 
$\gamma$-ray pulsar visibility calculations. We present the results in the final Section \ref{PopResults}.

\section{Neutron star characteristics}
\label{NSchoice}
The neutron star mass, radius, and moment of inertia used in this paper have been chosen according to the experimental mass measurements in binary
NS-NS systems, X-ray binaries, and NS-white dwarf binaries shown in Figure 3 of \cite{lp07}.

The assumed NS mass and radius are $M_\mathrm{NS}=1.5~\msun$ and $R_\mathrm{NS}=13$ km. 
The mass value lies between the weighted average and average values of X-ray and white dwarf-NS binaries estimates and, with the
$R_\mathrm{NS}=13$ km, represent, a possible solution for the EOS that describe the NS interior (Figure 2 of \cite{lp07}).

The moment of inertia of a NS is evaluated by Equation 35 of \cite{lp07}.
For the 13 km radius and the 1.5$\msun$ mass of our standard NS, we obtain
$I\sim1.8 \times 10^{38}$ kg m$^2$. Because of the uncertainty on the mass and radius estimates, this value has an uncertainty of about 70\%. 

For each simulated NS we have generated a value of the magnetic obliquity $\alpha$ (angle between the pulsar rotation and magnetic axes) and 
of the observer line of sight $\zeta$ (angle between the pulsar rotation axis and the observer line of sight). 
After the supernova explosion that generates the neutron star, the magnetic axis $\alpha$ has equal probability to point in any direction of a 3 dimensional space.  
This is also true for the observer line of sight direction $\zeta$ with respect to the pulsar rotational axis.
The $\alpha$ and $\zeta$ distributions are isotropic. 

The spin-down power $\dot{E}$ is defined as the rate with which the pulsar loses rotational kinetic energy, as 
\begin{equation}
\label{Edotcomputation}
\dot{E}\equiv 4\pi^2 I \dot{P}P^{-3} = \frac{-\mbox{d}E_\mathrm{rot}}{\mbox{d}t}\sim 7.1 \times 10^{39}\dot{\frac{P}{1 \mathrm{s/s}}}\frac{P}{1 \mathrm{s}}^{-3}~\mathrm{W}.
\end{equation}
The latter equation is based on the NS structure assumptions through the moment
of inertia $I$. Since mass and radius are chosen inside intervals
of allowed values, the $\dot{E}$ estimate is affected by an uncertainty of at least a factor 3.  

The choice of mass, radius, and moment of inertia formulation yields a moment of inertia value that is 1.5 times higher 
than in the ATNF catalog. This helps to reduce the discrepancy found between the simulated and observed $\dot{E}$ 
distributions (Section \ref{SPDWN}), while remaining well within the range of parameters allowed by the binary 
data and equations of state in \cite{lp07}.
The choice of different values for mass and radius would also impact the range of the $P$ and $\dot{P}$ distributions of the
evolved pulsar population (Section \ref{Comparison of the total simulated and observed samples}).

\section{Neutron stars at birth and their evolution}
\label{birthAndEvolution}

We synthesised $\sim 2.7\times10^8$ NSs with mass, radius, and moment of inertia as described in Section \ref{NSchoice},
and assuming a constant birth rate over the last 1 Gyr.
It yields $2.5\times10^6$ isolated ordinary pulsars to the left of the radio death line (see below).
In order to match the observed radio pulsars $P$ and $\dot{P}$ distributions,
an exponential magnetic field decay with a time scale of $2.8\times 10^6$ yr has been assumed  \citep{gvh04}.
The choice of such a short timescale decay is justified by the need to slow 
down the birth population enough to reproduce the characteristics of the observed radio sample. It provides a simple
mathematical solution to a more physical model of the rotational evolution of the NS, yet to be developed.
For our study, since we are dealing with young ordinary pulsars, this choice has been checked not to affect the obtained results.

The radio death line we used is defined as
\begin{equation}
\label{death1}
\log \dot{P}<a + b \log P.
\end{equation}

It is composed by three different segments \citep{sgh07,zhm00}, each one refers to a specific period interval characterised 
by the following $a$ and $b$ values 

\begin{equation}
\begin{gathered}
\label{death2}
P\le15\mathrm{ms}~~~~~~~~~~~~~~~~~~~a=-19.00~~b=0.814\\ 
15\mathrm{ms}< P\le300\mathrm{ms}                    ~~~~a=-17.60~~b=1.370\\
P>300\mathrm{ms}~~~~~~~~~~~~~~~~~~a=-16.69~~b=2.590
\end{gathered}
\end{equation}

\subsection{Birth spinning and magnetic characteristics}
\label{evolution}

\begin{figure}[htbp!]
\begin{center}
\includegraphics[width=0.24\textwidth]{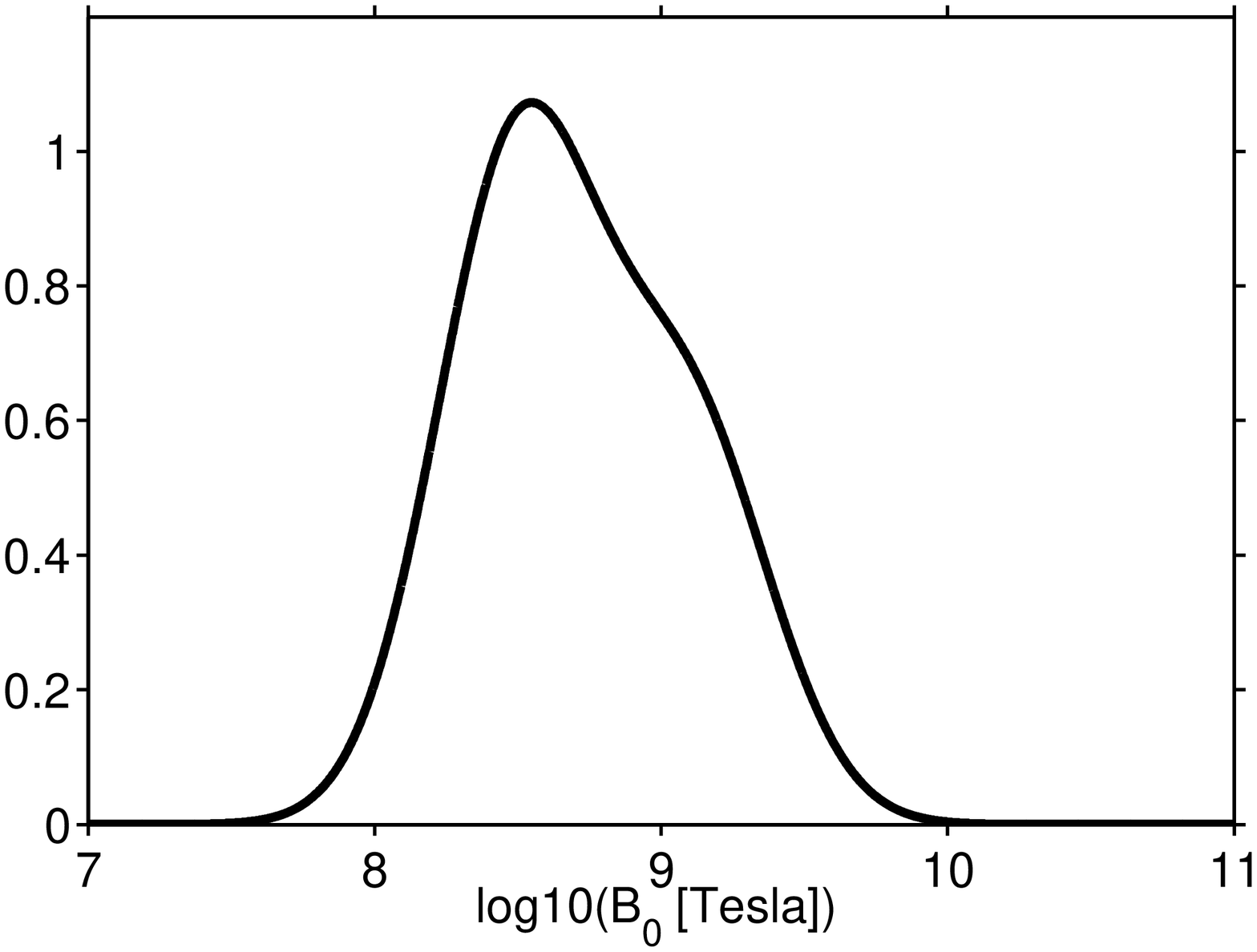}
\includegraphics[width=0.24\textwidth]{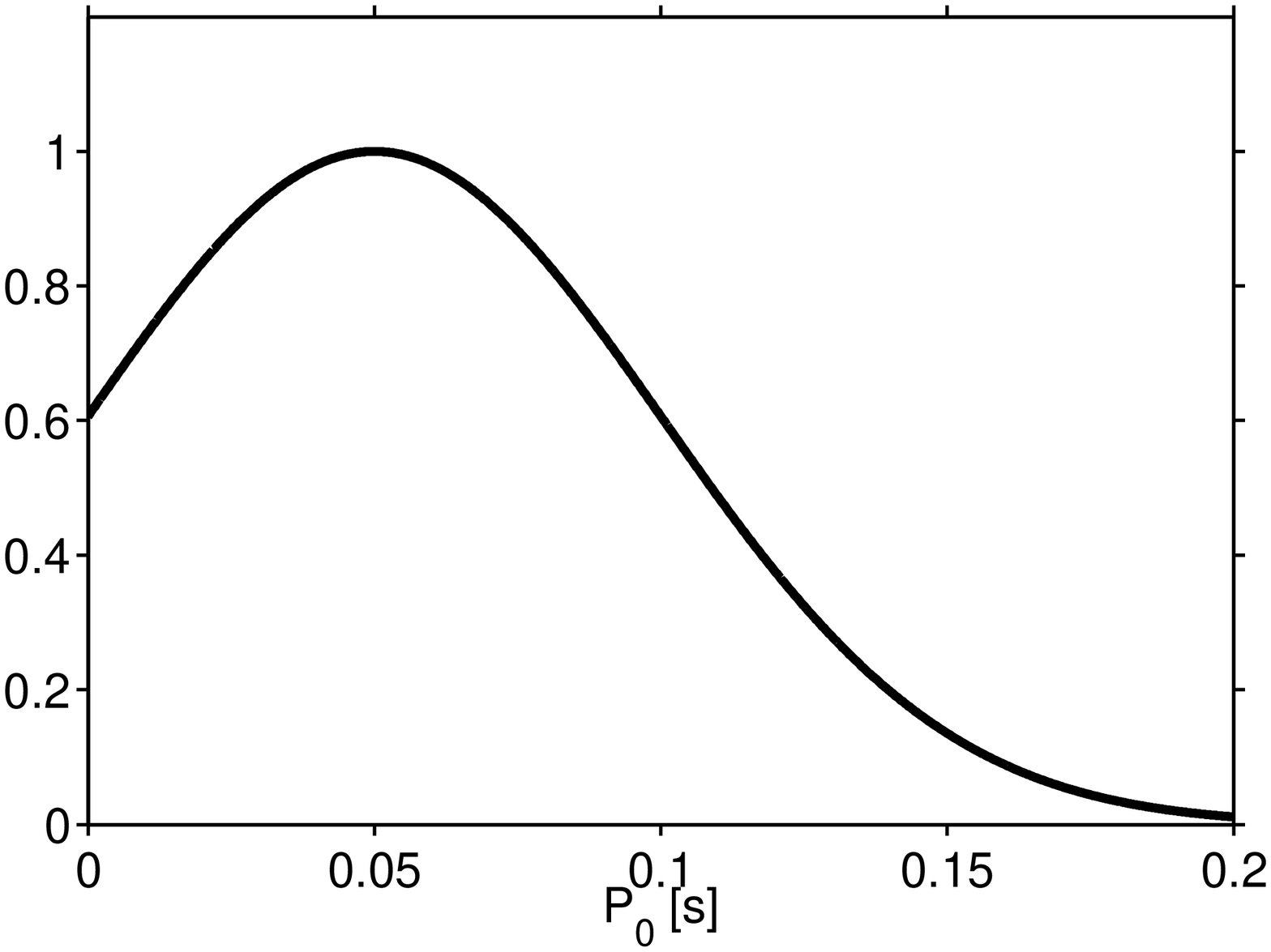}
\caption{\emph{Left:} The assumed surface magnetic field distribution at birth. \emph{Right:} The assumed spin period distribution at birth.}
\label{BirthEvolElectroDistr}
\end{center}
\end{figure}

The distribution of period at birth, ${\bf P_0}$, plotted in the right panel of Figure \ref{BirthEvolElectroDistr}, 
follows a single gaussian of width 50 ms, centred at 50 ms,  and truncated at 0 to avoid negative periods. 
The same distribution was adopted by \cite{wr11} on the basis
of radio luminosity arguments but it differs from the choice of \cite{twc11} who selected the birth period randomly in the
range $20\le P_0 \le 30$ ms.

The magnetic field birth distribution ${\bf B_0}$ shown in the left panel of Figure \ref{BirthEvolElectroDistr} has been
built as the sum of two gaussians in $\log_{10} B_{0}$ [Tesla], both 0.4 in width, respectively centred at 8.5 and 9.1, and with an amplitude ratio of 1:7/12. 
Our choice represents a compromise between that of \cite{wr11}, a single gaussian centred  at 8.65 and width 0.3, and the 
\cite{twc11} one, a single Gaussian centred  at 8.6 and width 0.1. The high-$B_{0}$ Gaussians provide energetic pulsars when evolved. 

Both the ${\bf P_0}$ and ${\bf B_0}$ distributions have been optimized  \emph{a posteriori} to obtain, after evolution, a simulated pulsar sample as close 
as possible to the observed one by minimizing the observed lack of high $\dot{E}$ objects (Section \ref{SPDWN}).
The ${\bf \dot{P}_0}$ birth distribution has been derived from ${\bf P_0}$ and ${\bf B_0}$ by using the equation 
\begin{equation}
\label{spndown}
B_{S}=\left(\frac{3c^3}{8\pi^2}
\frac{I_{NS}}{R_{NS}^6}P\dot{P}\right)^{1/2}.
\end{equation}

This formulation includes no dependence on the magnetic obliquity alpha, as proposed by \cite{rs75} for the 
spinning down of magnetospheres carrying current flows. More recently, \cite{spi06} numerically showed 
for force-free magnetospheres that the spin down of orthogonal rotators is twice that of aligned rotators. In the 
non-ideal case of a magnetosphere accelerating charges to produce pulsed emission the impact of alpha on 
$\dot{E}$ is still under discussion, 
so we chose for this paper the alpha independent prescription of \cite{rs75}. Hereinafter, all 
luminosities are given as a function of $\dot{E}$ to judge how the uncertainty on the spin-down rate  propagates.

\subsection{Birth location and velocity in the Galactic plane}
\label{Birth distribution in the Galactic plane}

To follow the dynamical evolution of the pulsars in the Galactic reference frame, we synthesised their
birth position $x$, $y$, $z$ in the Galaxy as well as their kick velocity and direction.

\begin{figure}[htbp!]
\begin{center}
\includegraphics[width=0.49\textwidth]{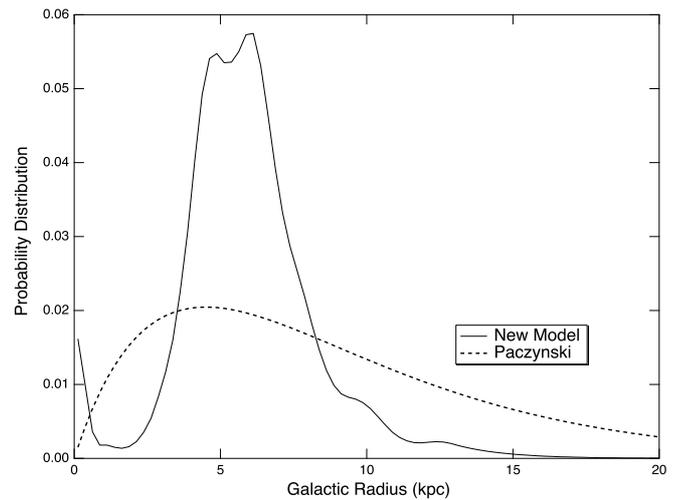}
\caption{Surface density of the new born neutron stars. The dashed curve represents the Paczy\'nski distribution  \citep{pac90}, while the adopted one
following the distribution of radio HII regions, is shown as a solid curve. Both curves are normalised.}
\label{BirthEvolDistDistr}
\end{center}
\end{figure}

We emulated the distribution of the NS progenitors by using
the location of the HII regions in the Galaxy. The latter are good tracers of massive stars because O-B stars are required to ionise the hydrogen bubbles.
For the number density of pulsars at birth as a function of Galactocentric distance, we used the HII region profile recently obtained by 
\cite{babr10} from radio observations that can probe HII regions to large distance with little absorption.
Figure \ref{BirthEvolDistDistr} shows the comparison between the \cite{pac90} birth distribution used in earlier publications (\cite{gvh04,twc11})
and the HII region profile used here. Both distributions extend from the Galactic centre up to 40 kpc and have been 
normalised to 1 over the Galaxy.

We assume that all the NSs are born in the Galactic disk, with an exponential thin disk distribution with a scale height of 50 pc
(consistent with \cite{wr11} that adopted an exponential thin disk with a 75 pc scale height) and with a surface density distribution defined in Figure  
\ref{BirthEvolDistDistr}. Due to the large supernova kick velocity, the neutron stars evolve quickly out of the plane of the Galaxy.
The assumed kick velocity distribution is the same as in \cite{wr11} and \cite{twc11}.
It is described by a Maxwellian distribution, characterised by a mean of 400 km s$^{-1}$ and a width of 256 km s$^{-1}$  \citep{hllk05}.

\subsection{Evolution}

We have evolved both the pulsar position and velocity in the Galactic gravitational potential (described in \cite{pac90} and  
\cite{gob+02} Equations 17, 18, and 19, and \cite{twc11}).
The spin characteristics have been evolved to the present time assuming a magnetic dipole.

\begin{figure}[htbp!]
\begin{center}
\includegraphics[width=0.48\textwidth]{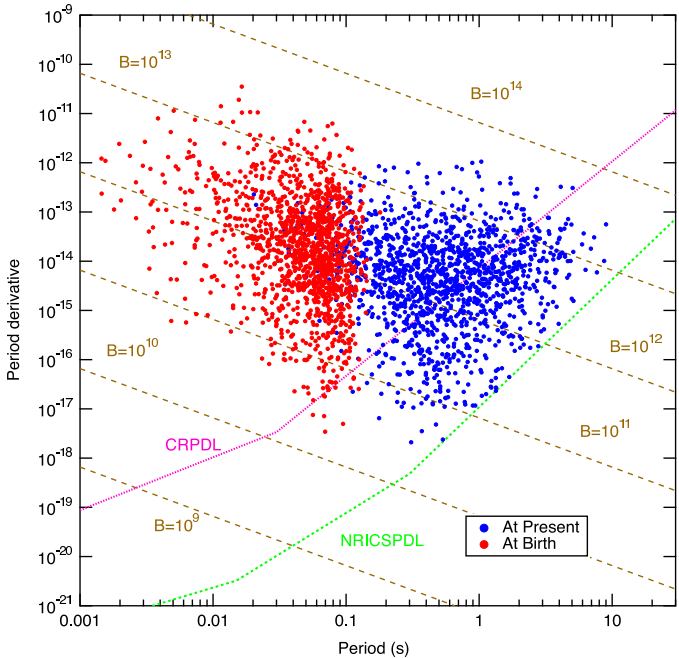}
\caption{$P$-$\dot{P}$ diagram of the pulsar population  at birth (in red), and the population evolved to the present (blue).}
\label{evolPopFig}
\end{center}
\end{figure}

The simulated pulsar population at birth is shown, in red, in the ${\bf P}$-${\bf \dot{P}}$ diagram of Figure \ref{evolPopFig}.
Following  \cite{gob+02}, by knowing the analytical expression for $B(t)=f(B_0,t)$, it is possible to follow the evolution of the spin parameters from the birth time
$t_0$ to the present time $t_p$. The magnetic decay is described by  
\begin{equation}
B(t)=B_\mathrm{0,8} e^{-t/\tau_\mathrm{D}}
\end{equation}
where $\tau_\mathrm{D}$ = 2.8 Myr is the decay timescale, and $B_\mathrm{0,8}$ is the birth magnetic field in units of $10^{8}$ Tesla.

Assuming magnetic dipole spin-down and initial period $P_0$, the period and the period first time derivative at the present time can be obtained 
from Equations 7 \& 8 of \cite{gob+02}. The simulated pulsar population after evolution is shown, in blue, in Figure \ref{evolPopFig}.

\section{Radio emission model}
\label{RadioM}

After evolving the neutron stars in the Galactic frame, values of the radio dispersion measure (DM), and the radio scattering
measure (SM), are assigned to each star using the NE2001 model  \citep{cl01}.  
The sky temperature at 408 MHz ($T_\mathrm{sky,408}$) for each star is obtained using the all-sky map from the study of \cite{hssw82}.

The empirical radio emission model we have implemented in our simulations follows the work of  
\cite{gvh04} and  \cite{hgg04}. We assume that the radio beam is composed of a core component 
originating relatively near the neutron star surface and a conical component radiated at higher altitude, both centered on the 
magnetic axis in the co-rotating frame. The adopted form of this model is similar to that proposed by \cite{acc02},
based on the work of \cite{ran83} and \cite{kj03} and modified to include frequency dependence 
by \cite{gvh04}. The total flux at a given frequency from the two components seen at angle $\theta$ to the magnetic field axis is
\begin{equation}
\label{eq:Stheta}
S(\theta, \nu ) = F_\mathrm{core} e^{ - \theta ^2 /\rho _\mathrm{core}^2 }  + 
F_\mathrm{cone} e^{ - (\theta  - \bar \theta )^2 /\omega _e^2 } 
\end{equation}
where 
\begin{equation}
\label{eq:Fi}
F_i(\nu) = {-(1+\alpha_i) \over \nu}\left({\nu\over 50 {MHz}}\right)^{\alpha_i+1}{L_i\over \Omega_i D^2}.
\end{equation}
The index $i$ refers to the core or cone, $\alpha_i$ is the spectral index of the total angle and frequency integrated flux for each component, $L_i$ 
is the component luminosity, and $D$ is the distance to the pulsar.  
The total solid angles of the Gaussian beams describing the core and cone components are
\begin{gather}
\Omega_\mathrm{core}=\pi\rho_\mathrm{core}^2\\ 
\Omega_\mathrm{cone}= 2\pi^{3/2}w_e \bar \theta 
\end{gather}
where the latter can be written as
\begin{equation}
\Omega_\mathrm{cone}= \omega_\mathrm{cone}\left(\nu_\mathrm{GHz}\right)^{-0.26}
\end{equation}
where $\nu_\mathrm{GHz}$ is the frequency expressed in Giga-Hertz. 
The factor $\omega_\mathrm{cone}$ represents the
portion of $\Omega_\mathrm{cone}$ that is independent of the frequency and used later in Equation (\ref{eq:rhocone}).
The width of the Gaussian describing the core beam is
\begin{equation}
\label{eq:rhocore}
\rho _\mathrm{core}  = 1.5^{\circ} \left(\frac{P}{1\mathrm{s}}\right)^{ - 0.5} 
\end{equation}
where $P$ is the pulsar period in seconds.  
The annulus and width of the cone beam are
\begin{equation}
\label{eq:thetabar}
\bar \theta  =  (1-2.63\,\delta_w) \rho_\mathrm{cone}
\end{equation}
\begin{equation} 
\label{eq:widann}
w_e = \delta_w \rho_\mathrm{cone}
\end{equation}
where $\delta_w = 0.18$ \citep{gsg+06}, and
\begin{equation}  
\label{eq:rhocone}
\rho_\mathrm{cone} = 1.24^\circ r_\mathrm{KG}^{0.5} \left(\frac{P}{1\mathrm{s}}\right)^{ - 0.5}
\end{equation}
is the radius of the open field volume at the emission altitude derived by \cite{kj03}, and
\begin{equation} 
\label{eq:rKG}
r_\mathrm{KG}  \approx 40\, \left({\dot P\over 10^{ - 15}{s\,s^{-1}}}\right)^{0.07} \left(\frac{P}{1\mathrm{s}}\right)^{0.3} \left(\nu_\mathrm{GHz}\right)^{ - 0.26} 
\end{equation}
$r_\mathrm{KG}$ is in units of stellar radius.
The ratio of the core-to-cone peak flux $r$ is expressed as 
\begin{equation}
\label{r_first}
r=r_1\left(\frac{\nu}{\nu_1}\right)^{\alpha_\mathrm{core}-\alpha_\mathrm{cone}-0.26}
\end{equation}
and requires $\alpha_\mathrm{core}-\alpha_\mathrm{cone}-0.26=0.9$, $\alpha_\mathrm{core}-\alpha_\mathrm{cone}=-0.64$ where
$\nu_1 = 1\mathrm{MHz}$.
\cite{gsg+06}, who carried out a study of 20 pulsars having three peaks in their average-pulse profiles
at frequencies 400, 600 and 1400 MHz, found a core-to-cone peak-flux ratio 
\begin{equation}  
\label{eq:Gratio}
r = \frac{{F_\mathrm{core} }}{{F_\mathrm{cone} }} = \left\{ {\begin{array}{*{20}c}
   {10^{4.1}\left(\frac{P}{1\mathrm{s}}\right)^{1.3} \nu_\mathrm{GHz}^{ -0.9}, ~~~~~~~~~~~~ P < 0.7\mathrm{s}}  \\
   \\
   {10^{3.3}\left(\frac{P}{1\mathrm{s}}\right)^{ - 1.8}  \nu_\mathrm{GHz}^{ -0.9}, ~~~~~~~~~~~~ P > 0.7\mathrm{s}}.  \\
\end{array}} \right.
\end{equation}
It is consistent with the ratio of \cite{acc02} at periods above about 1 s, and predicts that pulsars with 
$P\ \lsimeq\  0.05$ s are cone dominated.
Such a picture is supported by the study of \cite{cmk01} who measured the polarisation of a number of pulsars
younger than 100 kyr, finding that they possess a high degree of linear polarisation and very small circular
polarisation, typical of cone beams.
The luminosities of the core and cone components are
\begin{equation}  
\label{eq:Lcc}
L_\mathrm{cone} = {L_\mathrm{radio}\over 1+(1/r_0)},  ~~~~~~ L_\mathrm{core} = {L_\mathrm{radio}\over 1+r_0},
\end{equation}
where
\begin{equation}  
\label{eq:r0}
r_0 = \frac{1}{r_1} \left(\frac{1+\alpha_\mathrm{core}}{1+\alpha_\mathrm{cone}}\right) \left(\frac{\omega_\mathrm{cone}}{\Omega_\mathrm{core}}\right)
\nu_{1000}^{0.26} ~\nu_{50}^{\alpha_\mathrm{cone} - \alpha_\mathrm{core}}~\nu_{1}^{-0.9},
\end{equation}
$r_1$ is evaluated from Equations \ref{r_first} and \ref{eq:Gratio}, $\alpha_\mathrm{core} = -1.96$, $\alpha_\mathrm{cone} = -1.32$, 
$\nu_{1000}=1000$ MHz and $\nu_{50}=50$ MHz, and
\begin{equation}  
\label{eq:Lradio}
L_\mathrm{radio}  = 2.805 \times 10^{9}\left(\frac{P}{1\mathrm{s}}\right)^{ - 1} \left(\frac{\dot{P}}{1 \mathrm{s/s}}\right)^{0.35}\, \mathrm{mJy~kpc^2~ MHz} 
\end{equation}
as modified from \cite{acc02}.

\section{\emph{PC \& SG}: particle luminosity and gap width}
\label{SG}

\subsection{Particle luminosity}
\label{PCSGenergetic}
The \emph{slot gap} region is defined between the last open magnetic field line, defined by the colatitude 
$\theta_0 \simeq (\Omega R/c f(1))^{1/2}$, and the magnetic field line with a colatitude 
value $(1-w_\mathrm{SG})$ where $w_\mathrm{SG}$ is the SG width expressed in units of 
the dimensionless colatitude of a PC magnetic field line, $\xi \equiv \theta/ \theta_0$.

It is possible to define the emission component from the PC pair cascades along the PFF that forms on the inside 
surface of the SG by assuming that mono-energetic radiation is emitted tangent to field lines  \citep{mh03}.
$w_\mathrm{SG}$ is a function of pulsar period, $P$, and surface magnetic field, $B_\mathrm{NS}$ 
 \citep{mh03}.
The photons from the polar cap pair cascade 
are emitted in the region defined by $1-w_\mathrm{SG}$.
The luminosity of the SG  from each pole is
\begin{equation} \label{eqn:Lpole}
L_\mathrm{e}^{SG} =  \alpha c \int_0^{2\pi}  d\phi_\mathrm{PC}  \int_\mathrm{\theta_0(1-w_\mathrm{SG})}^{\theta_0}
\rho(\xi,\eta) \Phi(\xi,\eta) r^2 \sin\theta d\theta
\end{equation}
where  $\rho(\xi,\eta)$ and $\Phi(\xi,\eta)$ are the primary charge density
and potential as a function of the emission altitude $\eta \equiv r/R_\mathrm{NS}$ and of $\xi$, in units of NS radius, and 
$\phi_\mathrm{PC}$ is the magnetic azimuthal angle. 
Using the expressions for $\Phi$ and for $\rho$ from  \cite{mh03}, the PC particle luminosity (from the low-altitude SG) is
\begin{equation}
\begin{gathered} 
\label{eqn:Lglow}
L_\mathrm{e}^{PC} =  \dot{E} w_\mathrm{SG}^3  (1 - {w_\mathrm{SG}\over 2})[\kappa (1-\kappa)(1-{1\over \eta^3})\cos^2\alpha + {9\over 8} \theta_0^2 \times \\
\times  (1 - {w_\mathrm{SG} + {3 \over 10} w_\mathrm{SG}^2 }) H^2(1) \left[{H(\eta)\over H(1)}\sqrt{\eta{f(1)\over f(\eta)}}-1\right]\sin^2\alpha]
\end{gathered}
\end{equation}
where $\dot{E} = \Omega^4 B_\mathrm{NS}^2 R_\mathrm{NS}^6/ (6c^3 f(1)^2)$ is the spin-down power, 
$\kappa = 0.15 I_\mathrm{38}/R_6^3$, $I_\mathrm{38}$ is 
the NS moment of inertia in unit of 10$^{38}$ kg m$^2$, $R_6$ is the NS radius $R_\mathrm{NS}$ in unit of $10^6$ m, 
$H(\eta)$ is a relativistic correction factor of order 1, $f(\eta)$ is the correction factor for the dipole component 
of the magnetic field in a Schwarzschild metric, and $\alpha$ is the pulsar obliquity 
 \citep{mt92,hm98}. 

Using the equations for $\Phi$ and for $\rho$ from  \cite{mh04a},
the high-altitude SG particle luminosity from each pole can also be determined from Equation (\ref{eqn:Lpole}) as
\begin{equation}
\begin{gathered} 
\label{eqn:Lghigh}
L_\mathrm{e}^{SG} = \dot {E} w_\mathrm{SG}^3  \beta \left(1 - {w_\mathrm{SG}\over 2}\right) \mathcal{A} +{1\over 2}\left(1-w_\mathrm{SG} + {3 \over 10} w_\mathrm{SG}^2\right) \mathcal{B} 
\end{gathered}
\end{equation}
where $\beta = (1 - 3\eta/4\eta_\mathrm{lc})^{1/2}$ and $\eta_\mathrm{lc} = r_\mathrm{lc}/R_\mathrm{NS} = c/\Omega R_\mathrm{NS}$.  
The parameters $\mathcal{A}$ and $\mathcal{B}$, are defined as:
$$
\mathcal{A}= -\left(1 - {\kappa\over \eta^3}\right)\left[\kappa \left(\beta-{1 \over \eta_\mathrm{c}^3}\right)+1-\beta\right]\left(1+{\eta\over\eta_\mathrm{LC}}\right)\cos^2\alpha 
$$
$$
\mathcal{B}= -{9 \over 4}H(1)H(\eta)\left[{H(\eta_\mathrm{c}) \over H(1)}\sqrt{\eta_\mathrm{c}{f(1)\over f(\eta_\mathrm{c})}}-\beta\right]\left[\eta{f(1)\over f(\eta)}\right]^{1/2}\theta_0^2\sin^2\alpha 
$$
where $\eta_\mathrm{LC}=R_\mathrm{LC}/R_\mathrm{NS}$, $\beta = (1 - 0.75\eta/\eta_\mathrm{LC})$, and $\eta_\mathrm{c}=1.3$.
According to \cite{mh04a}, the energies of the primary electrons in the SG quickly become radiation-reaction limited, with the rate of 
acceleration balancing the curvature radiation loss rate, resulting in 100\% efficiency with $L_{\gamma} = L_\mathrm{e}^{SG}$  
in this case.

\subsection{Gap width}
\label{SGwidth}

In the SG model, the width of the slot gap $w_\mathrm{SG}$ can be estimated as the magnetic colatitude where the variation in height of the 
curvature radiation PFF $z_0$ (in units of stellar radius) becomes comparable to a fraction $\lambda$ of the stellar radius $R_\mathrm{NS}$ \citep{mh03}:
\begin{equation}
\label{smalllambda}
\left(\frac{\partial z_0}{\partial\xi}\right)_\mathrm{\xi=\xi_\mathrm{SG}}\sim\lambda.
\end{equation}
In Equation  \ref{smalllambda}, $z_0$ represents the dimensionless altitude, above the polar cap, of pair formation due to
curvature radiation
\begin{equation}
\label{zeta0}
z_0 = 7\times10^{-2}\frac{P_\mathrm{0.1}^{7/4}}{B_\mathrm{8}I_\mathrm{38}^{3/4}}\frac{1}{\xi^{1/2}(1-\xi^2)^{3/4}}
\end{equation}
where $P_\mathrm{0.1}=P/0.1$ s, and $B_\mathrm{8}$ is the magnetic field in units of 10$^{8}$ Tesla.
By solving numerically Equation  \ref{smalllambda} with $z_0$ defined in Equation \ref{zeta0}, one obtains $\xi_\mathrm{SG}$
for a specific pulsar. The  $w_\mathrm{SG}$ gap width value is then obtained as
\begin{equation}
\label{SGgapW}
w_\mathrm{SG}= 1-\xi_\mathrm{SG}.
\end{equation}

The $\lambda$ parameter constrains both the energetics and emission pattern of the SG emission and
impacts both the SG and PC luminosity (Section \ref{Luminosity}) and light-curve sharpness 
and shape. For large $\lambda$ values the light-curve peaks appear too sharp compared with the observed LAT
profiles, therefore the slot gap is too narrow and not energetic enough to explain the observed LAT fluxes. 
On the other hand, smaller $\lambda$ values imply wider slot gaps, sufficiently luminous when 
compared with the observations, but light-curve peaks too broad when compared with the observed ones.

As a result we compromise between the narrow
light-curve structures and the $\gamma$-ray luminosity through a reasonable radiation efficiency $\epsilon_\mathrm{\gamma}$. We tried 
two different approaches to constrain $\lambda$: one based on energetic arguments, and one based 
on the optimisation of the expected light-curves for some of the LAT pulsars.

Since $L_\mathrm{\gamma}$ scales as $w_\mathrm{SG} ^3 \times \dot{E}$ and since we want the luminosity to be close to 
$L_\mathrm{\gamma} \propto \dot{E}^{1/2}$ (\cite{aaa+10}, First pulsar catalog) we need to have
$w_\mathrm{SG} \propto \dot{E}^{-1/6}$ to obtain a reasonable agreement with the LAT data. The luminosity remains close to
$\dot{E}^{0.5}$ for all the tested $\lambda$ values, but favours $\lambda < 0.4$ to explain the bright LAT pulsars.
A good compromise is found for $\lambda=0.35$. 

One can calculate numerically the pair formation front shape for the P and B values of some of the best known pulsars, Crab, Vela, CTA1, and Geminga, 
to obtain an approximate $w_\mathrm{SG}$ value (\cite{mh03}). The results  yield $w_\mathrm{SG,Crab}$=0.03, $w_\mathrm{SG,Vela}$=0.1, 
$w_\mathrm{SG,CTA1}$=0.16, $w_\mathrm{SG,Geminga}$=0.3 for $\lambda$ values between 0.02-0.6. 

In order to investigate how the pulsar light-curve changes as a function of $\lambda$, we performed a fit to some LAT light-curves with the SG 
phase-plots (see Section 7), evaluated for  a set of $w_\mathrm{SG}$ values obtained for different $\lambda$ values. 
We studied the behaviour of the best-fit likelihood value as a function of  $\lambda$ for Vela, Crab, 
J1028-5820, J1048-5832, J2021+3651, and J2229+6114. 
For all the studied pulsars, in the $\lambda<0.4$ range that allows bright enough pulsars, the maximum-likelihood 
value presents a local maximum between 0.2 and 0.4. This result is consistent with the $\lambda$ estimate obtained from the luminosity study and the 
pair formation front evaluation from Crab, Vela, CTA1, and Geminga.

In this paper, we set $\lambda$=0.35. This value reproduces the bulk of the light-curve structure of the observed 
objects and yields a reasonable estimate of the SG luminosity. In choosing $\lambda$, we put more emphasis on matching the 
sharply peaked light-curves often recorded by the LAT than on achieving bright luminosities. This selection of $\lambda$ was driven by the need 
to preserve realistic beam patterns (thus their brightness and visibility across the beam) and is a key assumption that contains the results of our population studies.
We mitigated the low SG gamma-ray luminosities by using a radiative efficiency greater than 1 as discussed in section 
\ref{Flxcomp}.

\section{OG and OPC: particle luminosity and gap width}
\label{OG}

\subsection{Gap width}
\label{OGOPCw}

To determine the gap width, we consider two different 
prescriptions. The first one \citep{wrwj09} simply assumes that the gap width is equal to the $\gamma$-ray radiation 
efficiency. Because of the $L_\mathrm{\gamma}\propto \dot{E}^{0.5}$ relation observed in the first LAT pulsar 
catalog (\cite{aaa+10}) , the gap width should follow as
\begin{equation}
\label{OPCwidth}
w_\mathrm{OPC}=\left(\frac{10^{26} \mathrm{W}}{ \dot{E}}\right)^{1/2}.
\end{equation}

Our second prescription follows the calculations of the self sustaining  
OG model presented in \cite{zcjl04}. In this formulation, the X-rays that trigger the pair production come from the 
bombardment of the NS surface by the full return current from the OG. 
The bright X-ray luminosity allows active OGs and $\gamma$-ray emission for many old pulsars.
The outer gap width across magnetic field 
lines is determined by computing the location of the pair formation surface.
From \cite{ks98}, the polar angle $\theta_\mathrm{c}$ corresponding to the magnetic field line tangent to the light cylinder is:
\begin{equation}
\tan\theta_\mathrm{c} = -\frac{3}{4\tan\alpha}\left[1+(1+\frac{8}{9}\tan^2 \alpha)^{0.5}\right]
\end{equation}
with the light cylinder radius given by
\begin{equation}
R_\mathrm{L}=\frac{r_\mathrm{c}}{\sin\theta_\mathrm{c}}.
\end{equation}
Here $r_\mathrm{c}$ is the distance between the pulsar and the point where the light cylinder is tangent to the magnetic field line
corresponding to $\theta_\mathrm{c}$.
The lower boundary of the outer gap is estimated from the null-charge surface, ${\bf \Omega} \cdot {\bf B}=0$, that in two dimensions 
is described by $(r_\mathrm{in},\theta_\mathrm{in})$. 
By definition, the polar angle at the inner edge of the outer gap is
\begin{equation}
\tan \theta_\mathrm{in}=\frac{1}{2}\left(3\tan\alpha + \sqrt{9\tan^2 \alpha + 8}\right).
\end{equation}
The computation of $r_\mathrm{in}$ is obtained from the relation
\begin{equation}
\frac{\sin^2(\theta-\alpha)}{r}=\frac{\sin^2(\theta_\mathrm{c}-\alpha)}{r_\mathrm{c}}
\end{equation}
which results in 
\begin{equation}
r_\mathrm{in}=\frac{R_L \sin^2 (\theta_\mathrm{in} - \alpha)}{\sin{\theta_\mathrm{c}}\sin^2 (\theta_\mathrm{c} - \alpha)}.
\end{equation}
The relation that defines the fractional OG size in this case is:
\begin{equation}
\label{OGwidth}
w_\mathrm{OG}=5.2 B_\mathrm{8}^{-4/7} P^{26/21} R_\mathrm{6}^{-10/7}G(\langle r \rangle,\alpha) = f(\langle r \rangle,\alpha).
\end{equation}
where $G(r,\alpha)$ is a factor that is numerically solved for each pulsar by taking into account the average  distance $\langle r \rangle$ 
at which primary $\gamma$-rays are produced and along which magnetic field line they pair produce when they interact with 
an X-ray coming radially from the NS surface. The average distance $\langle r \rangle$ is defined in \cite{zcjl04} as
\begin{equation}
\langle r \rangle =\frac{\int _\mathrm{r_{in}}^\mathrm{r_{max}}f(\langle r \rangle,\alpha)rdr}{\int _\mathrm{r_{in}}^\mathrm{r_{max}}f(\langle r \rangle,\alpha)dr}
\end{equation}
where $r_\mathrm{max}=\mathrm{min}(r_\mathrm{c},r_\mathrm{b})$ and $r_\mathrm{b}$ is the radius at which the fractional size of the outer gap stops to grow:
$f(r_\mathrm{b},\alpha$)=1.

A full calculation of the width of the OG radiating layer is complicated \citep{h06,h08} since both the screening and the radiation occur in the same location.
For this paper, we assume that this is an infinitely thin layer on the gap inner edge and that it is uniform in azimuth around the magnetic axis whereas 
\cite{h06} finds a significant azimuthal dependence.

\subsection{Particle luminosities}

The assumed gap width $w_\mathrm{OPC}$ defined in section \ref{OGOPCw} is not based on any physical prescription and is very different from the usual dependence 
luminosity $\propto$(gap width)$^3$ (both SG and OG) based on the electrodynamics.

The gap width luminosity is evaluated as
\begin{equation}
\label{eqn:LOPC}
L_\mathrm{\gamma,OPC} = w_\mathrm{OPC} \dot{E}_\mathrm{sd}.
\end{equation}

In the OG case, from \cite{zcjl04} and previous papers dealing with OG gap geometry, the total $\gamma$-ray luminosity is 
\begin{equation}
\label{eqn:LOG}
L_\mathrm{\gamma,OG} = w_\mathrm{OG}^3(\langle r \rangle ) \dot{E}_\mathrm{sd}
\end{equation}
where $w_\mathrm{\rm OG}$ is the fractional width of the gap at the average gap radius $\langle r \rangle$.

\section{Phase-plot and light-curve generation}
\label{Phase-plot calculation and normalisation}

\subsection{Assumptions and photon distributions}
\label{PhPlotDefinition}

To provide the $\gamma$-ray emission pattern 
for each emission mechanism, we used the geometric emission model from  \cite{dhr04} based on the following assumptions: \emph{(i) the 
pulsar magnetic field is dipolar and swept-back by the pulsar rotation (retarded potentials) \citep{deu55}}, 
\emph{(ii) the $\gamma$-ray emission is tangent to the magnetic 
field line and oriented in the direction of the accelerated electron velocity in the star frame}. Relativistic aberration and time of flight delays are taken into account. 

\begin{figure}[htbp!] 
   \centering
 \includegraphics[width=0.49\textwidth]{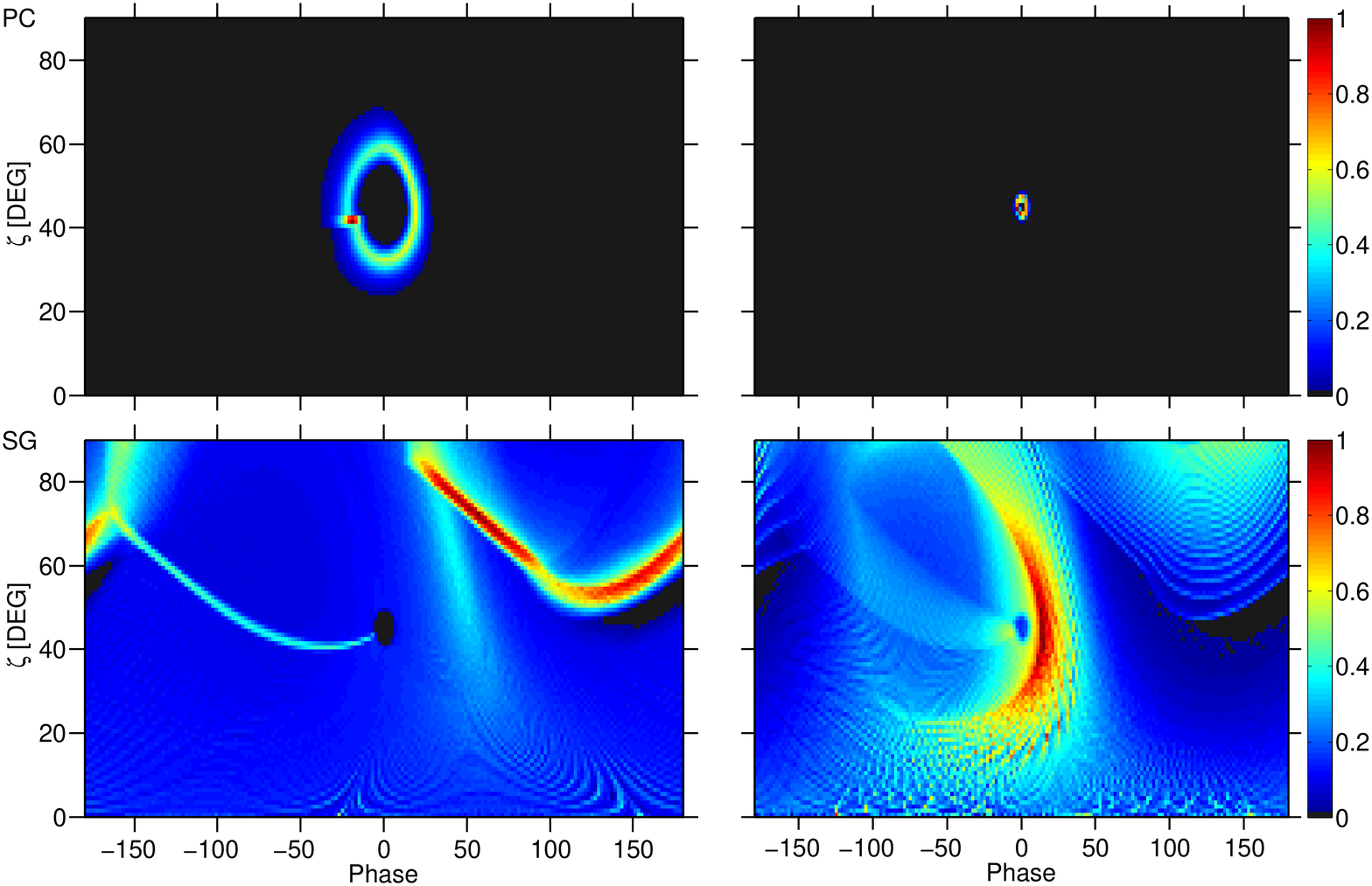} 
  \includegraphics[width=0.49\textwidth]{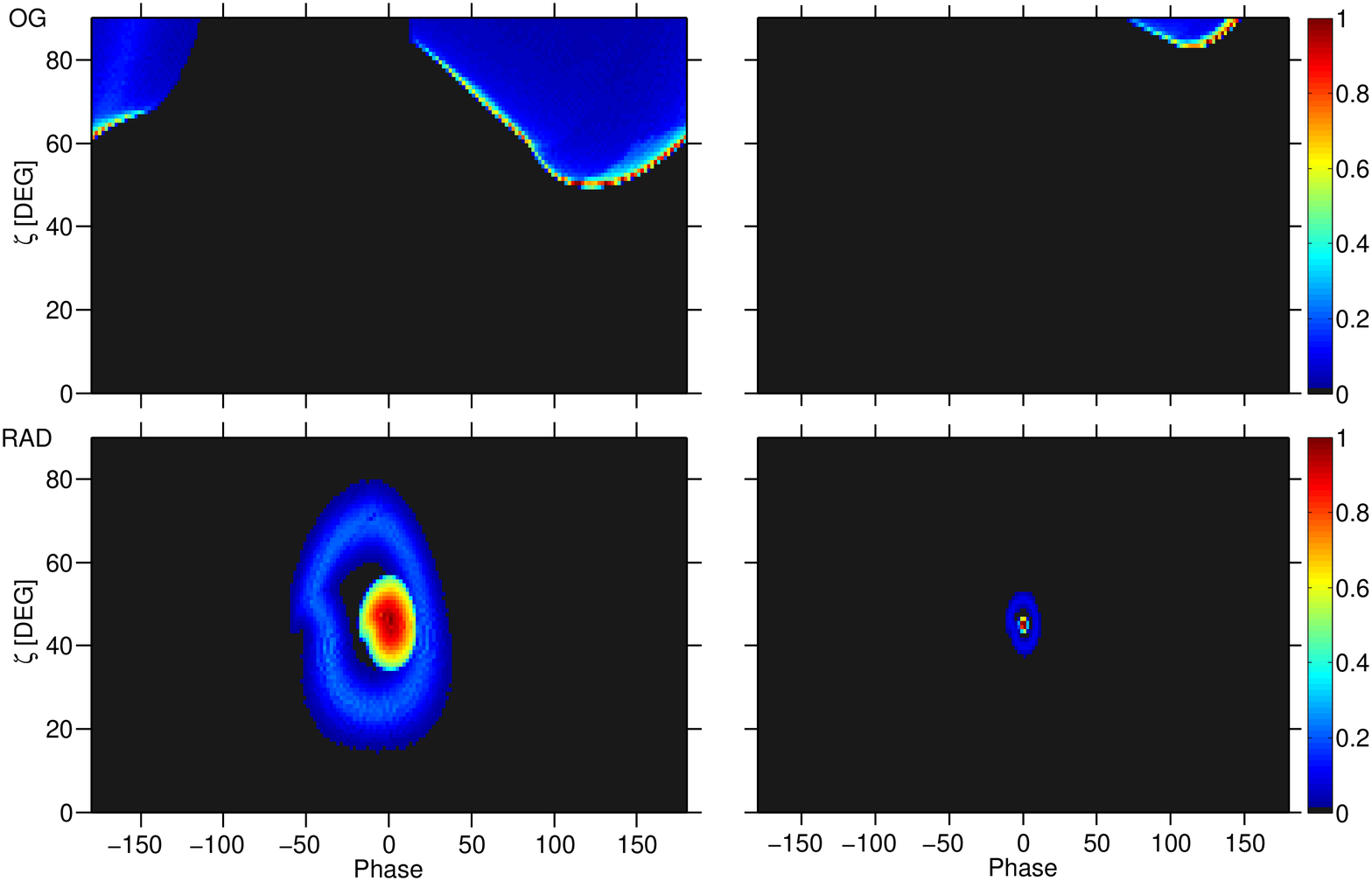} 
   \caption{The top to bottom panels illustrate the $\gamma$-ray emission pattern obtained for a young (left) and old (right)
   	         pulsar, respectively for the PC, SG, OG/OPC, and radio (core plus cone) models. For the 
	         PC and Radio models, the time evolution is obtained for $B=10^{8}$ T and period increasing from 30 to 1000 ms.
	         In the SG and OG/OPC models, the time evolution is obtained for a gap width $w_\mathrm{SG}$ increasing from 0.04 to 0.5 and 
	         $w_\mathrm{OG/OPC}$ increasing from 0.01 to 0.4 respectively.
	         All the plots are given for an obliquity $\alpha=45^{\circ}$.}
   \label{phpltExamples}
\end{figure}

In the computation of the emission pattern, the first step consists in localising the position of the magnetic field line from which the radiation 
is produced. Each field line is then divided into segments and for each segment the tangent direction and height with respect to 
the NS surface is evaluated. Since the emission gap is located, for each model, in a different magnetospheric region, the emission patterns are obtained by 
selecting the \emph{segments} corresponding to the gap position in each model. The $\gamma$-ray emission is assumed to be uniform along the field lines 
in the co-rotating frame. The phase $\phi$ of the pulsar emission is defined by the direction of the emitted photons with respect to the corotating frame.
The result of this computation is the two-dimensional emission pattern in the plane ($\phi$,$\zeta$), shown for each implemented emission model
in Figure \ref{phpltExamples}, which we refer to as a {\bf phase-plot}. Figure \ref{phpltExamples} also shows the evolution of the emission pattern as a pulsar ages.

To incorporate the radio emission geometry we modulate the field lines with the flux 
$S(\theta,\nu)$ given by Equation \ref{eq:Stheta}.  The differential flux radiated from a 
bundle of field lines centred at open-volume coordinates ($r$, $l$) (see \cite{dhr04}) is
\begin{equation}  
\label{eq:L_r}
dS(\theta, \nu) = S_i(\theta, \nu)\sin\theta \, d_{\rm ovc}\, \theta_0 \,r^\mathrm{max}\, {2\pi\over N_\mathrm{l}} \,d\nu
\end{equation}
where $N_l$ is the number of azimuthal divisions of each polar cap ring, $r^\mathrm{min}=0.1$ and $r^\mathrm{max}=1.0$ 
are the lower and upper boundary of the emission region, and $d_{\rm ovc}$ (= 0.1 for the radio phase-plot) is the spacing of the 
rings on the PC in open volume coordinates. For the SG model $d_{\rm ovc}$ is adjusted to have 20 rings within the gap.
The flux is assumed to be emitted at an altitude of $1.8 R_\mathrm{NS}$ for the core component and at an altitude given by Equation \ref{eq:rKG} 
for the cone component.

In the PC model, the emission profile in colatitude is infinitely thin along the inner edge of the slot gap (with  $w_\mathrm{SG}$ defined in section 
\ref{SGwidth}) and the intensity of the emission along the field line, $I_\mathrm{PC}$, exponentially decreases from the polar cap edges to the magnetic pole
\begin{equation}
I_\mathrm{PC}=
\left\{ {\begin{array}{*{20}c}
   \exp\left(\frac{s-s_f}{\sigma_\mathrm{in}}\right),~~~~s \le s_\mathrm{f} \\
   \\
   \exp\left(\frac{s-s_f}{\sigma_\mathrm{out}}\right),~~~~s > s_\mathrm{f}.  \\
\end{array}} \right.
\end{equation}
Here $s_\mathrm{f} = 2.5, \sigma_\mathrm{in} = 1.0, \sigma_\mathrm{out} = 2.0$, and $s$ is the curvilinear distance along the field line starting from the NS surface.
Both $s$ and $s_\mathrm{f}$ are in unit of $R_\mathrm{NS}$.

To model the emission component from primary electrons in the SG model, we assume that radiation is emitted
along the field lines in the slot gap, up to altitude $\eta = \eta_\mathrm{max}$ (where $\eta=r/R_\mathrm{NS}$). 
We assume an emissivity distribution across the SG as:
\begin{equation}
N(\xi_*) = (1 - \xi_*^2)
\end{equation}
where $\xi_* = 0$ at the center of the SG and
\begin{equation}
\xi_* = 1 - {2(1 - \xi)\over w_\mathrm{SG}}.
\end{equation}
Such a distribution, that peaks in the centre of the gap and decreases to zero at the gap edges, follows from the $\xi_*$ distribution of the SG potential  \citep{mh04a}. 

For the OG/OPC model, we describe the emitting region as an
infinitely thin layer along the inner surface of the gap.
The radio and PC phase-plots show the hollow cone patterns centered on the magnetic pole, while the SG and OG phaseplots show the caustic emission
patterns characteristic of outer magnetosphere emission \citep{dhr04}. 

\subsection{Light-curve generation}
\label{Calculation and light-curve generation}
The ($\phi$,$\zeta$) phase-plot space has been sampled in $180\times90$ bins.
Each bin $n(\phi,\zeta)$ of the phase-plot gives the number of photons per solid angle per primary particle that can be observed in the $\zeta$ direction 
at the rotational phase $\phi$:
\begin{equation}
n(\phi,\zeta) = \frac{dN_\mathrm{ph}}{\sin\zeta d\zeta d\phi}.
 \label{EnFlxperbin}
\end{equation}
Each phase-plot is obtained for a specific set of pulsar parameters that define its magnetospheric
structure: the spin period ${\bf P}$, the surface magnetic field ${\bf B_\mathrm{NS}}$, and the magnetic obliquity $\alpha$. 
For the studied models, the phase-plot has the following dependencies: 
$$n_\mathrm{\gamma,PC/Radio}=f(P,B,\alpha)$$ $$n_\mathrm{\gamma,SG}=f(w_\mathrm{SG},\alpha)~~and~~w_\mathrm{SG}=f(P,B)$$
$$n_\mathrm{\gamma,OG/OPC}=f(w,\alpha)~~and~~w_\mathrm{OG/OPC}=f(P,B).$$

For each emission model, we have evaluated phase-plots for $\alpha$ values, from 
5$^{\circ}$ to 90$^{\circ}$, with a step of $5^{\circ}$. For each $\alpha$ value, the phase-plots have been evaluated for 
2 magnetic field values and 9 spin period values for the PC and radio models, and for 16 gap width values in the SG and OG/OPC cases.
The complete set of sampled parameters is listed in Table \ref{Ph-Pl_values}.

To obtain the light-curve of a given NS, with a particular set of ${\bf P}$, ${\bf B_\mathrm{NS}}$, $\alpha$, and gap width parameters,
we interpolated the phase-plots noted in Table \ref{Ph-Pl_values}.  When comparing phase profiles for a different set of parameters, typically one profile will be narrower than 
another one making it nontrivial to interpolate between them.
We adopted a non-linear interpolation which expands the narrower light-curve covering the smallest phase range  
up to the phase extent of the wider profile, then applies a linear interpolation, and contracting the expanded profile back down to the extent of the original parent profiles.
This strategy preserves the thin peaks and high degree of modulation that characterises the pulsar emission profiles at radio and $\gamma$-ray
wavebands.
\begin{table*}[htbp!]
\begin{center}
\begin{tabular}{|c||c|c|c|c|c|}
\hline
                   & {\bf B}                             & {\bf P}                                                                       & $\alpha$      & Gap Width values  \\
                   & Tesla                            & milliseconds                                                           &Degrees       &                                   \\
                   \hline
                   \hline
PC & $10^{8}$, $10^{9}$ & 30, 40, 50, 75, 100                                                & 5-90 & 0.04, 0.06, 0.08, 0.1, 0.13, 0.16, 0.2, 0.225 \\
                   &                                         & 300, 500, 750, 1000                                             &5$^{\circ} \mathrm{step}$  & 0.25, 0.275, 0.3, 0.34, 0.38, 0.42, 0.46, 0.50\\
                    \hline
SG             &                                         &                                                                               & 5-90 & 0.04, 0.06, 0.08, 0.1, 0.13, 0.16, 0.2, 0.225 \\
                   &                  none              &              none                                                              &5$^{\circ}  \mathrm{step}$  & 0.25, 0.275, 0.3, 0.34, 0.38, 0.42, 0.46, 0.50\\
                   \hline
OG/OPC  &                                           &                                                                              & 5-90 & 0.01, 0.025, 0.04, 0.05, 0.067, 0.084, 0.1, 0.2 \\
                  &              none                    &             none                                                            &5$^{\circ}  \mathrm{step}$  & 0.3, 0.4, 0.5, 0.53, 0.56, 0.59, 0.62, 0.65\\
                  \hline
Radio & $10^{8}$, $10^{9}$ & 30, 40, 50, 75, 100                                                & 5-90 &  \\
                   &                                         & 300, 500, 750, 1000                                             &5$^{\circ}  \mathrm{step}$  & none\\
                    \hline
\end{tabular}
\end{center}
\caption{Magnetic field, period, and gap width values for which the phase-plots have been evaluated for each emission model. The SG and 
OG/OPC emission patterns do not depend directly on the pulsar period and magnetic field.}
\label{Ph-Pl_values}
\end{table*}

\section{Flux calculations}
\label{Flux calculations}
\subsection{Phase-plot normalisation and energy flux}

Let us define $L_\mathrm{pole}$ as the radiative luminosity from each pole, either in the $\gamma$-rays or in the radio.
Assuming a value for the primary particle production rate $\dot{N}_\mathrm{e}$ and
the energy $E$ of each photon, one obtains a radiation luminosity per phase-plot bin:
\begin{equation}
dL = \dot{N}_e E n(\phi,\zeta) \sin\zeta d\zeta d\phi=An(\phi,\zeta) \sin\zeta d\zeta d\phi
\end{equation}
where $A$ is a proportionality constant.
One can normalise the phase-plot to the total radiation luminosity over the two poles according to:
\begin{equation}
2 L_\mathrm{pole}=A \int_0^\pi \sin\zeta d\zeta \int_0^{2\pi}  n(\phi,\zeta) d\phi
\end{equation}
We define the specific intensity $I$ as
\begin{equation}
I=\frac{dL}{d\Omega}~~~~\rightarrow~~~~I(\phi,\zeta)=\frac{A n(\phi,\zeta) \sin\zeta d\zeta d\phi}{ \sin\zeta d\zeta d\phi}=A n(\phi,\zeta).
\label{Intensity}
\end{equation}

It is now possible to obtain the average energy flux observed by an Earth observer for a line of sight $\zeta_\mathrm{obs}$:
\begin{equation}
\langle \nu F_\mathrm{\nu} \rangle=\frac{\int_0^{2\pi}I(\zeta_\mathrm{obs},\phi)d\phi}{2\pi D^2}.
\label{nuFnu1}
\end{equation}
Here, $D$ is the pulsar distance.

From the Equations \ref{Intensity} and \ref{nuFnu1}, we can write the average energy flux observed at the Earth as:
\begin{equation}
\label{fluxEquation}
\langle \nu F_\mathrm{\nu} \rangle = \frac{ L_\mathrm{pole}}{\pi D^2} \frac{\int_0^{2\pi}n(\zeta_\mathrm{obs},\phi)d\phi}{\int_0^\pi \sin\zeta d\zeta \int_0^{2\pi}  n(\phi,\zeta) d\phi}
\end{equation}
The latter Equation establishes the relation between the luminosity derived in the framework of a given model, $L_\mathrm{pole}$,
and the integral of the pulsar light-curve $\int_0^{2\pi}n(\zeta_\mathrm{obs},\phi)d\phi$, obtained, from the phase-plot,
for $\zeta=\zeta_\mathrm{obs}$. This is related to the beaming factor $f_{\Omega}$ discussed in Section \ref{BFac} 

\subsection{Computations: gap width and energy flux}
\label{Flxcomp}

We calculated the $\gamma$-ray and radio light-curve for 
each pulsar of the sample, storing the value of the integral $\int_0^{2\pi} n(\zeta_\mathrm{obs},\phi) d\phi$ for the flux computation. 
The width of the emission gaps is computed using Equation \ref{SGgapW} for the SG, and equations \ref{OGwidth} and \ref{OPCwidth} for the OG and OPC.
 Because the PC and SG models do not apply when the gap becomes too large (pair-starved gaps should then be used), 
the flux for gap widths larger than 0.5 has been set to 0. 
Because no emission remains visible from the thin inner edge of OG/OPC gaps when the gap width exceeds 0.7
the flux for gap widths larger than 0.7 has been set to 0. 
So all the pulsars with a gap width above these threshold levels are assumed to not produce
any $\gamma$-ray emission. 
\begin{figure}[htbp!]
\begin{center}
\includegraphics[width=0.49\textwidth]{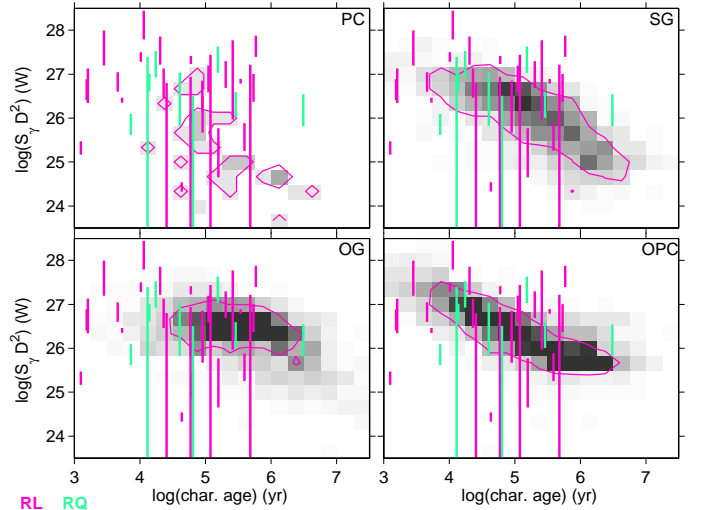}
\caption{Number density of the visible $\gamma$-ray pulsars obtained for each model as a function of characteristic age and energy flux 
times the square of the pulsar distance. These parameters can all be measured from the observations. The linear grey scale saturates at 1 
star/bin for the polar cap and 2.5 star/bin for the 
other models. The pink contours outline the 
region where simulated radio-loud $\gamma$-ray pulsars are found 
(at 20\% of the maximum density). The pink and green lines
show the data for the radio-loud and radio-quiet LAT pulsars, respectively.}
\label{NvisA_large_effi1p012p01p00p5_bldhor_nFnD2vsage_histo}
\end{center}
\end{figure}

For the radio luminosity computation, $L_\mathrm{pole}$, we have used Equation 
\ref{eq:Lradio} to evaluate the total radio luminosity and Equation \ref{eq:Lcc} to evaluate the luminosities of each core and cone component.
The $\gamma$-ray luminosity has been obtained by scaling the particle luminosities derived in Equations \ref{eqn:Lglow}, \ref{eqn:Lghigh}, \ref{eqn:LOG}, 
and \ref{eqn:LOPC}, for the PC, SG, OG and OPC models respectively by using a radiative eficiency $\varepsilon_\mathrm{\gamma}$. 
The latter has been chosen to provide a good agreement between the observed and simulated $S_\mathrm{\gamma}D^2$ distributions 
as a function of characteristic age ($S_\mathrm{\gamma}$ is the photon flux and $D$ the pulsar distance). This distribution involves only readily observable quantities.
The solution adopted for each model is shown in Figure \ref{NvisA_large_effi1p012p01p00p5_bldhor_nFnD2vsage_histo}. The choice of radiative efficiencies
are: $\epsilon_\mathrm{PC}=1.0$, $\epsilon_\mathrm{SG}=12.0$, $\epsilon_\mathrm{OG}=1.0$, and
$\epsilon_\mathrm{OPC}=0.5$. 
The high value of  $\epsilon_\mathrm{SG}$ needed for the SG requires either a
 \emph{super Goldreich-Julian current} or a stronger value of the accelerating electric field in the gap compared to the original calculation by \cite{mh04a}. 
 This is quite possible if the polar cap is slightly offset, i.e., non-symmetrical around the magnetic axis, as one expects from the shape of the magnetic field
 lines distorted by the stellar rotation. \cite{hm11} show that this distortion leads to a larger pair multiplicity as well as an increased electrical field along the 
 field lines, thus an enhanced luminosity. 
 Offset polar caps can sustain the modest increase in particle energy that is required in the present population study to account for the flux and pulsar counts 
 observed by the LAT without invoking 
 a radiation efficiency larger than one. The offset polar cap prediction was not available at the time of the population synthesis work, so we keep here the 
 original polar cap luminosity and $\epsilon_\mathrm{SG}=1200\%$.

\subsection{Gamma-ray energy and photon flux}

In order to evaluate the photon flux from the energy flux and to compare it with the LAT sensitivity in photon flux, we need to assume 
an emission spectrum for the  pulsars.
The typical photon spectrum of a LAT pulsar is well fitted by a power-law with an exponential cutoff, like
\begin{equation}
\label{enflxtoint}
\frac{dN_{\gamma}}{dE}=k\left(\frac{E}{E_0}\right)^{-\Gamma} e^{-E/E_\mathrm{cut}}.
\end{equation}
In the first LAT pulsar catalog \citep{aaa+10}, the distribution of the parameters $\Gamma$ and $E_\mathrm{cut}$ can be described by two Gaussians:
\begin{equation}
\begin{gathered}
\label{ConstSpectralstuff}
Gaussian~X:~~mean=1.97;~~variance=0.18  \\
\\
Gaussian~Y:~~mean=3.06;~~variance=0.37  
\end{gathered}
\end{equation}
The spectral index $\Gamma$ and the $\log_\mathrm{10}(E_\mathrm{cut})$ are defined as:
\begin{equation}
\begin{gathered}
\label{twogaussSpec}
\theta = 0.5982~~~[rad]   \\
\Gamma = X\cos\theta - Y\sin\theta \\
\log_\mathrm{10}(E_\mathrm{cut})=X\sin\theta + Y\cos\theta. 
\end{gathered}
\end{equation}
The Gaussian widths, centroids, and correlation angle had been derived from the analysis of the spectral parameters 
measured for the 1st LAT pulsar catalogue. We took here the same values.

The photon flux computation has been done 
using Equation \ref{fluxEquation}, \ref{enflxtoint}, \ref{ConstSpectralstuff}, and \ref{twogaussSpec} by assuming that the 
luminosity is mainly produced at photon energies $\ge 100$ MeV. The choice of this threshold and the choice 
of radiative efficiencies to match the data in Figure \ref{NvisA_large_effi1p012p01p00p5_bldhor_nFnD2vsage_histo} are linked.

\section{Gamma-ray and radio visibilities}
\label{Radio pulsar visibility}
\subsection{$\gamma$-ray pulsar visibility}
\label{LATvisMap}

In order to select the simulated pulsars that could be detected by the LAT during two years of observation, we 
made use of the 6 month pulsar visibility map published for the 1st LAT pulsar catalogue \citep{aaa+10} and of 
the 1 year pulsar visibility of blind pulsar searches \citep{dja+11}. The two maps have been used to estimate the 
$\gamma$-ray detectability of the radio-loud 
pulsars (corresponding to the LAT radio selected objects) and the radio-quiet 
ones (corresponding to the LAT blind search objects) respectively. The maps give the minimum visible photon flux 
$S_\mathrm{min.ph}$ and have been obtained taking into account the real LAT observation time in the sky, the photon 
energy, and the effective collection area corrected for the different incidence directions. 
Since the sky survey mode for LAT observations has been continued after 6 months,
the maps have been scaled to 2 years as the square root of time for the radio-selected sensitivity map and linearly with 
time for the blind search sensitivity map. Very few pointed observations were programmed that would significantly alter 
the shape of the visibility map. Photons collected in survey mode largely dominate and the flux threshold for detectability 
is primarily limited by the intensity of the interstellar background.

\subsection{Radio visibility}
The synthesis of the population is not based on any assumed NS birth rate;
 we assume a flat star formation rate over the last 1 Gyr. 
 Instead, the simulated sample was scaled 
to the real number of pulsars detected in the Galaxy. The scaling factor has been evaluated by selecting all the ATNF radio pulsars 
present in a select group of ten surveys and comparing this number with simulated radio pulsars visible in the same region. 
We have generated a large enough population to reduce the Poisson fluctuations and to improve the 
statistics in the analysis results.
\begin{table*}[htbp!]
\centering
\begin{tabular}{| p{3cm} || c | c | c | c | c | c | c | c |}
\hline 
   &  $\frac{Gain}{\beta}$ (K Jy$^{-1}$) & $\sigma_\mathrm{S/N}$ & $T_\mathrm{rec}$ (K)  &    $\nu_\mathrm{surv}$ (MHz)  &  $T_\mathrm{int}$ (s) & $t_\mathrm{samp}$ (ms) & $\Delta\nu_\mathrm{b}$ (MHz) & $\Delta\nu_\mathrm{ch}$ (MHz) \\
\hline \hline
Molonglo 2         &  5.100            & 5.4     & 225             & 408   & 40.96    & 40.0     & 4.0          & 4.0      \\
\hline
Green Bank 2          & 0.886            &  7.5   & 30.0             & 390   & 136        &  33.5    & 16.0        & 2.0     \\
\hline
Green Bank 3          & 0.950            &  8.0   & 30.0             & 390   & 132        &  2.0       & 8.0          & 0.25   \\
\hline
Parkes 2         & 0.430             &  8.0   & 50.0             & 436   & 157.3     & 0.6        & 32.0       & 0.125   \\
\hline
Arecibo 2         & 10.91$^{1}$ &  8.0   & 100$^{2}$   & 430   & 39.3       &  0.4       & 0.96       & 0.06    \\
\hline
Arecibo 3         & 13.35$^{3}$ &  8.5   & 70.0$^{4}$  & 430   & 67.7        &   0.5     &10.0        & 0.078   \\
\hline
Parkes 1         & 0.256             &  8.0   & 45.0            & 1520 & 157.3     &  2.4      & 320.0     & 5.0      \\
\hline
Jodrell Bank 2         & 0.400             &  6.0   & 40.0             & 1400 & 524.0     &  4.0      & 40.0       & 5.0         \\
\hline
Parkes MB      & 0.460             &  8.0   & 21.0            & 1374 &  2100.0  & 0.25     & 288.0    & 3.0       \\ 
\hline
Swinburne       & 0.427             &  10.0 & 21.0            & 1374 & 265.0     &  0.25    & 288.0    & 3.0         \\ 
\hline
\end{tabular}
\centering
\begin{list}{}{}
\item[$^{1}$] Computed using $\frac{Gain}{\beta}=\frac{19-(0.42\times\left|19-\delta\right|)}{1.1375}$. 
\item[$^{2}$] Computed using $T_\mathrm{rec}=90+2.083\times\left|19-\delta\right|)$
\item[$^{3}$] Computed using $\frac{Gain}{\beta}=\frac{19.7-(0.42\times\left|19-\delta\right|)}{1.2236}$.
\item[$^{4}$] Computed using $T_\mathrm{rec}=65+2.083\times\left|19-\delta\right|)$.
\end{list}
\caption{\footnotesize Instrumental parameters of the radio surveys. For the Arecibo surveys we chose 
               to adopt a more accurate definition for the gain and the receiver temperature that are function of the declination $\delta$.
               Respectively from the left to the right column, are indicated: telescope gain divided by a system losses factor, minimum
               signal to noise detected, receiver temperature, central observation frequency, integration time, sampling time, total
               bandwidth, and channel bandwidth.}
\label{Surveys2}
\end{table*}

We selected the simulated pulsars within the visibility criteria of 10 radio surveys from the ATNF 
database\footnote{http://www.atnf.csiro.au/research/pulsar/psrcat/} for which the survey parameters are well known and that cover
the largest possible sky surface while minimising the overlapping regions. These surveys are: Molonglo2  \citep{mlt+78}, 
Green Bank 2 \& 3   \citep{dtws85,stwd85}, 
Parkes 2 (70 cm)  \citep{lml+98}, Arecibo 2 \& 3  \citep{sstd86,nft95}, Parkes 1  \citep{jlm+92}, Jodrell Bank 2 \citep{cl86}, 
Parkes Multi-beam  \citep{mlc+01} and the extended Swinburne surveys  \citep{ebsb01,jbo+09}.
The ratio between the number of simulated pulsars meeting the surveys' visibility criteria and the number of objects actually 
detected is
\begin{equation}
\label{scala}
S_\mathrm{f}=\frac{N_\mathrm{obs,10~surv}}{N_\mathrm{sim,10~surv}}
\end{equation}
This is the factor we used to scale the simulated pulsar sample. 

\subsubsection{Radio pulsar selection}
\label{Radio pulsar selection}

During a radio survey, the edges of the survey region are defined by the position of the radio-telescope beam centre. 
Nevertheless, because of the solid angle extension and complexity of the beam, it is possible to observe 
a pulsar slightly out of the declared survey region. Thus, to say that all the pulsars observed during a survey fall inside the declared survey coordinates 
edges is not totally correct. 
The first parameter we re-evaluated for each survey is the number of pulsars seen inside a given region. 
\begin{table*}[tbp!]
\centering
\begin{tabular}{| p{3cm} || c | c | c | c | c | c | c | c | c |}
\hline 
   &   $l_\mathrm{st}$ $(^\circ)$ & $l_\mathrm{ed}$ $(^\circ)$ & $b_\mathrm{st}$  $(^\circ)$  &  $b_\mathrm{ed}$  $(^\circ)$ & $dec_\mathrm{st}$ $(^\circ)$  & $dec_\mathrm{ed}$ $(^\circ)$ & $\epsilon_\mathrm{surv}$  & $\epsilon_\mathrm{Dewey}$ & $Duty\ cycle$\\
\hline \hline

Molonglo 2         & -          & -          & -            &  -         & -85.0    & 20.0    & 0.62   &   0.4  &  0.03    \\
\hline
Green Bank 2    & -          &  -         & -            & -          & -18.0    & 90.0    & 0.32   &  0.7   &  0.03   \\
\hline
Green Bank 3    & 15.0   &   230   & -15       & 15       & -           & -           & 0.41    &  0.75   &  0.03   \\
\hline
Parkes2              & -          &    -        & -            & -          & -90.0   & 0.00     & 0.90    &   0.75   &  0.03  \\
\hline
Arecibo 2            & 40      &    66     & -10       & 10       & 9.50    & 25.0     & 0.54    &   1.0   &  0.05  \\
\hline
Arecibo 3            & 38      &    66     & -8.1      & 8.2      & 5.00    & 26.5     & 0.66    &   0.7   &  0.05   \\
\hline
Parkes 1             & -92     &    20     & -4         & 4           & -           & -            & 0.41    &   0.6   &  0.03  \\
\hline
Jodrell Bank 2   & -5       &   105    & -1.3      & 1.3       & -           & -             & 0.50    &   0.8   &  0.03  \\
\hline
Parkes MB         & -105   &   52      &  -6.03     & 6.35  & -           & -             & 0.98    &    0.9  &  0.05  \\ 
\hline
Swinburne         & -100   &   50      &  4.5      & 30        & -           & -             & 0.87    &  1.0    &  0.05  \\ 
\hline
\end{tabular}   
\caption{\footnotesize Estimated survey parameters. Respectively from the left to the right column are indicated:
               longitude start \& end, latitude start \& end, declination start \& end, new survey efficiency, Dewey scaling factor, 
               and pulsar duty cycle, defined as the pulse width over the period and used in the computation of the intrinsic pulse width $W$ (Equation \ref{Wsmearing}).}
\label{Surveys1}
\end{table*}

The second important parameter  is the survey efficiency $\epsilon_\mathrm{surv}$. It is defined as a filling factor, e.g.
the ratio between the actual solid angle covered by the radio telescope beam during the observations, and the area within the declared survey 
boundaries. The survey efficiency can be considered as the probability of 
observing a pulsar present in the survey region only if the parent spatial distribution is uniform. 
To evaluate the boundaries of the survey region and to define the survey efficiency we decided:
\begin{enumerate}
\item to slightly extend the sky survey boundaries in order to include the largest number of pulsars actually detected by a survey, without changing 
          too much the original boundaries 
\item to evaluate the detection flux threshold for each pulsar within a survey by scaling the Dewey formula  \citep{dtws85} with a free parameter, $\epsilon_\mathrm{Dewey}$, to match 
the observations.
         \begin{equation}
         \label{FDGDeweyEq}
         S_\mathrm{min} = \epsilon_\mathrm{Dewey} \times S_\mathrm{threshold}
         \end{equation}
         where the threshold flux $S_\mathrm{threshold}$ is expressed by the Dewey formula
         \begin{equation}
         \label{DeweyEq}
         S_\mathrm{threshold}=\frac{\sigma_\mathrm{S/N}[T_\mathrm{rec}+T_\mathrm{sky}(l,b)]}{G\sqrt{N_\mathrm{p}Bt}}\sqrt{\frac{W}{P-W}} .
         \end{equation}
\end{enumerate}
The Dewey formula, or radiometer formula, takes into account the characteristics of a given radio telescope and detector
as well as a pulsar period and direction to give the minimum flux the survey would be able to detect.
In Equation \ref{DeweyEq} $\sigma_{S/N}$ is the minimum signal to noise ratio taken into account, $T_{rec}$ is the receiver temperature, $T_{sky}$
is the sky temperature at 408 MHz, $G=Gain/\beta$ is the ratio between the radio telescope gain and the dimensionless factor $\beta$ that accounts
for system losses,
$N_p$ is the number of measured polarisations, $B$ is the total receiver bandwidth, $t$ is the integration time, and $P$ is the pulsar period. 
$W$ is the effective pulse broadening, defined as
\begin{equation}
\label{Wsmearing}
W^2=W_{0}^2 + \tau_\mathrm{samp}^2 + \tau_\mathrm{DM}^2 + \tau_\mathrm{scat}^2 + \tau_\mathrm{trailDM}^2.
\end{equation}
Here, $W_0$ is the intrinsic pulse width (Duty Cycle), $\tau_\mathrm{samp}$ is a low-pass filter time constant applied before sampling 
(when this parameter is unknown, a value equal 
to twice the sampling time has been used), $\tau_\mathrm{DM}$ is the pulse smearing due to the DM over one frequency interval $\Delta\nu$, and 
$\tau_\mathrm{scat}$ is the pulse broadening due to interstellar scattering \citep{dtws85}.
The dispersion broadening time, $\tau_\mathrm{DM}$ (ms), across one frequency channel, $\Delta\nu$, is related to the dispersion measure (DM) as 
\begin{equation}
\tau_\mathrm{DM}=\frac{e^2}{2\pi m_\mathrm{e}c}\left(\frac{1}{\nu_{1}^2}-\frac{1}{\nu_{2}^2}\right)\mathrm{DM}
              \approx8.3\times10^6\frac{\Delta\nu_\mathrm{MHz}}{\nu_\mathrm{MHz}^3}\mathrm{DM}
\end{equation}
where $m_\mathrm{e}$ is the mass of the electron, $c$ is the speed of the light, and $\nu_{1}$, $ \nu_{2}$ are the edges of the frequency channel.
The dispersion measure, DM (pc cm$^{-3}$), is obtained using the \cite{cl01} NE2001 model. The same model 
provides the scattering measure, SM (kpc m$^{-20/3}$), which allows to estimate the broadening time due to interstellar scattering as
\begin{equation}
\tau_\mathrm{scat}=1000\left(\frac{\mathrm{SM}}{292}\right)^{1.2}d~\nu_\mathrm{GHz}^{-4.4} 
\end{equation}
where $d$ is the pulsar distance in kpc \citep{jlm+92,sd96}.
The last term of equation \ref{Wsmearing}, $\tau^2_\mathrm{trailDM}$, is an additional time broadening added when the sampling is performed for a DM
value different from the real one. It corresponds to the fourth term of equation 2 in \cite{dtws85} and becomes important just for low period pulsars.

The sky temperature at frequencies other than 408 MHz is obtained as:
\begin{equation}
T_\mathrm{sky}(\nu_\mathrm{MHz})=T_\mathrm{sky.408}\left(\frac{408~\mathrm{MHz}}{\nu_\mathrm{MHz}}\right)^{2.6} .
\end{equation} 

Tables \ref{Surveys2} and \ref{Surveys1} list all the radio telescope and detector characteristics of the 
surveys we took into account.
Some of the survey parameters in the literature listed as average values have been re-evaluated using the above mentioned prescription.

The scaling of the radiometer equation was motivated by the uncertainties 
related to the Dewey formula, because of 
flux oscillations due to scintillation. The scintillation is caused by the turbulent variation of 
the interstellar medium that the pulsar light has to cross before reaching the observer. The consequence is an oscillation (scintillation) of the pulsar 
flux that can introduce spurious detections of pulsars with a flux lower than the survey threshold or that can cause
the non-detection of pulsars with a flux higher than the survey threshold. So we scaled the 
$S_\mathrm{threshold}$ level in order to take 
into account possible spurious detections or missed detections due to scintillation. 
$S_\mathrm{threshold}$ should not be lower than the 
flux of the weakest pulsar of the survey.  A reasonable estimate is to employ the average of the low-flux
tail of the pulsars of the survey. 

In the ATNF database we can count how many pulsars fall within a survey boundary, how many 
would match the survey visibility criterion (flux $ >$ $S_\mathrm{threshold}$), 
and how many of these pulsars have really been observed by the survey. 
The comparison of the ratios of the radio flux recorded for each pulsar to the minimum visible flux $S_\mathrm{min}$ in its direction provides an estimate
of the Dewey scaling factor. The scaling values $\epsilon_\mathrm{Dewey}$ are given in Table \ref{Surveys1} 
and the distribution of the flux ratios is shown in Figures  \ref{SurveysFig1}, \ref{SurveysFig3}, and \ref{SurveysFig5}
(right plots) for each survey (only the ratios below 10 are displayed to focus near the visibility threshold). 

Then, for each survey, we derived the ratio between 
the number of pulsars really detected by the survey and 
the total number of observable ATNF pulsars (the sum of the detected ones plus those that match the position and flux survey criteria  but were not detected).  
We consider this last ratio as the new survey efficiency, $\epsilon_\mathrm{surv}$, 
i.e. the percentage of pulsars detected by the survey with respect to all the detectable ATNF pulsars in the survey region.  
The new efficiency $\epsilon_\mathrm{surv}$ is listed, for each Survey, in Table \ref{Surveys1}.

By using the newly estimated survey parameters listed in Tables \ref{Surveys2} and  \ref{Surveys1}, and by using the radiometer equation \ref{DeweyEq},
the number of real pulsars that meet the visibility criteria of our surveys is {\bf 1430} (ATNF database, January 2012). 
We use this number and the number of simulated pulsars that match the 
same criteria to scale the visible component of the  simulated $\gamma$-ray pulsar population in Equation \ref{scala}.
\begin{table*}[htbp!]
\centering
\begin{tabular}{| c || c | c | c | c | c | c | c | c | c |}
\hline 
   &  $PSR_\mathrm{(Radio~\lor~\gamma)}$   &  $PSR_\mathrm{(Radio~\land~\gamma)}$   & $PSR_\mathrm{ \gamma~only}$ & $PSR_\mathrm{(Radio~\land~\gamma)}/PSR_\mathrm{\gamma~all}$    \\
   \hline \hline
PC            &  1431  &     3.6     &   0.5  & 0.87  \\
\hline
SG            &  1508  &     75   &  79  &  0.49   \\
\hline
OG            &  1496  &    123   &  66 &  0.65  \\
\hline
OPC          &  1524  &   107   &  94   & 0.53  \\
\hline
LAT           & $\backslash$  &    25    &  30     & 0.45 \\
\hline
\end{tabular}
\caption{\footnotesize For each model and for the observed dataset, we give from left to right the scaled numbers of pulsars visible in the radio or 
$\gamma$-ray band, in the radio and $\gamma$-ray bands, in $\gamma$-rays only, and the fraction of radio loud objects in the $\gamma$ visible
sample. All the data refer to two years of LAT observations.}
\label{DetectionTab}
\end{table*}

\section{Results}
\label{PopResults}

\subsection{Detection statistics}

Table \ref{DetectionTab} indicates, for each model, the numbers of NSs that passed the radio and/or $\gamma$ visibility criteria and their 
comparison with the LAT detections after 2 years of observations.
The number of radio visible pulsars in the simulation has been scaled to the 1430 ATNF radio pulsars that passed the same selection criteria. 
\begin{figure}[htbp!]
\begin{center}
\includegraphics[width=0.49\textwidth]{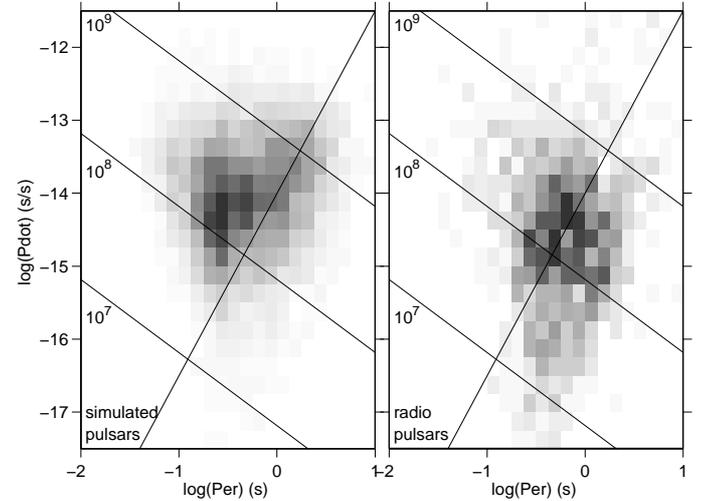}
\caption{Number density of the visible radio pulsars as a function of period and period derivative. The left and right plots respectively show the 
simulation and  observed data with the same grey scale saturating at 25 star/bin and the same visibility criteria. The rising grey line marks the 
slot-gap death line. The declining grey lines mark the iso-magnetic lines at $10^7$, $10^8$, and $10^9$ Tesla.}
\label{NvisA_large_effi1p012p01p00p5_bldhor_PdotPer_histo_radio}
\end{center}
\end{figure}
The scale factor of 0.136 is required to match the simulated and observed radio samples and has been applied to all star counts quoted hereafter, in
particular to the $\gamma$-ray simulated samples. This scale factor implies a NS birth rate of $\sim3.7$ NS/century over the last 1 Gyr.
The choice of radiative efficiencies driven by a reasonable agreement in the $S_\mathrm{\gamma}D^2$ evolutions with characteristic age 
shows that the wide beams produced in the intermediate-high (SG) and outer models provide enough detections to account the LAT findings.
The low-luminosity narrow PC beam fails in predicting the LAT detection number and the fraction of radio-quiet objects because of the large 
overlap between the $\gamma$-ray and radio beams.

\subsection{Comparison of the total simulated and observed samples}
\label{Comparison of the total simulated and observed samples}
\begin{figure}[htbp!]
\begin{center}
\includegraphics[width=0.49\textwidth]{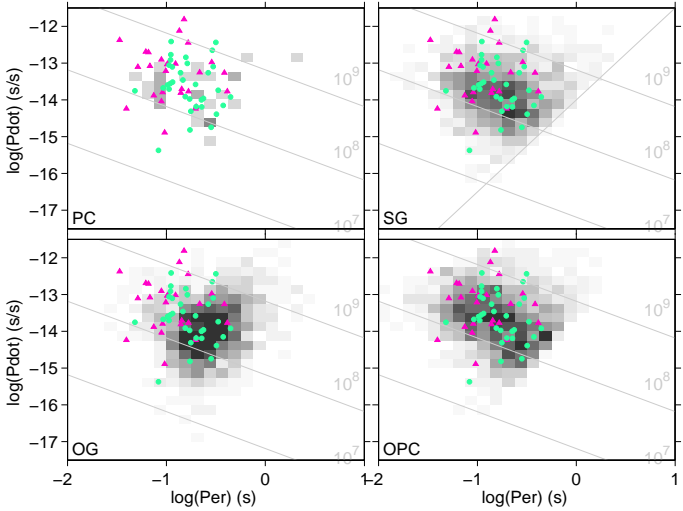}
\caption{Number density of the visible $\gamma$-ray pulsars obtained for each model as a function of period and period derivative. 
The linear grey scale saturates at 1.5 star/bin. The pink triangles and green dots show the radio-loud and radio-quiet LAT pulsars, 
respectively. The rising grey line in the slot-gap subplot marks the slot-gap death line. The declining grey lines mark the iso-magnetic 
lines at $10^7$, $10^8$, and $10^9$ T.}
\label{NvisA_large_effi1p012p01p00p5_bldhor_PdotPer_histo_gamma}
\end{center}
\end{figure}

Figures \ref{NvisA_large_effi1p012p01p00p5_bldhor_PdotPer_histo_radio}  and \ref{NvisA_large_effi1p012p01p00p5_bldhor_PdotPer_histo_gamma} 
show the comparison between the simulated distributions in the $P-\dot{P}$ diagram  for the radio visible component and for the $\gamma$-ray visible 
population for each model. The simulated distributions reasonably describe the observed samples and are in nice agreement with the same distributions
obtained by \cite{twc11}.
The simulated radio population is able to describe the observed $P-\dot{P}$ distribution for the fastest rotators that are likely to sustain substantial 
$\gamma$-ray emission and represent the LAT pulsar population. 
The PC model reproduces poorly the observed population. 
Both the SG and OG models over predict the number of middle
aged $\gamma$-ray pulsars and under predict the number of young $\gamma$-ray objects. 
Of those, the  OG shows the poorer description of the data;
the core of the distribution is too close to the pulsar death line and it lacks energetic pulsars. 
The OPC $\gamma$-visible population best describes 
the observed $P$ and $\dot{P}$ of the LAT population with a core centred on the observed objects and tails that cover the overall dispersion. 

\begin{figure}[htbp]
\begin{center}
\includegraphics[width=0.49\textwidth]{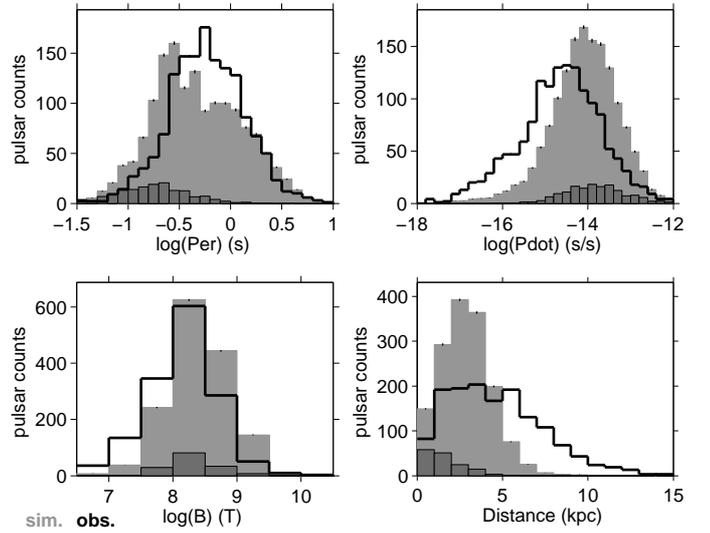}
\caption{Number distributions in period, period first time derivative, surface magnetic field strength, and distance obtained for the whole 
populations of radio \emph{or} $\gamma$-ray visible pulsars in the simulations (light grey histogram) and in the LAT and radio survey data (thick line). 
The ATNF radio sample has been restricted to the objects that pass the same position and flux selection criteria as in the simulation. The 
slot-gap model has been used as an example for the $\gamma$-ray simulation. The dark shaded histograms show the distributions of the gamma 
active subsample of the whole simulated population. The abundances of simulated objects at low $P$, high $\dot{P}$, and high $B$ are 
dominated by $\gamma$-ray active pulsars. The excess of energetic and nearby simulated objects reflects the set of assumptions adopted for 
the birth distributions to provide a better match to the LAT data.}
\label{NvisA_large_effi1p012p01p00p5_bldhor_PPdotBD}
\end{center}
\end{figure}

Figure  \ref{NvisA_large_effi1p012p01p00p5_bldhor_PPdotBD} compares the total simulated populations and its $\gamma$-ray sub-sample to the observed  
total sample of radio and/or $\gamma$-ray visible objects for key characteristics: period, period first time derivative, surface magnetic field, and distance.
The simulated spin period distribution is too broad to describe the observed proportion between the number of intermediate period objects
($\sim50$ ms) and the wings of the distribution. The range of spin periods is well covered 
and well centred, but we lack simulated objects in the 0.3-1.0 second range. 
The simulated distributions in $\dot{P}$, $B$, and $D$ are all shifted to an excess of young, energetic, and nearby pulsars compared to the observed ones.
This results from the choice of birth characteristics and NS intrinsic characteristics ($M_{NS}$, $R_{NS}$, and $I$ formulation) that emphasised nearby 
and high-$\dot{E}$ objects while 
preserving the bulk of the radio distributions. This choice has been made a posteriori to minimise the lack of high-$\dot{E}$ objects discussed in 
section \ref{SPDWN}.
Nevertheless,  the discrepancies observed in Figures \ref{NvisA_large_effi1p012p01p00p5_bldhor_PdotPer_histo_radio} and 
\ref{NvisA_large_effi1p012p01p00p5_bldhor_PPdotBD} are not only due to the choice of birth distributions, but also to a radio 
model ill adapted  to explain the observed radio population at the highest $\dot E$s.
Whereas this would be problematic to study radio beam models, the reasonable representation at $P<500$ ms and the excess of
objects with $\dot{P} > 3 \times 10^{-15}$ s/s,  $B > 10^{8}$ T, and $D \le 4$ kpc, where most of the LAT pulsars are found, supports 
the study of $\gamma$-ray models. The necessity of an improved radio model is a result of this paper and its formulation, beyond the 
purpose of this study, will be the subject of future work.
In the histograms shown in Figure \ref{NvisA_large_effi1p012p01p00p5_bldhor_PPdotBD}, the total distributions are 
dominated by the radio sample since the $\gamma$-ray pulsars' contribution is much smaller.

\subsection{The spin-down power}
\label{SPDWN}

\begin{figure}[htbp!]
\begin{center}
\includegraphics[width=0.49\textwidth]{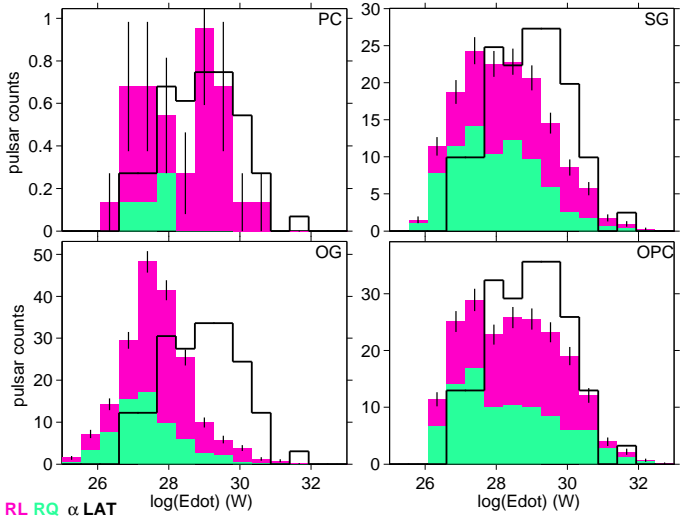}
\caption{Spin-down power distributions obtained for each model for the visible $\gamma$-ray pulsars. Pink and green refer to radio-loud and radio-quiet fractions of
the total population, respectively. The LAT distribution (in black) has been scaled to the total number of visible pulsars for each model to 
ease the comparison and show the relative lack of young energetic pulsars with $\dot{E} > 3 \times 10^{28}$ W.}
\label{NvisA_large_effi1p012p01p00p5_bldhor_Edot}
\end{center}
\end{figure}

\begin{figure}[htbp!]
\begin{center}
\includegraphics[width=0.49\textwidth]{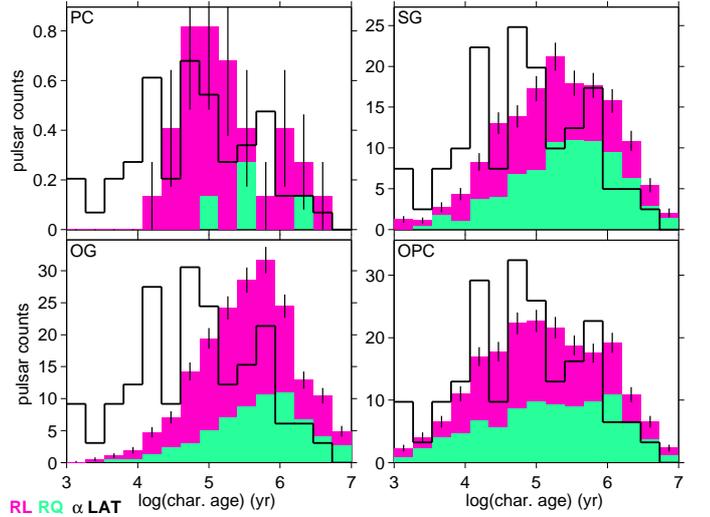}
\caption{Age distributions obtained for each model for the visible $\gamma$-ray objects. Pink and green refer to radio-loud and radio-quiet fractions of
the total population, respectively. The LAT distribution (in black) has been scaled to the total number of visible objects for each model to ease the comparison and 
show the relative lack of visible objects with age $<$100 kyr.}
\label{NvisA_large_effi1p012p01p00p5_bldhor_age}
\end{center}
\end{figure}

Figures \ref{NvisA_large_effi1p012p01p00p5_bldhor_Edot} and  \ref{NvisA_large_effi1p012p01p00p5_bldhor_age} compare the distributions in
spin-down power and characteristic age for the LAT and the $\gamma$-visible simulated pulsars. 
All models are significantly lacking simulated pulsars with spin-down power $\dot{E}> 3 \times 10^{28}$ W and characteristic age 
$t_\mathrm{char}<100$ kyr.  Additionally, all the models over-predict the number of low $\dot{E}$ pulsars and favour  a
pulsar population older than that observed by the LAT.
The difference in shape of the observed and simulated histograms suggests that the $\dot{E}$ inconsistency is not due to a simple scale mismatch, but
to a deficiency in modelling the pulsar evolution: even by scaling the spin down power upward none of the models would be able to describe 
the observed distribution.

Even though the PC model fails to produce enough visible gamma-ray pulsars because its narrow beam is under luminous and rarely visible, its evolution with $\dot{E}$
or age is less skewed to old age than the high-altitude SG or the outer-gap models.

The OG model provides the poorest description of the $\gamma$-ray evolution.
A strong evolution with age is predicted by the classical formulation of the OG because 
the gap size is controlled by the amount of X-rays emitted by the stellar surface heated by the back-flow
 of primary charges returning from the gap (self-sustaining OG model).
The strong evolution driven by this feedback is apparently not supported by the LAT data.
The OPC model gives slightly better results but still fails to predict enough high $\dot E$ pulsars. 
The similarity of the $\dot{E}$ profiles obtained for the SG and OPC models shows that the relative lack of energetic $\gamma$-ray pulsars is not related to 
the number of visible hemispheres (two pole caustic SG, one pole caustic OPC), or to the evolution of the emission region with age within the open magnetosphere   
(the emitting layer moves closer to the magnetic axis with increasing age in the OPC case while it remains near the last closed B line, but widens with age in the SG model).

The under-prediction of high-$\dot{E}$ visible $\gamma$-ray pulsars is rather puzzling since they are the intrinsically
brightest objects (high particle power and large $f_\mathrm{\Omega}$) with the widest beams (large open magnetosphere and thin gaps emitting near the closed field lines) 
sweeping widely across the sky. 
The problem affects all the models, so its origin does not depend much on the emission pattern or the luminosity trend with $\dot{E}$. 
For instance, the luminosity evolution of the OPC model was constructed to agree with the LAT data, yet the deficit of energetic $\gamma$-ray visible pulsars is still present. For a given luminosity the effective flux intercepted by the observer strongly depends on the gap thickness. 
For $\dot{E}>10^{28}$ W, the OPC gap width is
10 to 100 times smaller than the SG one, concentrating the photons in sharp caustics that remain visible to large distances and over many aspect angles, yet both the OPC and SG models produce a deficit of high $\dot{E}$ pulsars in a rather similar way. 
The discrepancy is also insensitive to the relative 
orientation of the radio and $\gamma$-ray beams since both radio-loud and radio-quiet  simulated pulsars are missing at high $\dot{E}$.
Nor is the problem related to the sensitivity horizon since all the models over-predict the fainter objects at low $\dot{E}$.
By testing different population configurations we have tried to understand which pulsar parameter has the largest impact 
on the high $\dot{E}$ tail of the $\gamma$-ray sample. Different birth distributions in period and magnetic field have been tested. 
Decreasing the birth spin period in order to increase the fraction of very young and energetic pulsars (section \ref{evolution}) 
yields a very small gain in the number of $\gamma$-visible energetic pulsars.  Scanning the allowed domain of intrinsic 
luminosities (e.g., SG $\lambda$ parameter) also failed to produce an increase in the young, energetic population.
\begin{figure}[htbp!]
\begin{center}
\includegraphics[width=0.49\textwidth]{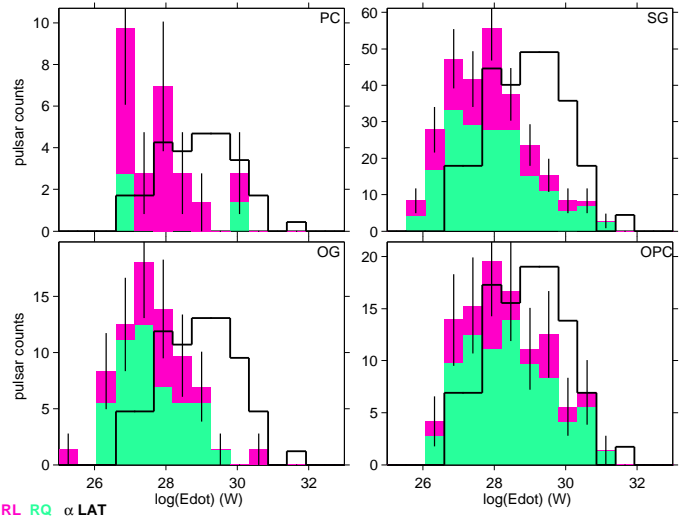}
\caption{Spin-down power distributions obtained for each model for the visible $\gamma$-ray objects for a different set of birth distributions: a Gaussian 
of width 200 ms centred at 0 for the periods; the sum of two Gaussians in $\log_{10}B_\mathrm{NS}$ [Tesla], both 0.6 in width, respectively 
centred at 8.4 and 9.1, and with an amplitude ratio 1:7/12 for the surface magnetic fields; and the \cite{pac90} surface density in the Galaxy.
Pink and green refer to radio-loud and radio-quiet fractions of the total population, respectively. The LAT distribution, 
scaled to the total number of visible objects, is plotted as a black contour.}
\label{Nvis_caseA_small_effi1p04p00p30p2_Edot}
\end{center}
\end{figure} 

A different choice of  $M_{NS}$, $R_{NS}$, or $I$ would shift the simulated distributions horizontally in $\dot{E}$, 
but would not alter their shape. 
\begin{figure}[htbp!]
\begin{center}
\includegraphics[width=0.49\textwidth]{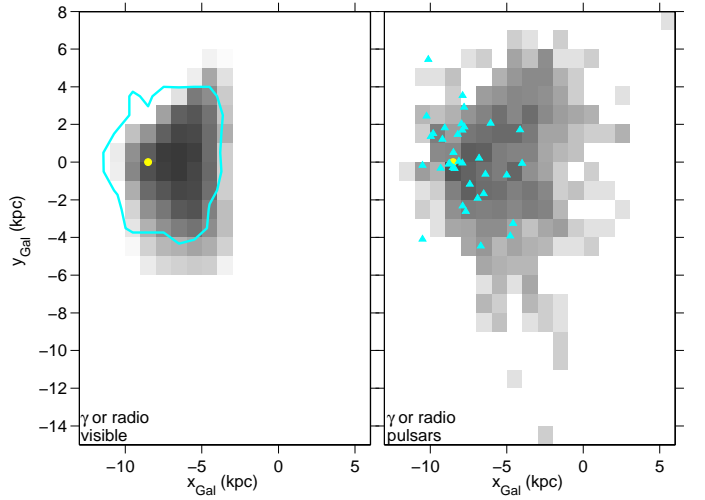} 
\caption{Number density of the visible radio and/or $\gamma$-ray pulsars in the Milky Way (polar view). The left and right plots respectively show the simulation and
 observed data with the same logarithmic gray scale saturating at 100 star/bin and the same visibility criteria. The cyan contour outlines the region where 
 simulated SG $\gamma$-ray pulsars are detectable. The cyan triangles show the location of the LAT pulsars. The yellow dot marks the Sun.}
\label{NvisA_large_effi1p012p01p00p5_bldhor_faceon_histo}
\end{center}
\end{figure}
The range of acceptable masses and radii given in \cite{lp07} limits an increase in the moment 
of inertia of the stars (thus $\dot{E}$) to within a factor of 2 or 3 beyond our present choice. 
This is too small a factor to address the lack of high-$\dot{E}$ pulsars in the simulations.
One of the tested configurations, illustrated in Figure \ref{Nvis_caseA_small_effi1p04p00p30p2_Edot}, shows how the lack of high-$\dot{E}$ pulsars remains, even after
choosing a more energetic pulsar population at birth and a much broader distribution for their birth places across the Galaxy, \cite{pac90},
as used in previous work e.g., \cite{gvh04} and \cite{twc11}.

Despite the stronger bias to energetic objects at birth adopted in Figure  \ref{Nvis_caseA_small_effi1p04p00p30p2_Edot} as compared to Figure 
\ref{NvisA_large_effi1p012p01p00p5_bldhor_Edot}, the lack of high-$\dot{E}$ $\gamma$-ray pulsars is less severe in the latter. 
\begin{figure}[htbp!]
\begin{center}
\includegraphics[width=0.49\textwidth]{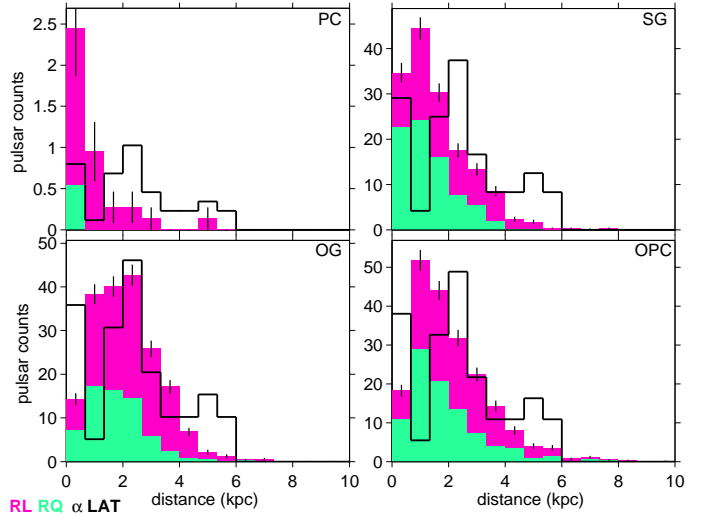}
\caption{Distance distributions obtained for each model for the visible $\gamma$-ray objects. Pink and green refer to radio-loud and radio-quiet fractions of
the total population, respectively. The LAT distribution (in black) has been scaled to the total number of visible objects 
for each model to ease the comparison and show the relative overabundance of nearby objects for the PC and SG models and under abundance of nearby 
objects for the OG and OPC models.}
\label{NvisA_large_effi1p012p01p00p5_bldhor_dist}
\end{center}
\end{figure}
This is due to the much larger fraction of 
births occurring in the inner Galaxy for the population shown in Figure \ref{NvisA_large_effi1p012p01p00p5_bldhor_Edot}. 
Because of the constraints on the supernova rate in the Galaxy, we cannot significantly increase the number of recent births, but the 
distribution provided by the HII region profile concentrates a larger fraction of the recent births in the inner Galaxy, within the LAT visibility horizon.
So, the $\dot{E}$ problem seems related both to the birth location and spin-down evolution of the pulsars.

It is possible that the magnetic obliquity $\alpha$ decreases with age, as suggested by  \cite{ycbb10}. First, the solid angle swept by the pulsar beam would decrease as $\alpha$ gradually decreases with age, so pulsars detected originally will later become invisible. 
Second, $\alpha$ has an impact on visibility through the gap width.  This is illustrated by the difference in the $\dot{E}$ histograms obtained for the OG and OPC cases. 
They share the same emission pattern, but the OG gap width depends on $\alpha$ while the OPC gap width is just proportional to $\dot{E}$.
Another speculative explanation would be a slower evolution of the dipole spin-down for very young and energetic objects, within the first 100 kyr.
This hypothesis would need to be justified on the basis of theory.
\begin{figure*}[htbp!]
\begin{center}
\includegraphics[width=0.8\textwidth]{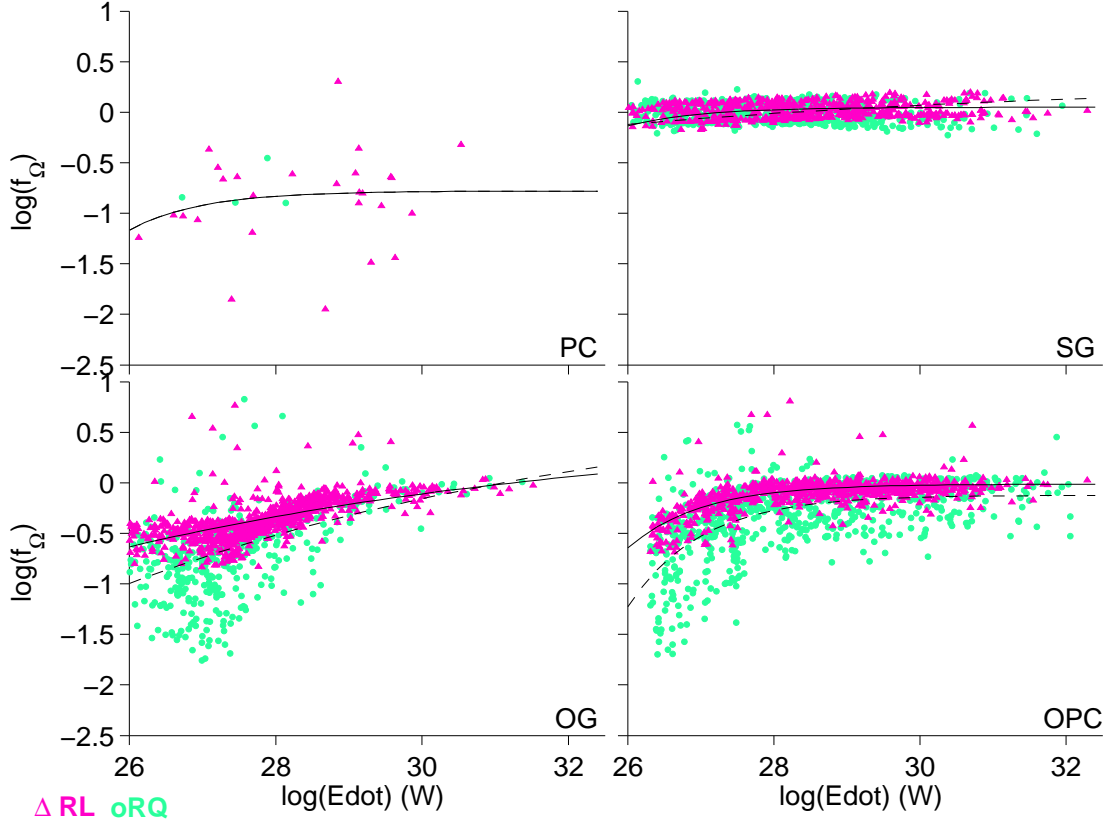}
\caption{Distribution of the $\gamma$-ray beaming factors obtained for each model as a function of the spin-down power. Pink and green dots refer to radio-loud 
and radio-quiet $\gamma$-ray visible objects, respectively. The solid and dotted lines give the best exponential fit to the radio-loud and radio-quiet data points, respectively.
For the PC case, a unique fit was applied to the whole sample of radio-loud and radio-quiet objects.}
\label{NvisA_large_effi1p012p01p00p5_bldhor_beamvsEdot}
\end{center}
\end{figure*}

\subsection{Spatial distribution in the Milky Way}

Figure \ref{NvisA_large_effi1p012p01p00p5_bldhor_faceon_histo} shows a polar view of the spatial density of visible radio or 
$\gamma$-ray pulsars in the Galaxy, resulting from the birth location profile described in section \ref{Birth distribution in the Galactic plane}. 

The majority of these pulsars are born within the solar circle and
the $\gamma$-ray visibility contour agrees well with the Galactic region where the bulk of the LAT pulsars have been detected.  
The radio visibility horizon is closer in the simulation than in reality, but the $\gamma$-ray visibility horizon spans the right distance range. 
The visibility
is therefore not the primary cause for the lack of high-$\dot{E}$ $\gamma$-ray predictions discussed in the previous section.

Figure \ref{NvisA_large_effi1p012p01p00p5_bldhor_dist} gives the distance distributions of the visible $\gamma$-ray pulsars. 
An interesting trend, observed in all the models except for the PC, is that the radio loud to radio quiet ratio increases up to 
4 or 5 kpc, and decreases down to zero at larger distance, implying that we lose the radio emission with distance 
more rapidly than we lose the $\gamma$-ray signal.

Since the measurement of the LAT pulsar distances is often affected by large uncertainties, the comparison
between models and data in Figures \ref{NvisA_large_effi1p012p01p00p5_bldhor_faceon_histo} and \ref{NvisA_large_effi1p012p01p00p5_bldhor_dist}
should be taken with care.

\subsection{The beaming factor $f_\mathrm{\Omega}$}
\label{BFac}

The beaming factor \citep{wrwj09} is defined as
\begin{equation}
\label{fOmega}
f_\mathrm{\Omega}=\frac{L_\mathrm{\gamma}}{4\pi D^2 \langle\nu F_\mathrm{\nu}\rangle}
\end{equation}
where $L_\mathrm{\gamma}$, $D$, and $\langle\nu F_\mathrm{\nu}\rangle$ are respectively the pulsar $\gamma$-ray luminosity, distance, and average energy flux.
The beaming factor is the ratio of the total energy flux radiated over the 4$\pi$ sr solid angle swept by the pulsar beam, 
after one complete rotation, to the phase-averaged energy flux observed at a given $\zeta_\mathrm{obs}$ angle.
Its value depends both on the intrinsic solid angle of the emission beam, on the beam inclination,
and on the amount of energy that it contains.
From Equations \ref{fOmega} and \ref{fluxEquation}, the  beaming fraction is calculated for each simulated light-curve as:
\begin{equation}
\label{fOphplot}
f_\mathrm{\Omega}=\frac{\int_0^\pi \sin \zeta d\zeta \int_0^{2\pi} n(\phi,\zeta)d\phi}{2\int_0^{2\pi} n(\zeta_\mathrm{obs},\phi)d\phi}.
\end{equation}

Figure \ref{NvisA_large_effi1p012p01p00p5_bldhor_beamvsEdot} shows
the behaviour of the beaming factor as a function of $\dot{E}$. For each model we have fitted the trend for the radio-loud and radio-quiet components of the 
population by using an exponential function. In the PC case there are too few visible $\gamma$-ray pulsars, both radio-loud or radio-quiet, to fit the evolution of the beaming factor with $\dot{E}$ 
for each type separately. The PC $f_{\Omega}(\dot{E})$ has been evaluated by merging the samples and fitting the global trend.
The best-fit functions ($f_\mathrm{\Omega ,RL}(\dot{E})$ and $f_\mathrm{\Omega ,RQ}(\dot{E})$)
are indicated in Figure \ref{NvisA_large_effi1p012p01p00p5_bldhor_beamvsEdot}.
 \begin{figure*}[htbp!]
\begin{center}
\includegraphics[width=0.8\textwidth]{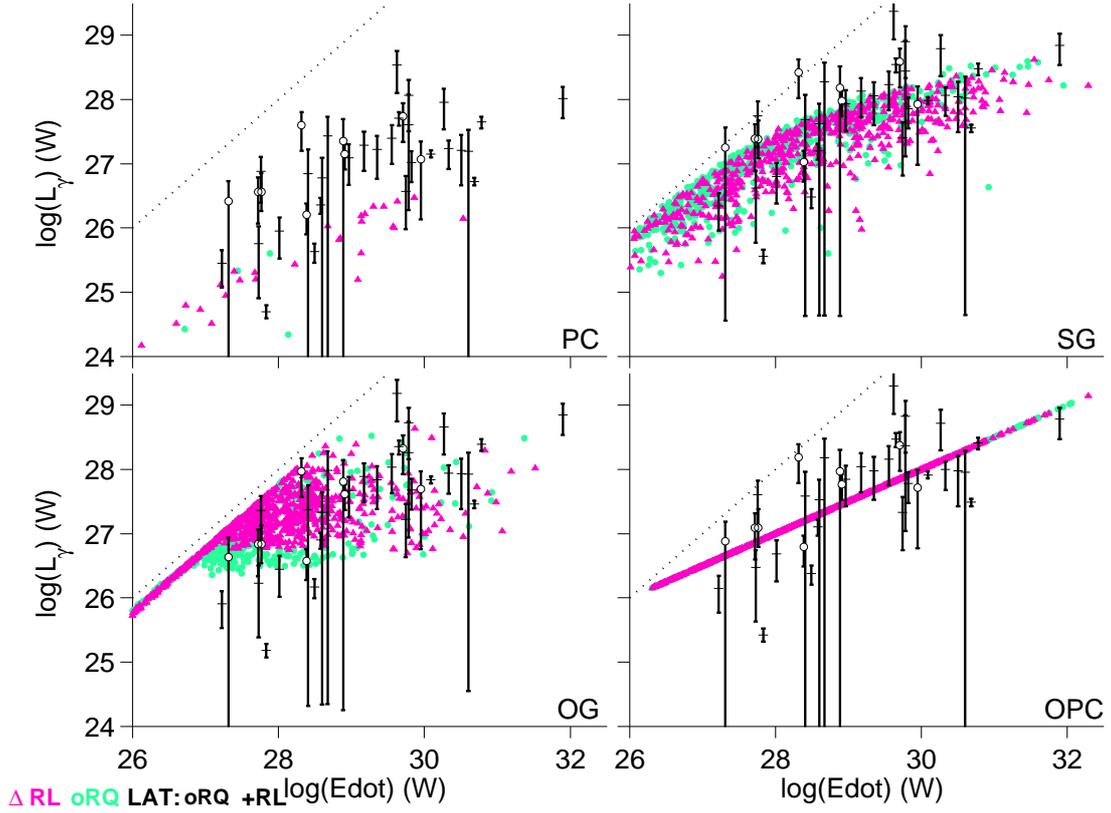}
\caption{Distribution of the $\gamma$-ray luminosities obtained for each model as a function of the spin-down power. Pink and green dots refer to radio-loud and 
radio-quiet $\gamma$-ray visible objects, respectively. The LAT luminosities (black circles and crosses for radio-loud and radio-quiet objects respectively) 
have been derived using the energy-flux 
measurement and the $f_\mathrm{\Omega}$ value estimated from the fit to the simulated data for the particular spin-down power and radio-loud or quiet state of the LAT 
pulsars. The dotted line shows the 100\% efficiency boundary.}
\label{NvisA_large_effi1p012p01p00p5_bldhor_trueLumvsEdot}
\end{center}
\end{figure*}

For all models, the small decrease of $f_\mathrm{\Omega}$ with age (decreasing $\dot{E}$) is due to 
the shrinking of the polar cap as the pulsar slows down. 
In the PC case, both the radio-quiet and radio-loud $f_\mathrm{\Omega}$ values are very dispersed and very small
because of the collimated PC beam.
For the SG, OG, and OPC, the $f_\mathrm{\Omega}$ distribution of the radio-quiet population component appears always more dispersed than the radio-loud one
and it spans lower $f_\mathrm{\Omega}$ values, while the radio-loud objects exhibit higher and more highly constrained beaming factors. 
Radio-quiet pulsars are generally seen at large $\mid \alpha - \zeta \mid$ impact angle where the $\gamma$-ray caustics are fainter, so $f_{\Omega}$ can reach low values.

The SG case shows a minimal change in beaming factor with age for both the radio-loud and radio quiet pulsars
because emission from the bright caustic can be seen over most $\zeta$ directions in the sky and because this model predicts a strong 
off-peak emission. The SG radio-quiet and radio-loud beaming factors are centred around $f_\mathrm{\Omega} = 1$.

In the OG and OPC cases, we note a pronounced dispersion in $f_\mathrm{\Omega}$, over 1 or 2 orders of magnitude,
for the radio-quiet pulsars. The OG and OPC models share the same emission pattern
(phase-plot), thus the dispersion covers the same range of values. In the OPC case, the beaming factor increases up to 
$\dot{E} \sim 10^{28}$ W and then stays constant around 0.8. 
Since all the OPC simulated pulsars at a given $\dot{E}$ have the same $\gamma$-ray luminosity (by construction), the observed spread in the 
$f_\mathrm{\Omega}$ values reflects the spread in beam flux as seen from different perspectives. It amounts typically to less than a 
factor of 2 for radio-loud objects and more than one order of magnitude for radio-quiet ones. As the pulsars age in the OG and OPC 
models, there is an increasing separation of the gamma-ray and radio beams on the sky as the gamma-ray beam shrinks towards the 
the spin equator, producing a greater number of radio-quiet pulsars with small $f_{\Omega}$.

In the outer gap models, the core of the $f_\mathrm{\Omega}$ distributions is consistent with the beaming factor
obtained by \cite{twc11} $f_\mathrm{\Omega}\sim0.4$.

\subsection{$\gamma$-ray luminosity trend with $\dot{E}$}
\label{Luminosity}

Figure \ref{NvisA_large_effi1p012p01p00p5_bldhor_trueLumvsEdot} shows, for each emission model, the evolution of the $\gamma$-ray luminosity 
with the spin-down power and its comparison with the LAT results. 
The luminosity of the LAT pulsars has been computed from the measured pulsed
flux using Equation \ref{fOmega} with a beaming factor $f_\mathrm{\Omega}(\dot{E})$ obtained from the best fit plotted in Figure 
\ref{NvisA_large_effi1p012p01p00p5_bldhor_beamvsEdot}, according to their radio-loud or radio-quiet state.

The observed evolution is roughly predicted by all the 
models. Given the large dispersion in both the data and model predictions, the luminosity trend with $\dot{E}$ cannot be used to discriminate between the gap models. In the OPC case, the $L_\mathrm{\gamma}(\dot{E})$ evolution is a built-in assumption of the model chosen to follow the observations.

The comparison with the LAT population indicates that the PC model is not luminous enough to account for the observed pulsars. 
Because of the large radiative
efficiency (increased power in the gap), the SG luminosity reasonably follows the LAT data and the SG population best describes the observed trend. 
Since the $\gamma$-ray emission is sustained by the particles generating the same polar cap electromagnetic  
cascade, $L_\mathrm{\gamma}$ in both the PC and SG models follows the same trend,  steepening from $L_\mathrm{\gamma}\propto E^{1/2}$ to $L_\mathrm{\gamma}\propto E$ with
decreasing $\dot{E}$ when the pulsar puts out most of its spin-down power into $\gamma$-rays. This trend is predicted but not yet observed 
because of the large dispersion in the LAT data points and large uncertainties in LAT pulsar distances. It is possible, however, that a more pronounced break to lower $\gamma$-ray luminosities is required at low $\dot{E}$.

The OG luminosity evolution shows a different behaviour for high and low spin-down values. At $\dot{E} < 10^{28}$ W, 
the gap width quickly saturates to a constant value which covers about three quarters of the open field volume.
Only objects with large obliquities $\alpha$ remain visible and their luminosity scales linearly with $\dot{E}$. They exhibit a small dispersion 
that is not consistent with the LAT data. At higher $\dot{E}$ values, the LAT pulsars fall within the range of predicted luminosities but they exhibit less dispersion than predicted by the model.
\begin{figure*}[htbp!]
\begin{center}
\includegraphics[width=0.8\textwidth]{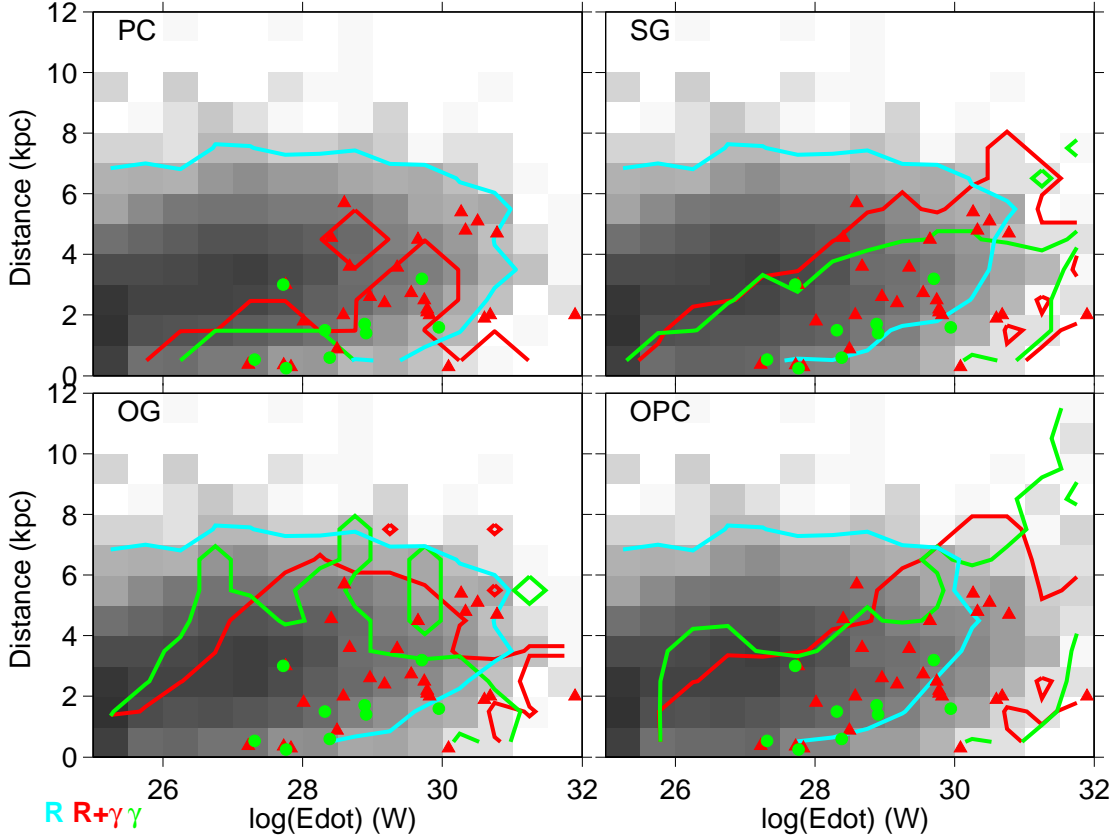}
\caption{For each model, the linear gray scale shows the number density of the radio- or $\gamma$-visible population, 
with a saturation at 10 star/bin. Contours are given for the radio-only visible (cyan), $\gamma$-only visible (green), 
and radio plus $\gamma$ visible (red) objects at 1\% of the maximum number density for each type.  The red triangles 
and green dots mark the location of the radio-quiet and radio-loud LAT pulsars, respectively. }
\label{NvisA_large_effi1p012p01p00p5_bldhor_bldhor_EdotDist_histo}
\end{center}
\end{figure*}

The $L_\gamma \propto \dot{E}^{0.5}$ evolution predicted by the OG and SG models at high $\dot{E}$ is driven primarily by the evolution of the 
Goldreich-Julian current across the open magnetosphere. This is true if the feedback between particle acceleration and electrical screening from the 
cascading yields a rather stable maximum energy for the primaries and a stable fraction of this energy is radiated away in the cascade. A 
steeper evolution ($L_\gamma \propto \dot{E}$) is expected for both polar-cap and outer-gap accelerators when the electrical screening 
becomes inefficient and the gap fills a large part of the open magnetosphere. The LAT data in Figure 
\ref{NvisA_large_effi1p012p01p00p5_bldhor_trueLumvsEdot} suggest a stronger luminosity 
evolution at $\dot{E} > 10^{29}$ W than proposed by the current models. This conclusion appears to be robust because it applies to the 
very different radiation patterns tested in the OG and SG cases and because the beaming factors that have been used to derive the LAT 
luminosities show little evolution and little scatter from one pulsar to the next for $\dot{E} > 10^{29}$ W for both models. 

One can note a larger scatter in the SG and OG luminosities plotted in Figure \ref{NvisA_large_effi1p012p01p00p5_bldhor_trueLumvsEdot} 
than in the beaming factors for the same $\dot{E}$ range. 
This is due to the intrinsic variation in gap width resulting from the variety of NS properties, amplified by the fact that the 
luminosity is proportional to the gap width cubed. The scatter in the SG luminosity distribution is driven by the spread in 
period and stellar magnetic field at each value of $\dot{E}$. The scatter in the OG luminosity further builds on a strong dependence of the gap 
width with obliquity. In that sense, looking forward to a time when more precise distance estimates are obtained and when tighter 
constraints on the gap location in the outer magnetosphere (from phase-resolved spectroscopy and light-curve studies) provide more 
reliable beaming factors for each pulsar, the dispersion in the luminosity plot can teach us about the diversity of young neutron stars that 
compose the LAT pulsar sample.

\subsection{Fractions of $\gamma$-ray loud and radio-loud pulsars}

Figure \ref{NvisA_large_effi1p012p01p00p5_bldhor_bldhor_EdotDist_histo} illustrates the change in visibility distance for pulsars 
of different ages ($\dot{E}$) and emission types (loud or quiet in the radio and $\gamma$-ray bands). 
A similar figure has been shown in \cite{wr11} to study the evolution of the pulsar visibility horizon.
Data points for the LAT pulsars have been overlaid, but one should note that very few distance estimates exist for the new LAT 
pulsars that have been found through blind periodicity searches, because of the lack of radio dispersion measures. Given the large 
uncertainties in distance, the agreement between the SG and OPC predictions and the LAT data is reasonable.

\begin{figure}[htbp!]
\begin{center}
\includegraphics[width=0.49\textwidth]{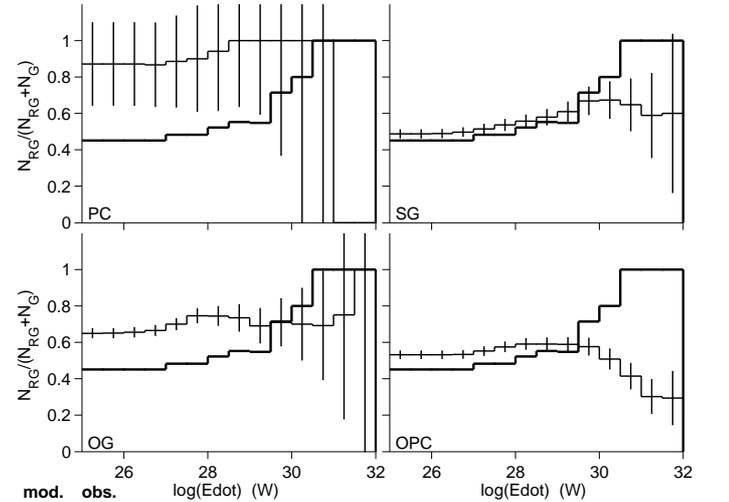}
\caption{Spin-down power evolution of the fraction of radio loud pulsars in the cumulative distribution of $\gamma$-ray visible 
pulsars with $\dot{E}$ larger than the plotted value. The thin lines give the simulation results for each model. The thick 
line gives the fraction evolution in the LAT sample.}
\label{NvisA_large_effi1p012p01p00p5_bldhor_gamLRG_fraction}
\end{center}
\end{figure}

\begin{figure}[htbp!]
\begin{center}
\includegraphics[width=0.49\textwidth]{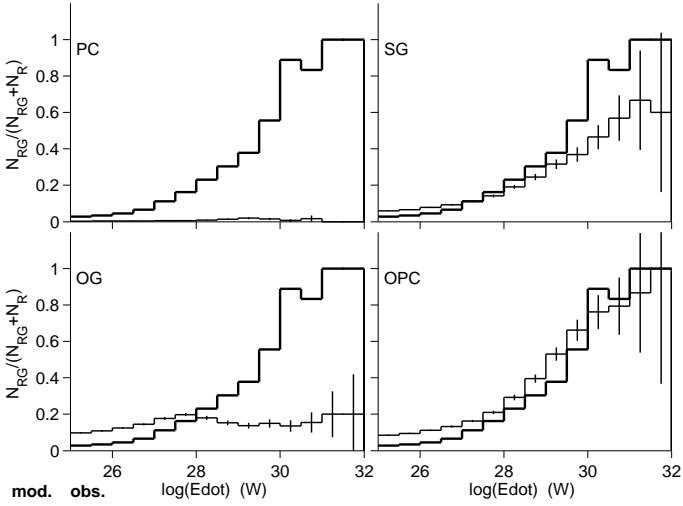}
\caption{Spin-down power evolution of the fraction of $\gamma$-loud pulsars in the cumulative distribution of radio-visible pulsars 
with $\dot{E}$ larger than the plotted value. The thin lines give the simulation results for each model. The thick line gives the fraction 
evolution in the LAT sample.}
\label{NvisA_large_effi1p012p01p00p5_bldhor_RGgamL_fraction}
\end{center}
\end{figure}

Figure \ref{NvisA_large_effi1p012p01p00p5_bldhor_gamLRG_fraction} shows the fraction of radio loud pulsars in the cumulative 
distribution of $\gamma$-ray visible pulsars with $\dot{E}$ larger than the plotted value, and Figure  
\ref{NvisA_large_effi1p012p01p00p5_bldhor_RGgamL_fraction} shows the fraction of $\gamma$-loud pulsars in the cumulative 
distribution of radio-visible pulsars above the given $\dot{E}$ (see \citealt{rmh10}).

They jointly illustrate the high probability of detecting both 
the radio and $\gamma$-ray beams in LAT objects with $\dot{E} > 10^{30}$ W, in contrast to the predictions of all the models. 
This fraction remains high at all $\dot{E}$ 
values for the PC prediction, at variance with the data, because the radio and $\gamma$-ray beams are produced in nearby regions of the magnetosphere.

Figure \ref{NvisA_large_effi1p012p01p00p5_bldhor_bldhor_EdotDist_histo} shows that the LAT visibility horizon for radio-loud and radio-quiet 
objects shrinks similarly with pulsar age (decreasing $\dot{E}$), so we lose both types at the same rate. 
The radio-loud to radio-quiet ratio evolves little for $\dot{E} < 10^{30}$ W, so the radio-loud fraction in the whole $\gamma$-ray sample 
(Figure  \ref{NvisA_large_effi1p012p01p00p5_bldhor_gamLRG_fraction}) flattens with age to about 0.5 in good agreement with the LAT data. 

The value of this fraction is controlled by the effective difference in flux sensitivity for the detection 
of a radio-loud or radio-quiet pulsar in the LAT data. Detecting a $\gamma$-ray pulsation after phase-folding with the radio ephemerides requires 
fewer photon counts and trials than for blind periodicity searches. 

The use of the first pulsar catalogue sensitivity map for the radio-loud objects \citep{aaa+10} and of the blind-search 
sensitivity map for the radio-quiet objects \citep{dja+11}, both scaled to 2 years, brings an excellent agreement between the model predictions and 
data at $\dot{E} < 10^{30}$ W. The use of the catalogue map (presenting the lowest flux thresholds) for both types of pulsars lowers the predicted fraction by 
a factor of 2. Figure \ref{NvisA_large_effi1p012p01p00p5_bldhor_bldhor_EdotDist_histo} shows that the $\gamma$-ray visibility horizon shrinks much more 
rapidly than the radio one with pulsar power. So $\gamma$-ray pulsars become undetectable faster than radio pulsars and the $\gamma$-loud fraction in the radio-visible
sample falls continuously down to a few percent at low $\dot{E}$ in Figure \ref{NvisA_large_effi1p012p01p00p5_bldhor_RGgamL_fraction}.

Figure  \ref{NvisA_large_effi1p012p01p00p5_bldhor_gamLRG_fraction}  shows that the radio-loud fraction in the LAT sample becomes 1 at high $\dot{E}$ 
while the SG and OPC predictions remain flat or decrease. The 
outer-gap predictions decline at high $\dot{E}$ because of a relative increase in the number of radio-quiet $\gamma$-ray pulsars seen at large  distances (see Figure 
 \ref{NvisA_large_effi1p012p01p00p5_bldhor_bldhor_EdotDist_histo} for the OPC). We checked that the SG prediction behaves similarly if we increase its 
 radiative efficiency to match that of the OPC model at high $\dot{E}$, so 
the radio-loud deficit is rather independent of the 1-pole versus 2-pole radiation pattern from the outer regions. 
The open magnetosphere widens with $\dot{E}$ 
and wide beams are produced near the edge of the open volume by the efficiently screened, thin gaps. One would thus expect a larger overlap between the 
radio and $\gamma$-ray beams (i.e. an increase in the radio-loud to radio-quiet ratio) as $\dot{E}$ increases. 
Another effect must overcome this trend. 
It is related to the detectability of the large reservoir of faint radio-quiet pulsars in the outer magnetosphere gap models. 
The latter exist at large  $|\alpha-\zeta|$ impact angles and they dominate in potential number. 
The $\gamma$-ray caustic emission extends to larger  $|\alpha-\zeta|$ angles as $\dot{E}$ increases, so the number of potentially 
visible $\gamma$-ray pulsars (radio-quiet) increases. Then, as the dimmer parts of the caustic emission intercepted at high impact angle gradually passes the 
sensitivity threshold as the luminosity increases, a larger fraction of the potential reservoir becomes $\gamma$-ray visible. In other words, the radio-loud probability 
increases because of the widening of the radio and $\gamma$-ray beams, but the flux detectability of the large reservoir of faint radio-quiet objects increases even 
more and the net result is that the radio-loud to radio-quiet ratio can remain constant or decrease at high $\dot{E}$. We checked, by lowering the luminosity of the SG 
and OPC models or by increasing the flux sensitivity threshold, that the radio beam widening effect becomes dominant when
 it is harder to detect faint radio-quiet objects.

The LAT visibility is good enough for the simulations to predict a small fraction of radio-loud objects at high $\dot{E}$, at variance with the LAT data. The use of higher 
flux thresholds for $\gamma$-ray detection would alleviate this deficit, but it would significantly deteriorate all the other observable distributions. The predicted deficit
is robust against different gap locations and extents in the outer magnetosphere (2 pole or 1 pole emission, infinitely thin emitting layer for OPC and emission 
across a gap thinning with increasing $\dot{E}$ for the slot gap, emission above the null surface or reaching to lower altitudes). It is also robust against the gap width 
estimation (which impacts the caustic extent) since the SG and OPC gap width values we obtained differ by 10 to 30 at high $\dot{E}$. The pronounced discrepancy 
suggests that the assumed radio beam is too narrow at young ages. A broader beam of radio emission at higher altitude has also been suggested by \cite{man05} 
and \cite{rmh10}.

We have tested flat distributions in $\alpha$ and $\zeta$ instead of isotropic ones. The flat distributions give more weight to small angles, therefore bias the samples toward more 
numerous radio-quiet pulsars.  Adopting isotropic distributions in $\alpha$ and $\zeta$ results in lower fractions of both radio and $\gamma$-ray, with even lower predictions at medium and high $\dot{E}$ for all models compared 
to the predictions shown in Figures  \ref{NvisA_large_effi1p012p01p00p5_bldhor_gamLRG_fraction} and  \ref{NvisA_large_effi1p012p01p00p5_bldhor_RGgamL_fraction}. A rapid decrease of the magnetic obliquity, over a timescale of 1 Myr, 
has been suggested by \cite{ycbb10} from their study of radio pulse widths. The discrepancy we find here applies to ages less than 30 kyr, so magnetic alignment 
is unlikely to play a key role in reconciling the observed overabundance of young radio and $\gamma$-ray pulsars and the model predictions.

\section{Summary}
\label{summary}

The exceptional results obtained with the \emph{Fermi} LAT telescope in the last few years offer the unique and exciting opportunity to 
constrain the physics of the pulsed $\gamma$-ray emission by studying the early evolution of the pulsar population
and its collective properties. We compared simulation predictions with \emph{Fermi} LAT observations  for this young ordinary pulsar population.

We synthesised a radio and $\gamma$-ray pulsar sample, assuming a core and cone model for the radio emission
and $\gamma$-ray emission according to four gap models, the Polar cap (PC), Slot Gap (SG), Outer Gap (OG), and 
an alternative outer gap, the One Pole Caustic Model (OPC), that uses the OG beam geometry and a simple luminosity evolution with 
$\dot{E}$ consistent with the LAT data \citep{wrwj09}. We compared 
model expectations and LAT data by applying $\gamma$-ray and radio visibility criteria to our sample and by scaling it to the number of radio 
pulsars observed in the Milky Way.

We found that the narrow beam of the low-altitude polar cap emission contributes at most a handful of pulsars in the LAT sample. The modelled luminosity 
is also too faint by an order of magnitude to account for the LAT data if one applies the average PC beaming factors we found for the given spin-down powers
of the LAT pulsars. The large dispersion found in PC beaming factors, however, can substantially solve the luminosity discrepancy.
We find that all the LAT pulsars are much more luminous (by 1 order of magnitude) than the PC expectations. Yet, there is a huge 
dispersion (1 or 2 orders of magnitude) in $f_{\Omega}$ for the PC beams, so applying the average $f_{\Omega}(\dot{E})$ trend to 
he LAT pulsars could be off by more than one order of magnitude in reality, so all the LAT points could go up and down by more than 
1 decade in Figure \ref{NvisA_large_effi1p012p01p00p5_bldhor_trueLumvsEdot} without problems.
The wide beams from the outer gaps and slot gap models can easily account for the \emph{Fermi} LAT detection number in 2 years,
provided an increase of a factor of $\sim 10$ of the standard slot-gap luminosity. The required increase may result
from an enhanced accelerating electric field in the context of offset polar caps \citep{hm11}.
The evolution of the enhanced SG luminosity with spin-down power is compatible with the large dispersion seen in the LAT data.

We took into account the difference in the LAT flux sensitivity for detecting pulsed emission from radio-selected pulsars and for blind 
periodicity searches. The use of the two different sensitivity maps explained the almost equal amounts of radio-loud and radio-quiet pulsars found by the LAT.
For all models, we found that the $\gamma$-ray visibility horizon extends to comparable distances 
for radio-loud and radio-quiet pulsars as a function of $\dot{E}$, from 6 to 8 kpc at the highest powers down to 2 kpc for the least energetic LAT pulsars. 
The radio visibility horizon compares well with the $\gamma$-ray horizon at high $\dot{E}$, but it extends to much larger distances for less energetic pulsars, 
except for the rapidly evolving OG case for which the pulsars with $\dot{E}  \lesssim 3 \times 10^{27}$ W put 58\% of their spin-down power into $\gamma$-rays
and remain visible to 5 kpc. 

All the $\gamma$-ray models fail to reproduce the high probability of detecting both the radio and $\gamma$-ray beams at high $\dot{E}$.
The OPC prediction for the fraction of $\gamma$-loud pulsars among the radio pulsars is consistent with the radio and LAT data, but the model 
significantly under-predicts the fraction of $\gamma$-ray pulsars that are radio-loud.  The SG model also over-predicts the number of radio-quiet 
$\gamma$-ray pulsars. 
These discrepancies may indicate that pulsar radio beams are 
larger than those we have modeled, either because they are intrinsically wider or because the emission occurs at higher altitude \citep{man05}, or both.
The same conclusion has been argued by  \cite{kj07}, that postulates emission over a wide range of emission heights rather than over a wide 
range of beam longitudes, and more recently by \cite{rmh10} and \cite{wr11} in the light of the \emph{Fermi} observations.

The beaming factor $f_\mathrm{\Omega}$ hardly evolves with $\dot{E}$ in the SG case. It is well constrained 
around 1 for both radio-loud and radio-quiet pulsars.
In the OPC case, $f_\mathrm{\Omega}\sim0.8$ for radio-loud objects with $\dot{E}> 10^{28}$ W,  and it decreases 
by a factor of 2 for the less energetic objects detected by the LAT.
In the OG case, $f_\mathrm{\Omega}$ decreases from 1 to 0.3 with $\dot{E}$ decreasing down to $10^{28}$ W and the 
evolution flattens around $\sim 0.2$ for lower powers. In all the models, the beaming factor of radio-quiet pulsars follow the average trend 
found for radio-loud pulsars, but with a large dispersion than spans 1 or 2 orders of magnitude.

The classical outer-gap model (OG) fails to explain many of the most important pulsar population characteristics, 
such as spin-down power distribution and luminosity evolution, whereas the outer-gap alternative (OPC), which is based on
a simple scaling of the gap width with $\dot{E}^{-1/2}$, provides the best agreement between model predictions 
and data, as concluded by \cite{wr11}. This agreement relies on the very narrow gaps assumed in the OPC case. They are 10 to 100 times thinner
than the values obtained for the SG for the same spin-down power, so the $\gamma$-ray luminosity is concentrated
in thin and wide beams along the edge of the open magnetosphere.
The OG model predicts a stronger luminosity evolution because it uses the polar cap 
heating by the returning particles to close the gap. The stronger evolution driven by this
feedback is apparently not supported by the LAT data.
\cite{twc11} studied the evolution of the two layer OG luminosity as a function of $\dot{E}$. 
Its result is consistent with the one we plot in Figure \ref{NvisA_large_effi1p012p01p00p5_bldhor_trueLumvsEdot}
for the OG model. The less pronounced dispersion observed in \cite{twc11} is due to the choice of a $f_{\Omega}=1$ 
for all the pulsars.

All models studied here significantly under-predict the number of visible $\gamma$-ray pulsars seen at high 
$\dot{E}$. This inconsistency does not depend on the modelling of the 
$\gamma$-ray and radio visibility thresholds. The discrepancy with the observations is 
significant despite our choice of birth distributions skewed to young energetic pulsars, 
at slight variance with the constraints imposed by the total radio and $\gamma$-ray pulsar sample 
observed. The fact that the four models have different $\gamma$-ray luminosity evolutions and 
different beam patterns suggests a different cause for the discrepancy. Concentrating the birth 
location in the inner Galaxy lessened but did not resolve the discrepancy. Further increasing the 
number of energetic pulsars near the Sun would conflict with the observed pulsar distances. 
The estimate of the visibility threshold in radio or $\gamma$-ray flux is not at stake since all models 
over-predict the number of older, fainter, visible objects. 

The set of present results suggests that the observations require rather luminous 
albeit thin gaps in the magnetospheres of young pulsars. 
It will be a challenge for models to match this behaviour. 
The impact of a magnetic alignment 
with age \citep{ycbb10}, of an azimuthal variation of the accelerating field, or the choice of different braking indices for the pulsar spin-down may 
be important and will be included in future population studies to explore the origin of the scarcity of young energetic $\gamma$-ray pulsars in 
the model predictions.

\begin{acknowledgements}
MP acknowledges Patrizia Caraveo and the IASF-INAF of Milan  for the hospitality and help, 
thanks Damien Parent for the helpful suggestions, and acknowledges Volker Beckmann and 
the Fran\c{c}ois Arago Centre for hospitality. MP acknowledges P and Isabel Caballero for the help.
AKH and PLG acknowledges support from the NASA Astrophysics Theory and \emph{Fermi} GI programs.
PLG also appreciates the support from NSF under the RUI program. We acknowledge the referee 
for the useful suggestions.
 \end{acknowledgements}

\bibliographystyle{bib&style/aa}
\bibliography{bib&style/journals,bib&style/psrrefs,bib&style/crossrefs,bib&style/modrefs,bib&style/pierba}
\appendix
\section{Radio survey sensitivity plots}
\begin{figure*}[htbp!]
\begin{center}
\includegraphics[width=0.49\textwidth]{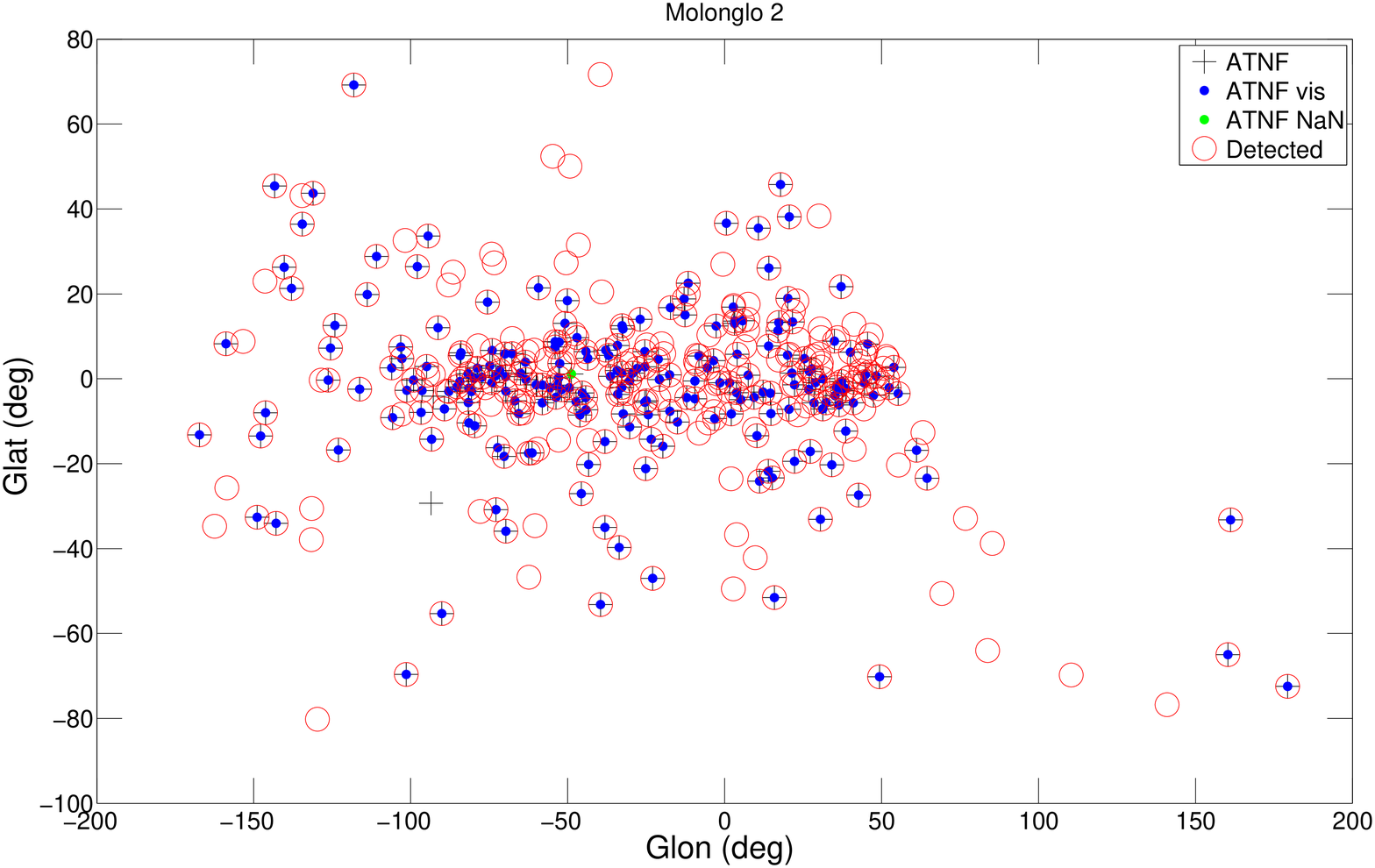} 
\includegraphics[width=0.49\textwidth]{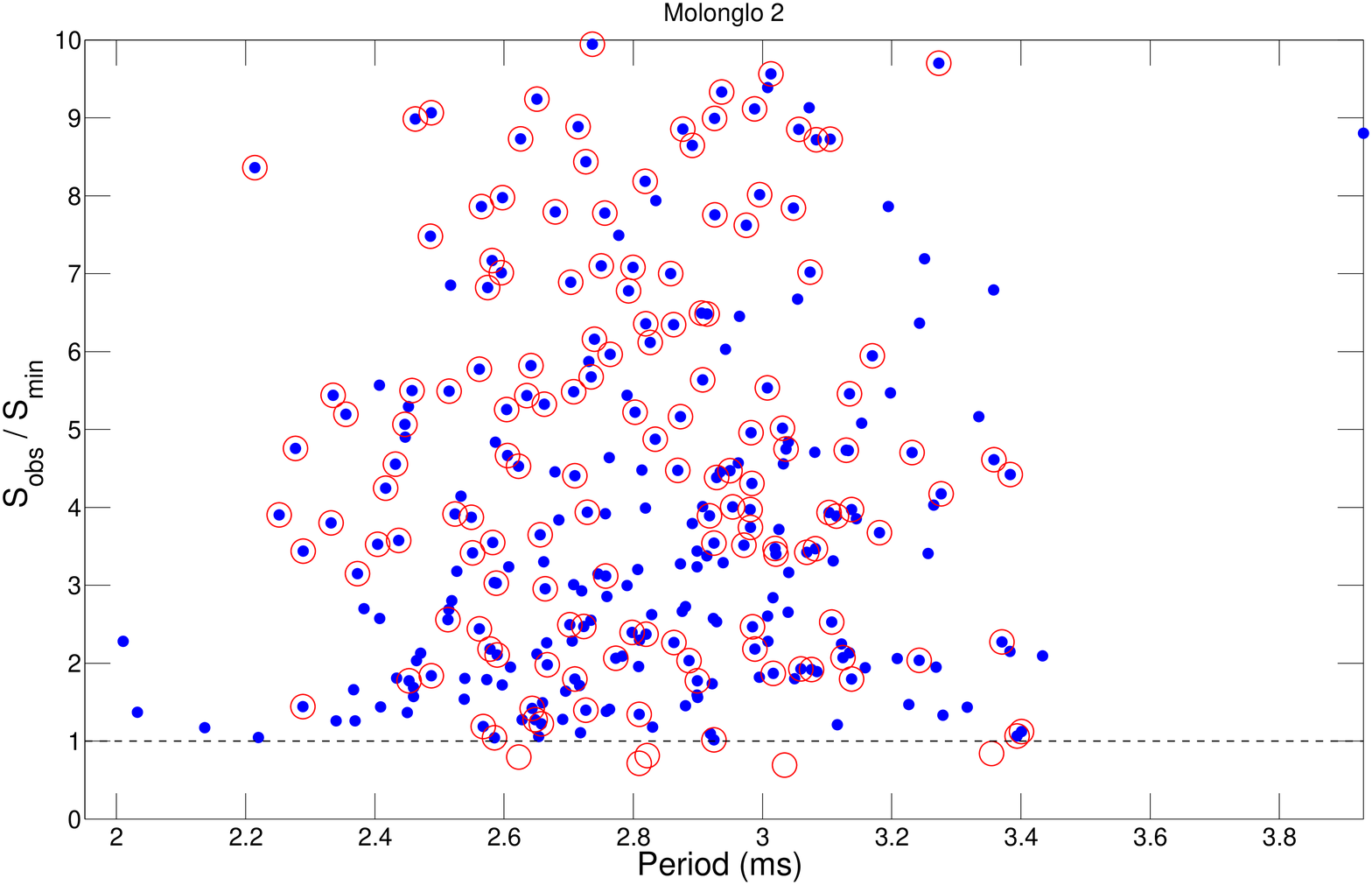} 
\includegraphics[width=0.49\textwidth]{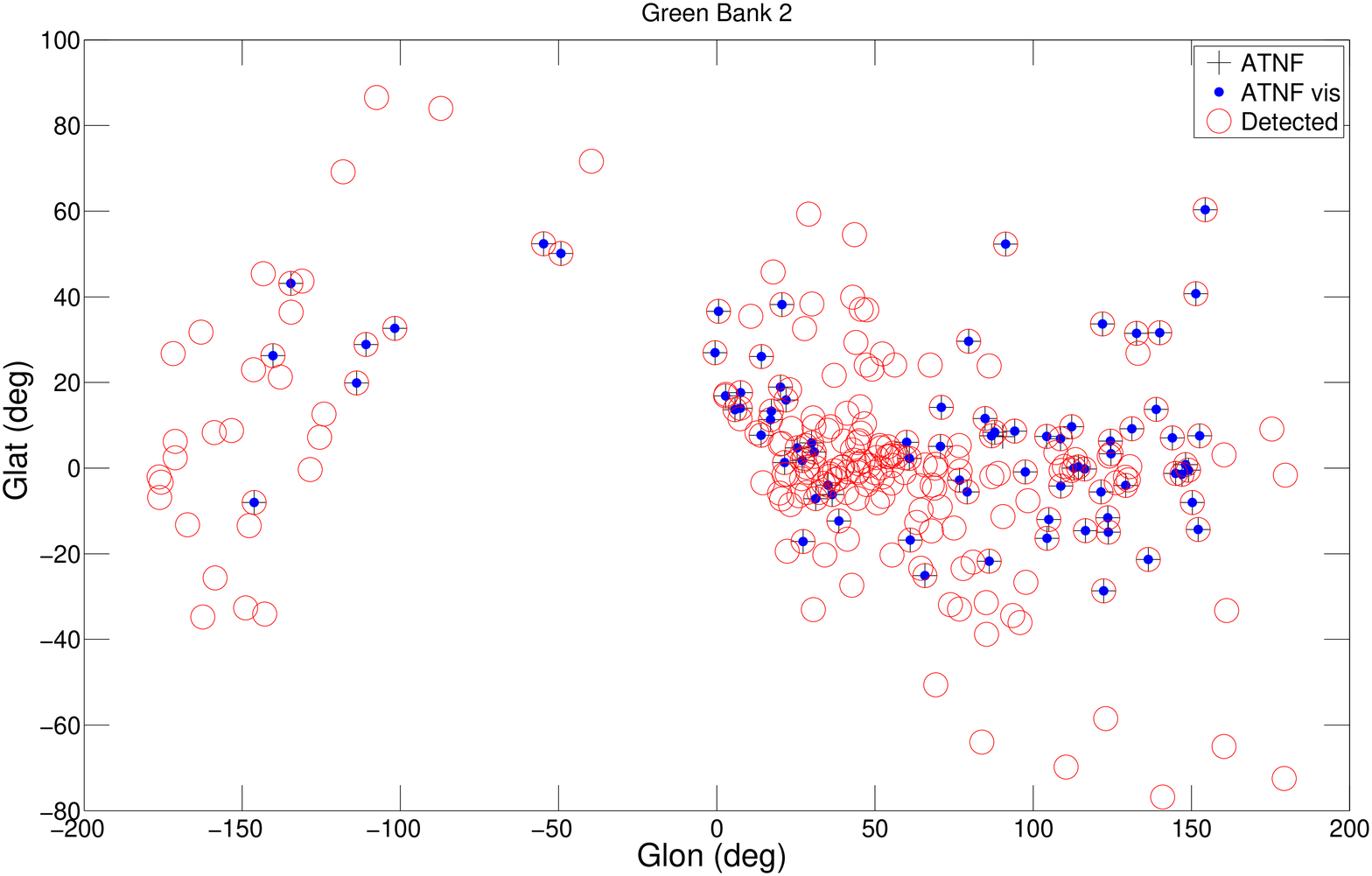}
\includegraphics[width=0.49\textwidth]{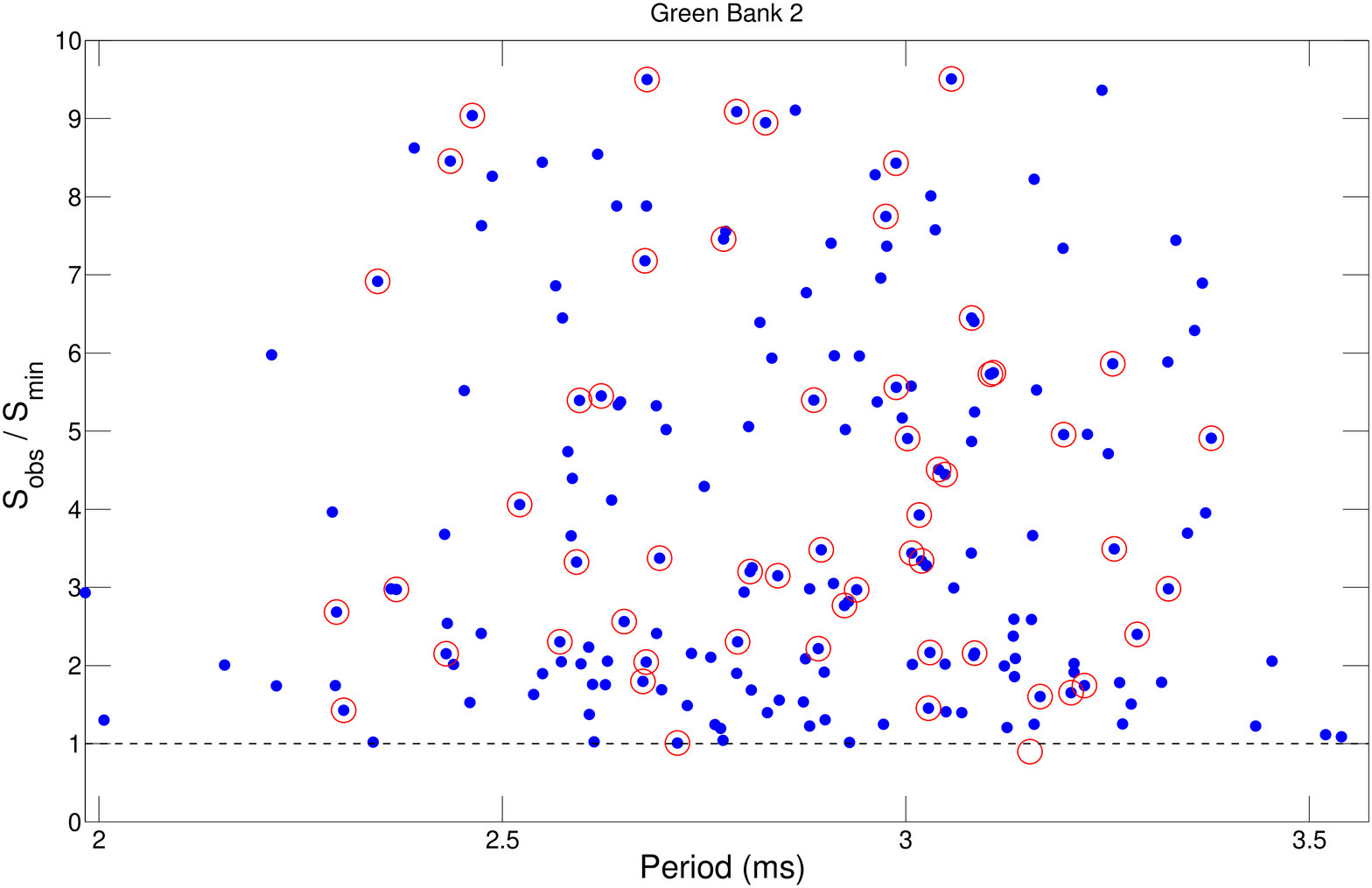}
\includegraphics[width=0.49\textwidth]{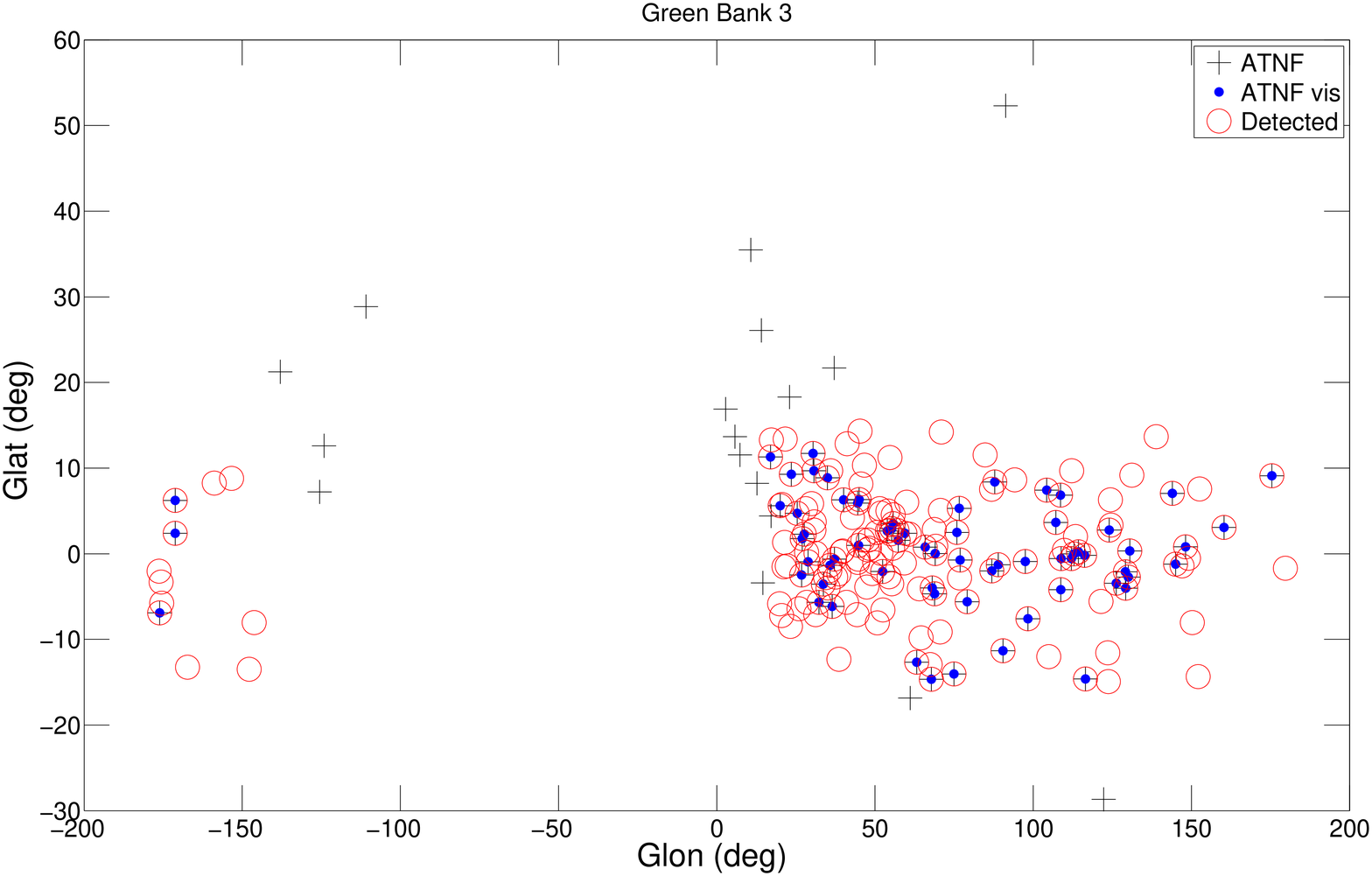}
\includegraphics[width=0.49\textwidth]{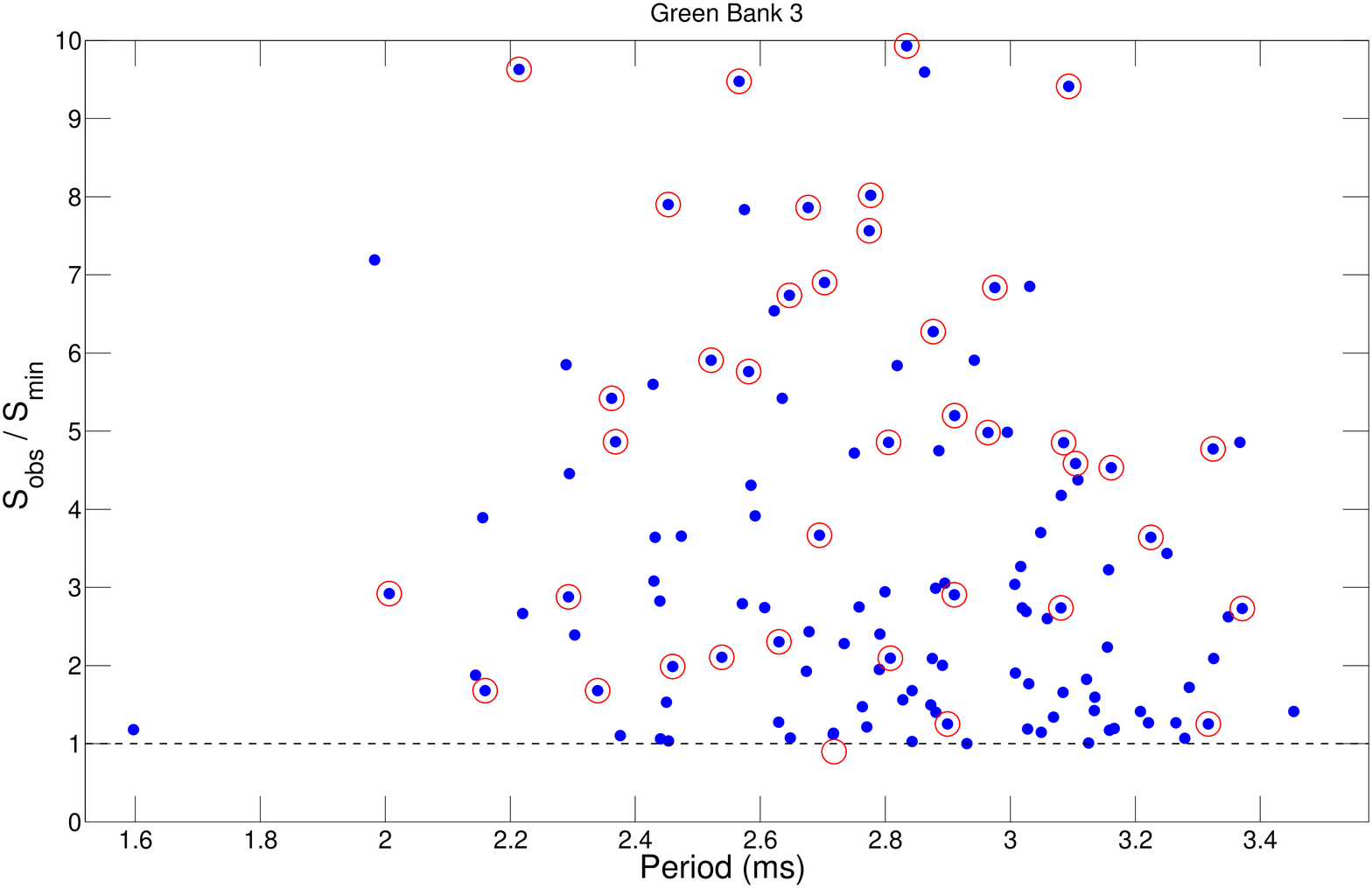}
\includegraphics[width=0.49\textwidth]{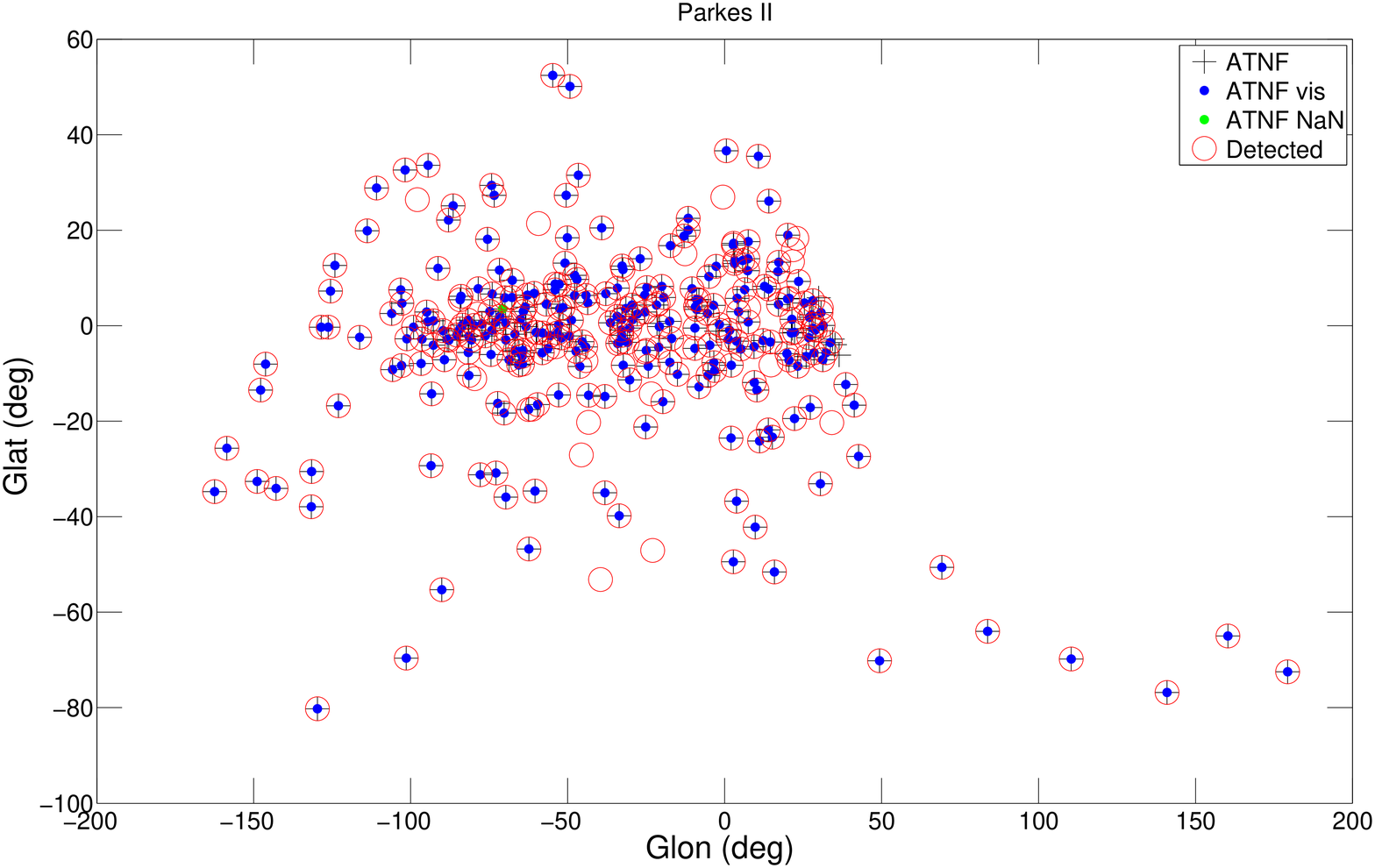} 
\includegraphics[width=0.49\textwidth]{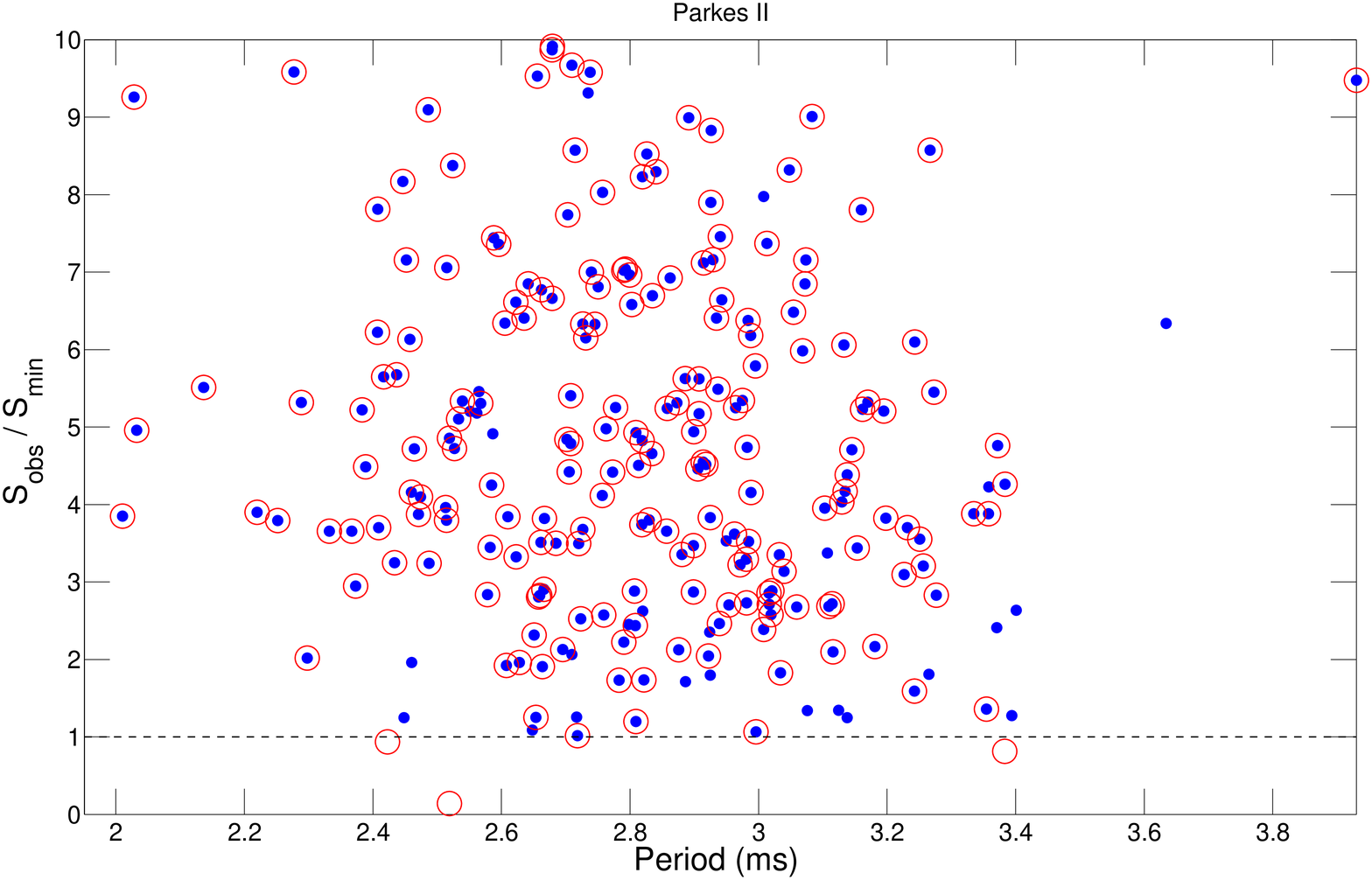} 
\caption{Definition of the radio visibility criteria for the surveys Molonglo 2, Green Bank 2, Green Bank 3, and Parkes 2. For each survey we show the pulsar selection,
             classification, and counting in the redefined survey coordinates region (table \ref{Surveys1}), and the ratio between the pulsar fluxes and the 
             threshold ones re-evaluated  by taking into account the fudge factor defined in table \ref{Surveys1}. The object categories listed in the legend are:
             \emph{ATNF} (black cross): pulsars inside the new survey coordinates region defined in Table \ref{Surveys1} that are listed in the ATNF catalogue;           
             \emph{ATNF vis (blue dot)}: pulsars that are visible according to the new survey parameters listed in Tables \ref{Surveys2} and \ref{Surveys1} and/or listed in the ATNF catalogue;
             \emph{ATNF NaN (green dot)}: pulsars listed in the ATNF catalogue for which it is not possible to evaluate the threshold flux;
             \emph{Detected (red circle)}: pulsars that have been detected according to the new survey parameters listed in Tables \ref{Surveys2} and \ref{Surveys1}.}
\label{SurveysFig1}
\end{center}
\end{figure*}
\newpage
\begin{figure*}[htbp!]
\begin{center}
\includegraphics[width=0.49\textwidth]{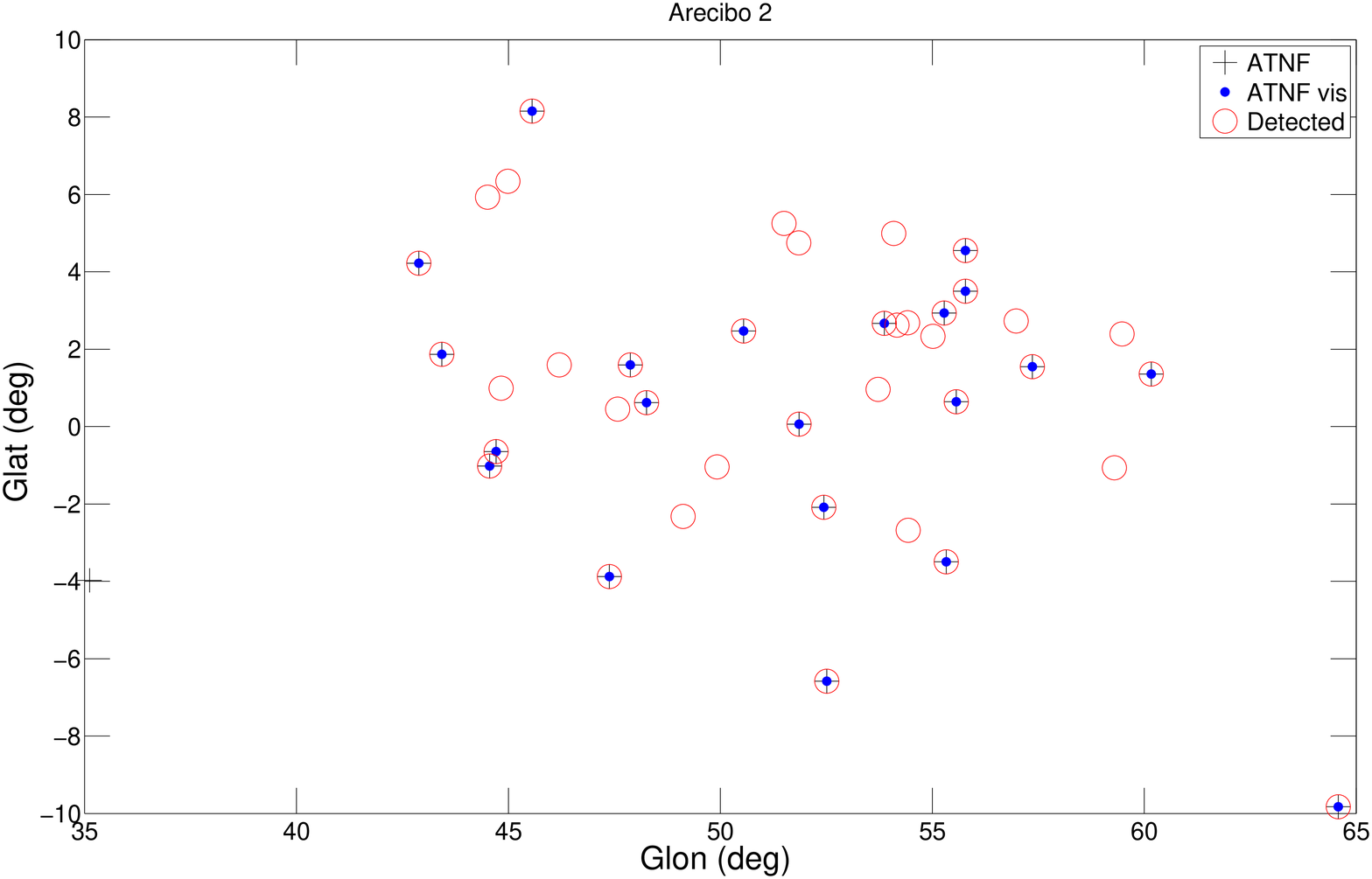}
\includegraphics[width=0.49\textwidth]{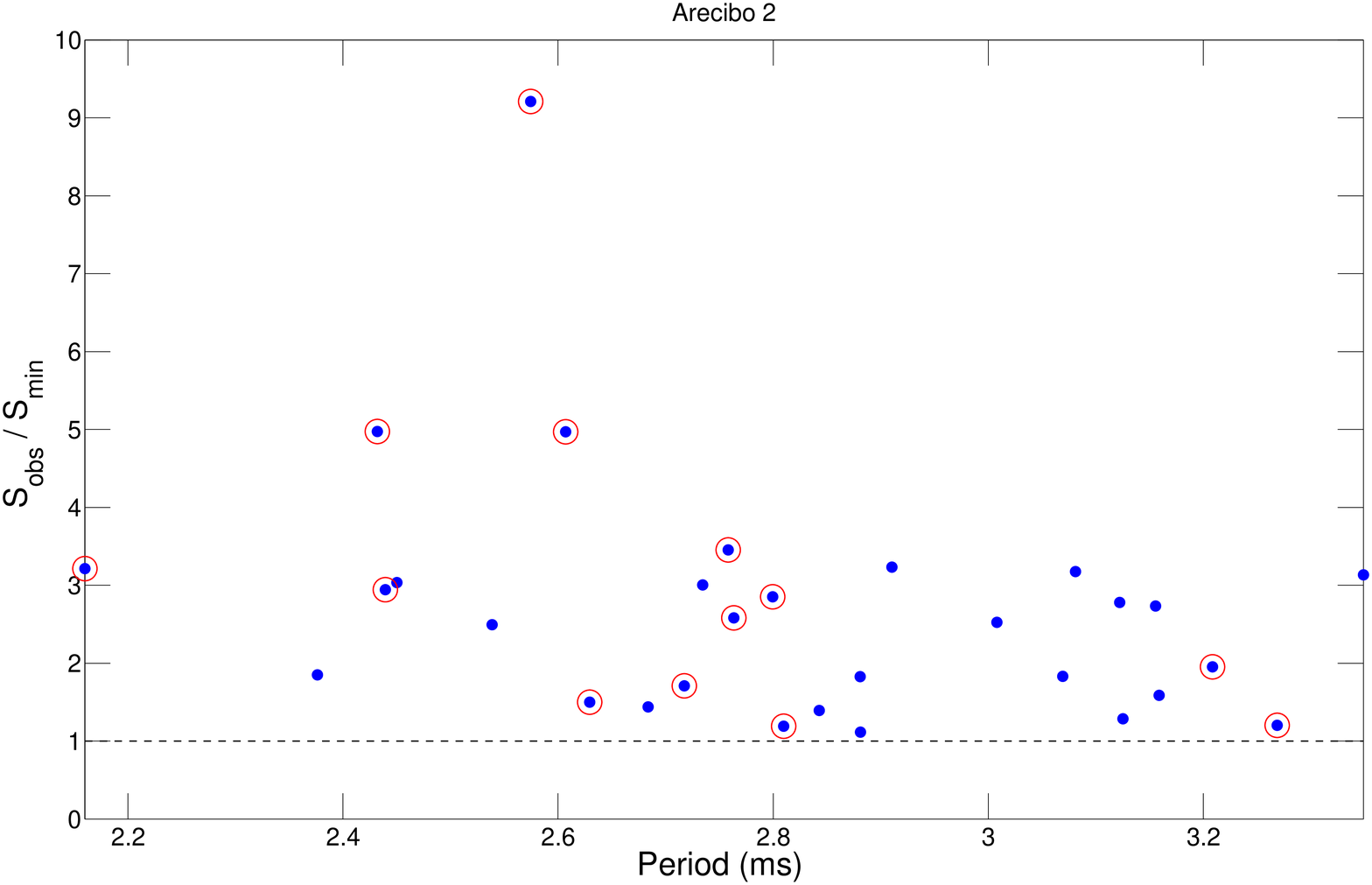}
\includegraphics[width=0.49\textwidth]{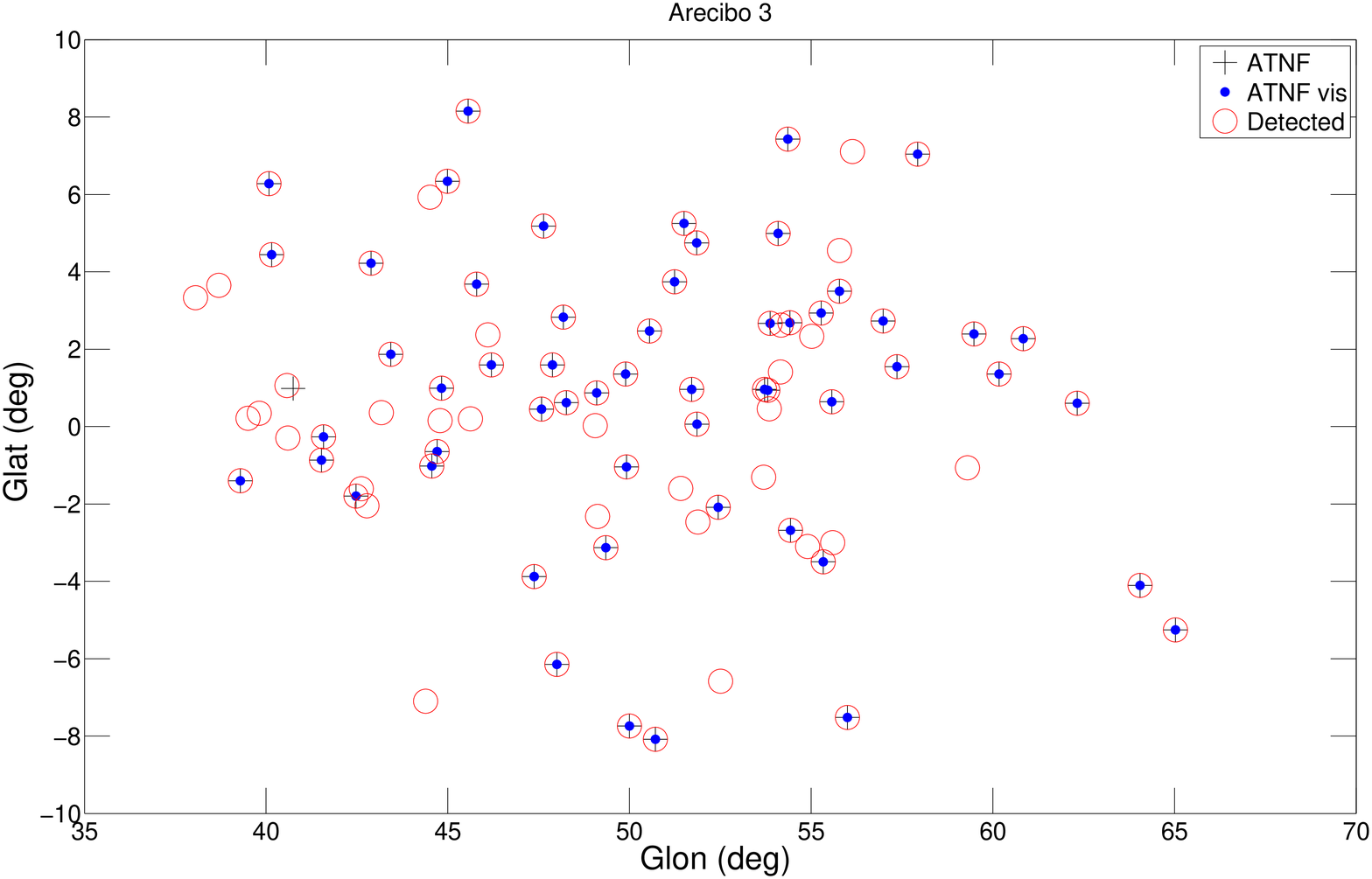}
\includegraphics[width=0.49\textwidth]{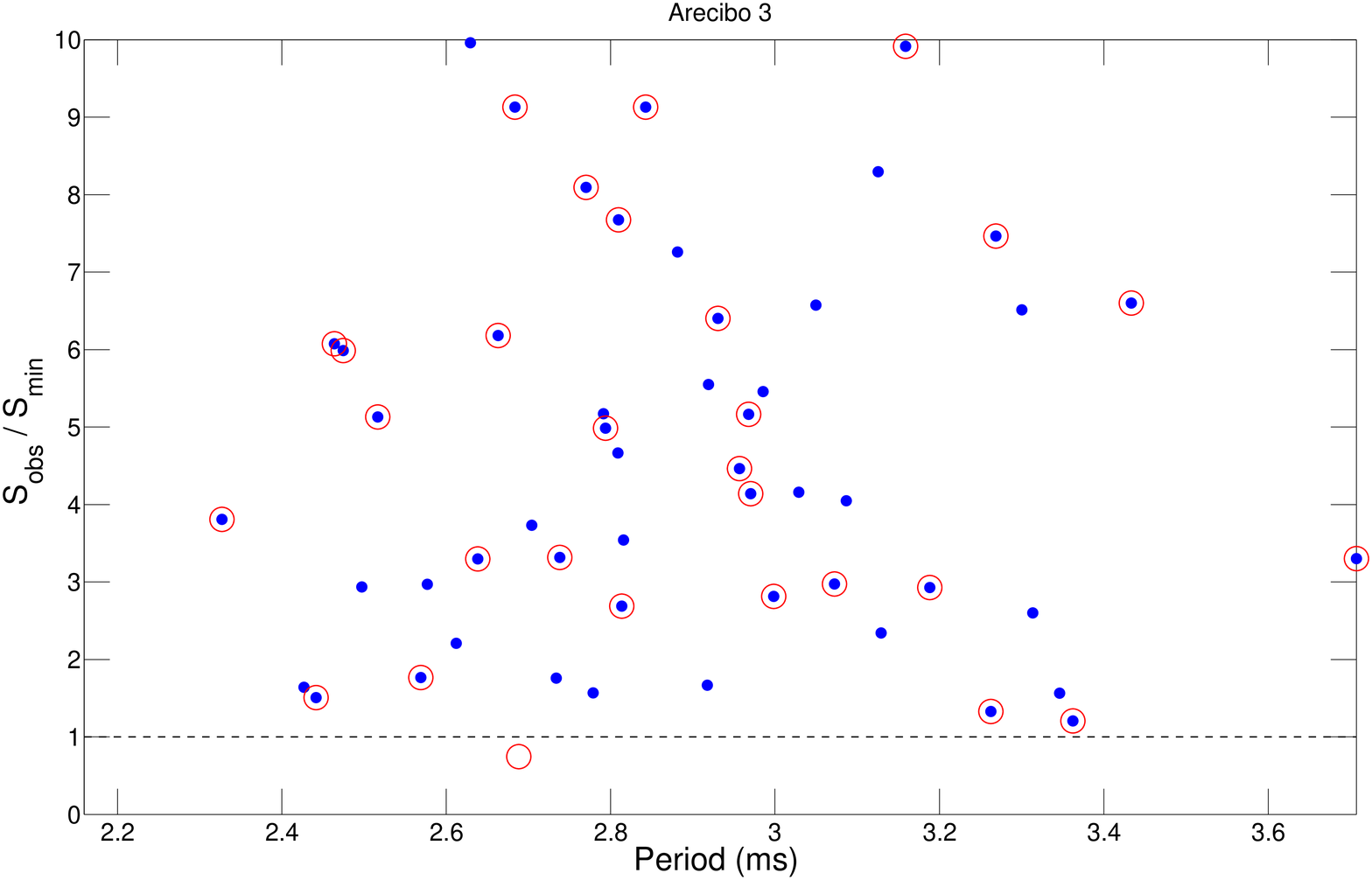}
\includegraphics[width=0.49\textwidth]{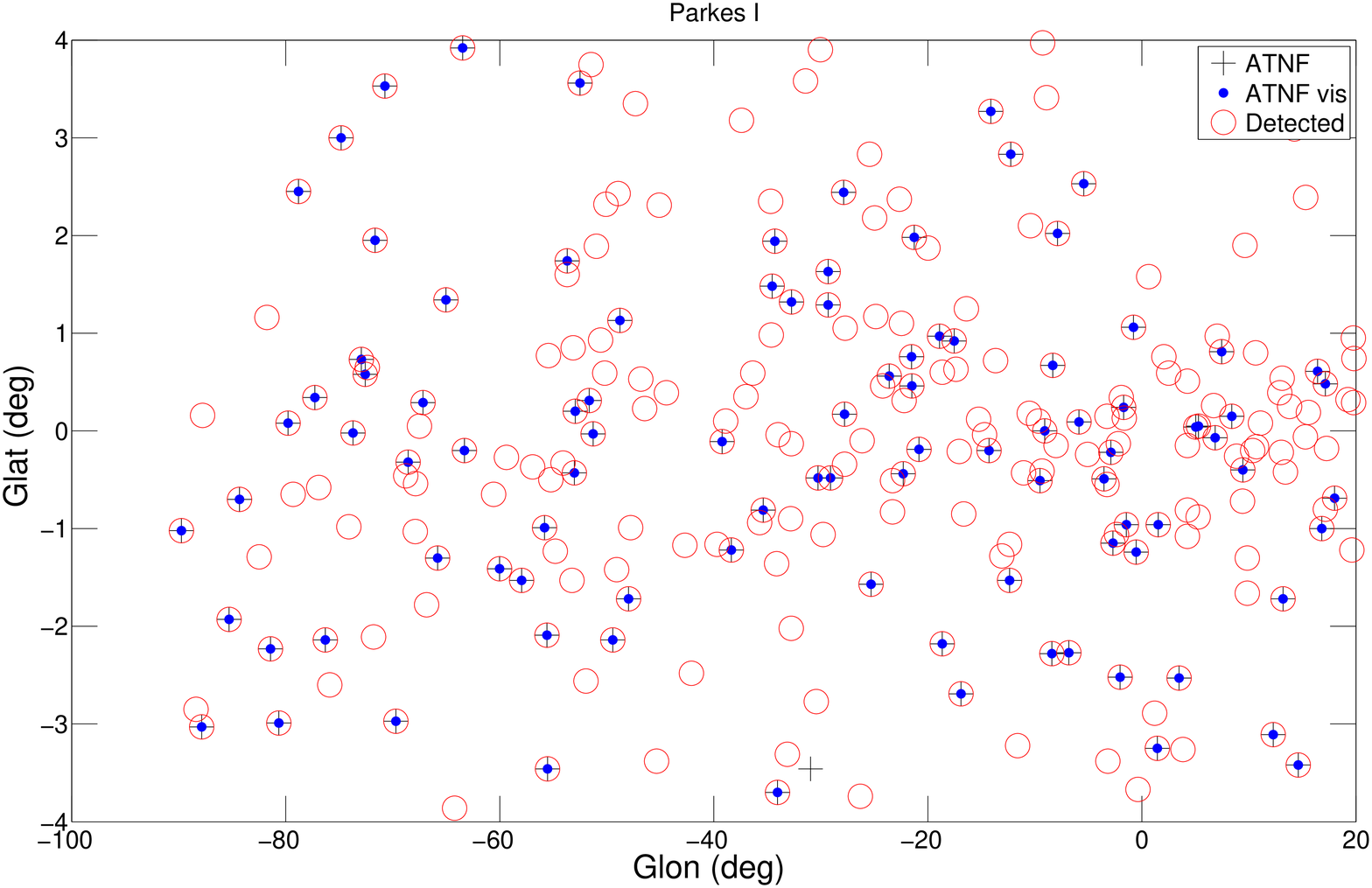} 
\includegraphics[width=0.49\textwidth]{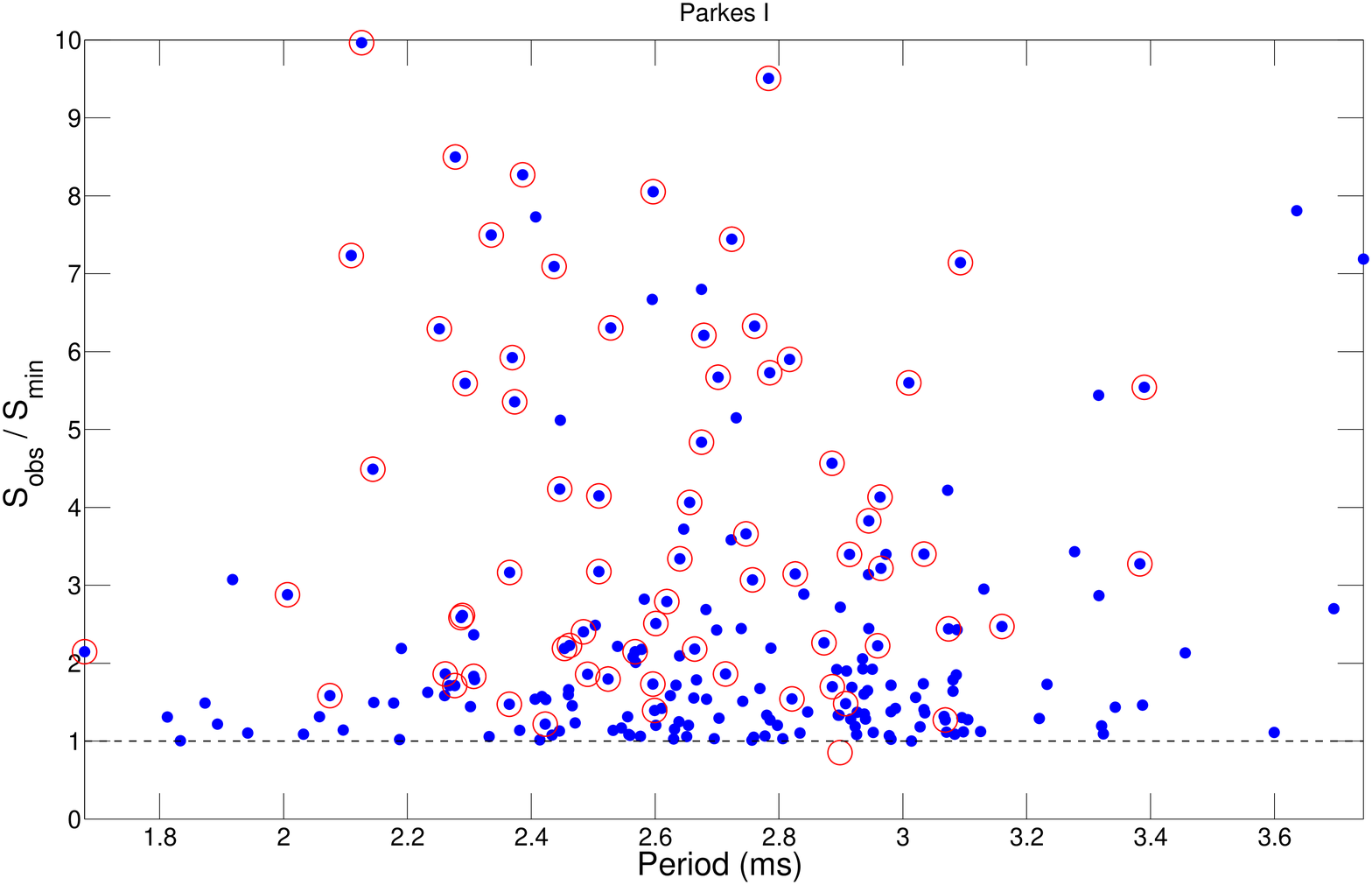} 
\includegraphics[width=0.49\textwidth]{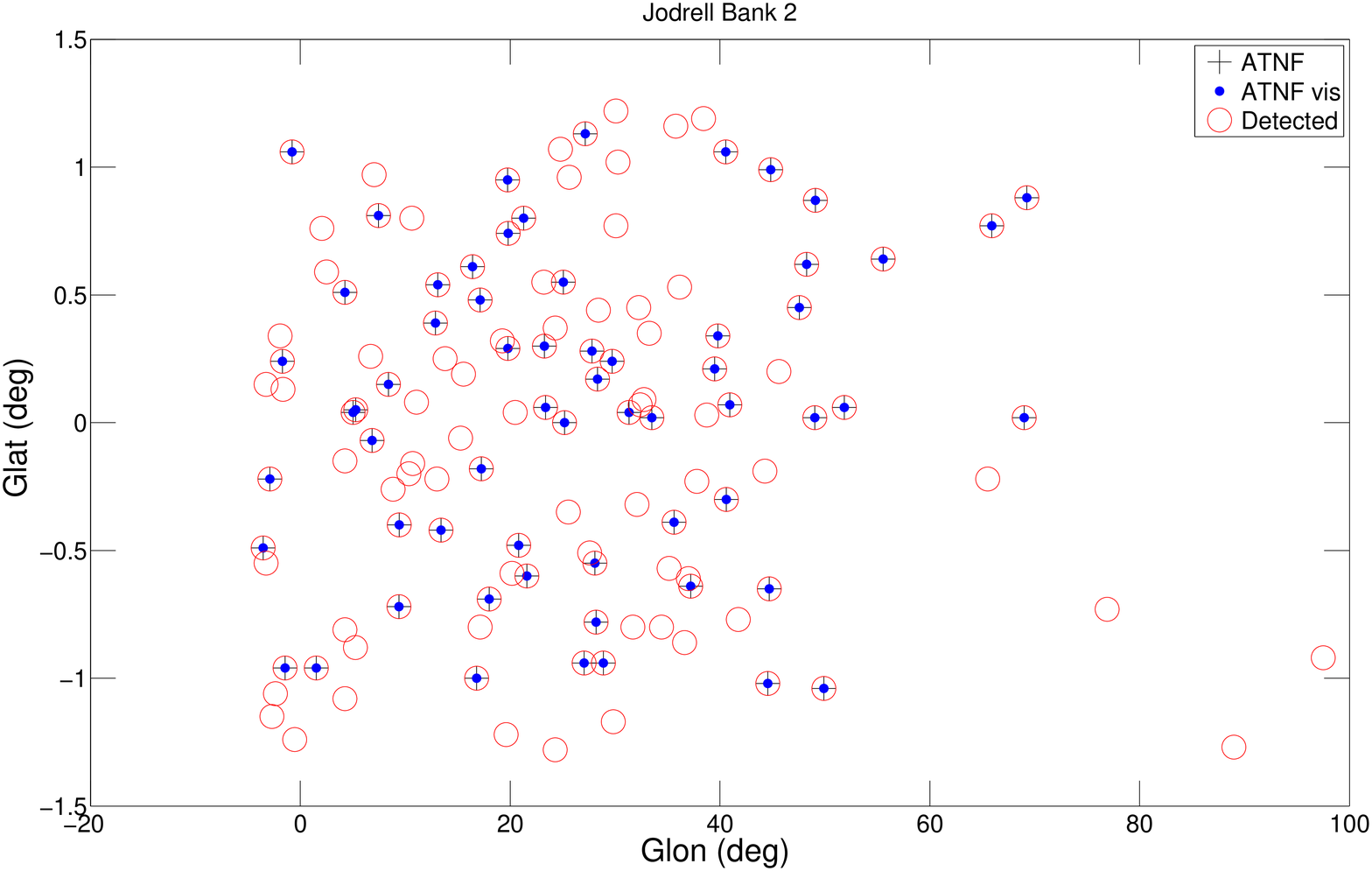}
\includegraphics[width=0.49\textwidth]{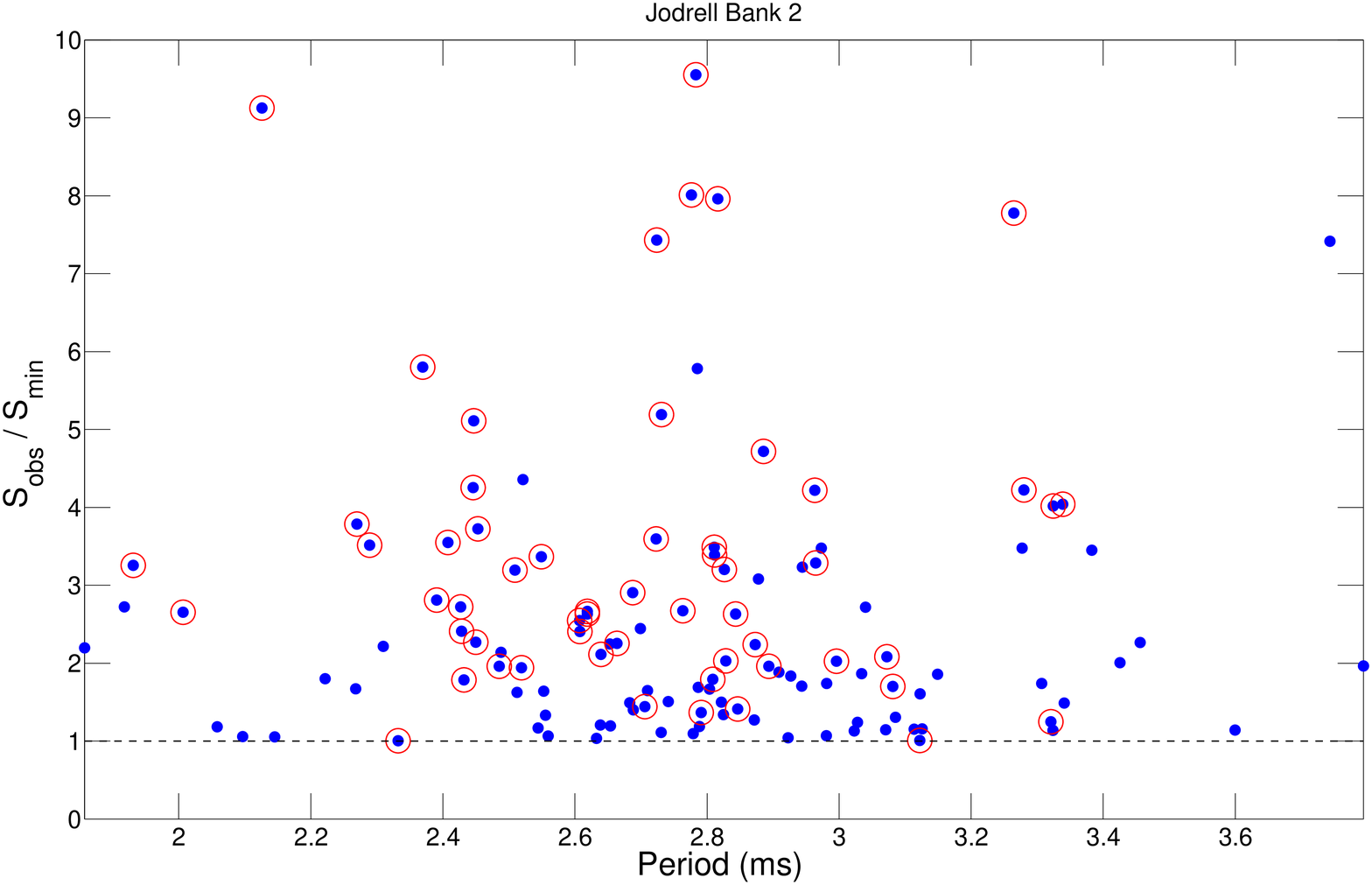}
\caption{Definition of the radio visibility criteria for the surveys Arecibo 2, Arecibo3, Parkes 1, and Jodrell Bank 2. For each survey we show the pulsar selection,
             classification, and counting in the redefined survey sky region (table \ref{Surveys1}), and the ratio between the pulsar fluxes and the 
             threshold ones re-evaluated  by taking into account the fudge factor defined in table \ref{Surveys1}. The object categories listed in the legend are:
             \emph{ATNF} (black cross): pulsars inside the new survey coordinates region defined in Table \ref{Surveys1} that are listed in the ATNF catalogue;           
             \emph{ATNF vis (blue dot)}: pulsars that are visible according to the new survey parameters listed in Tables \ref{Surveys2} and \ref{Surveys1} and/or listed in the ATNF catalogue;
             \emph{Detected(red circle)}: pulsars that have been detected according to the new survey parameters listed in Tables \ref{Surveys2} and \ref{Surveys1}.}
\label{SurveysFig3}
\end{center}
\end{figure*}
\newpage
\begin{figure*}[htbp!]
\begin{center}
\includegraphics[width=0.49\textwidth]{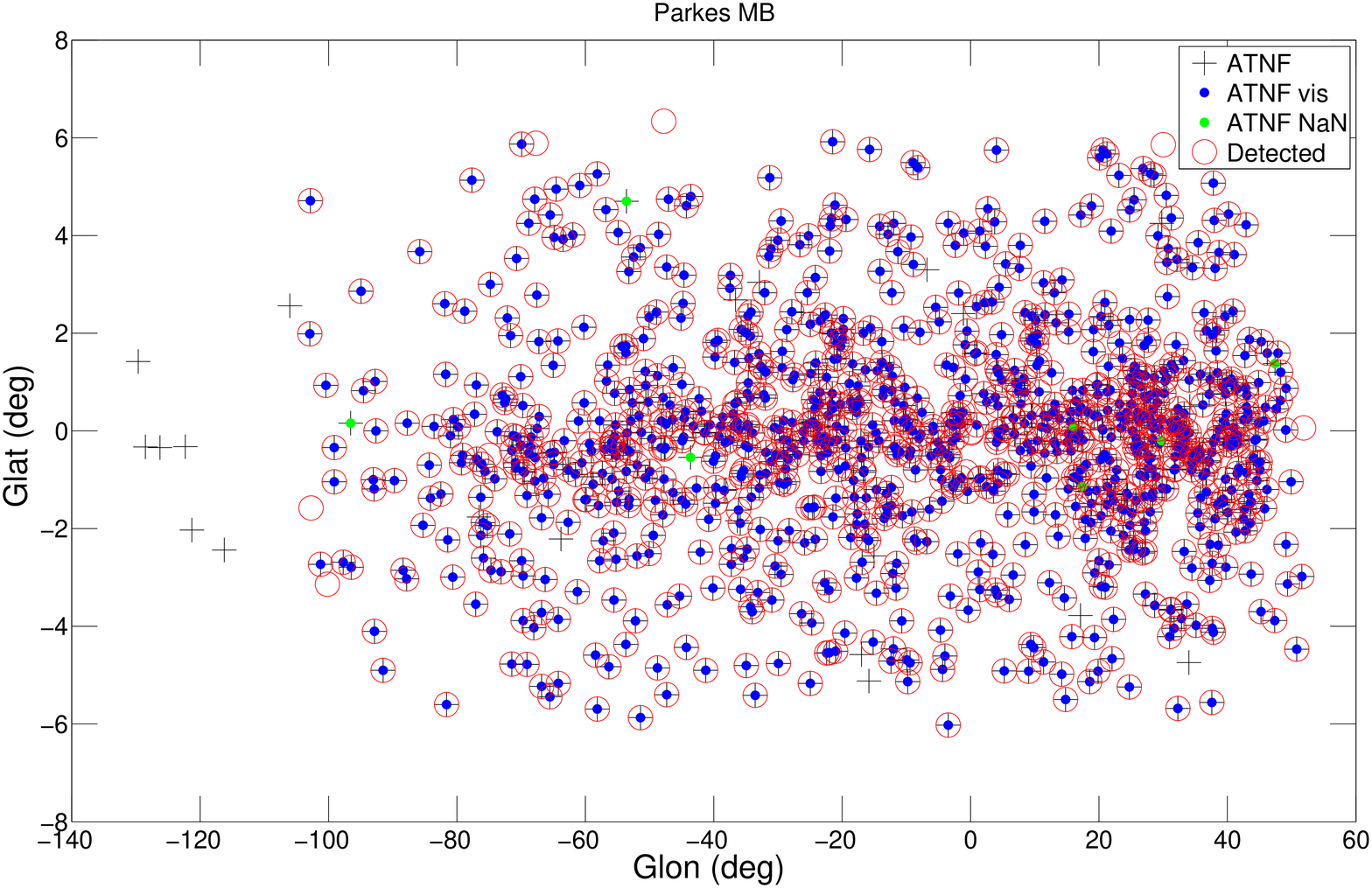}
\includegraphics[width=0.49\textwidth]{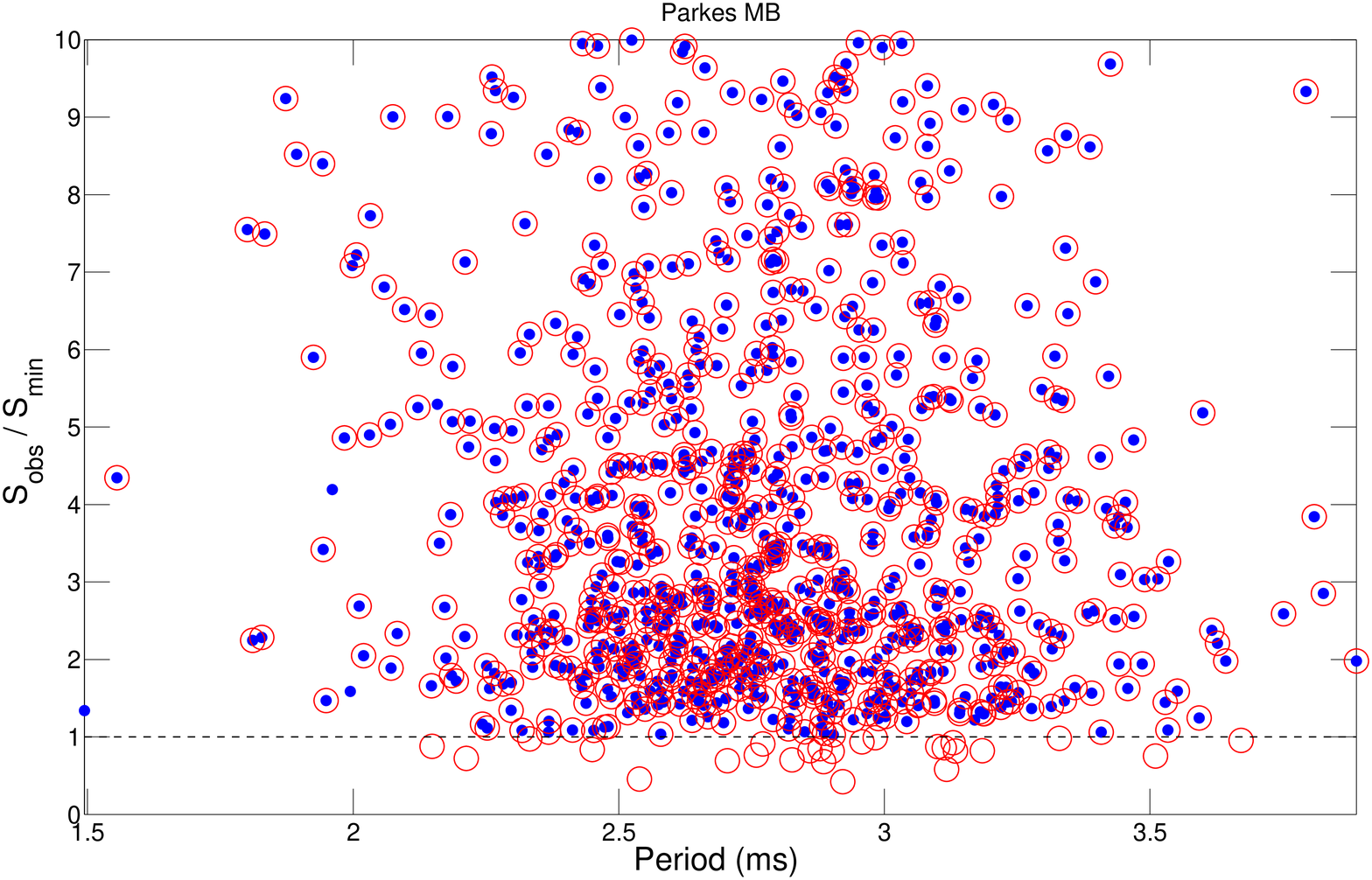}
\includegraphics[width=0.49\textwidth]{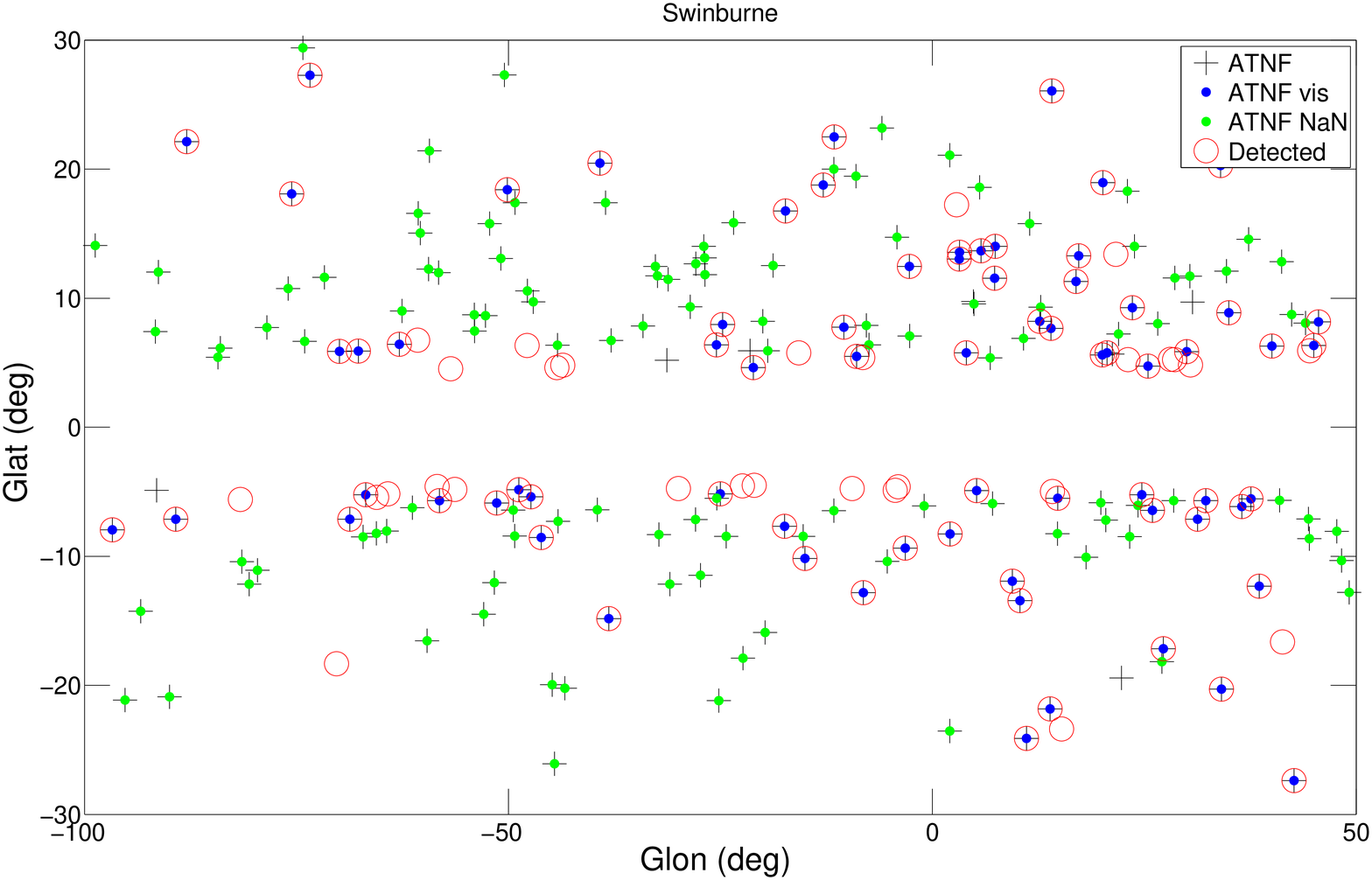} 
\includegraphics[width=0.49\textwidth]{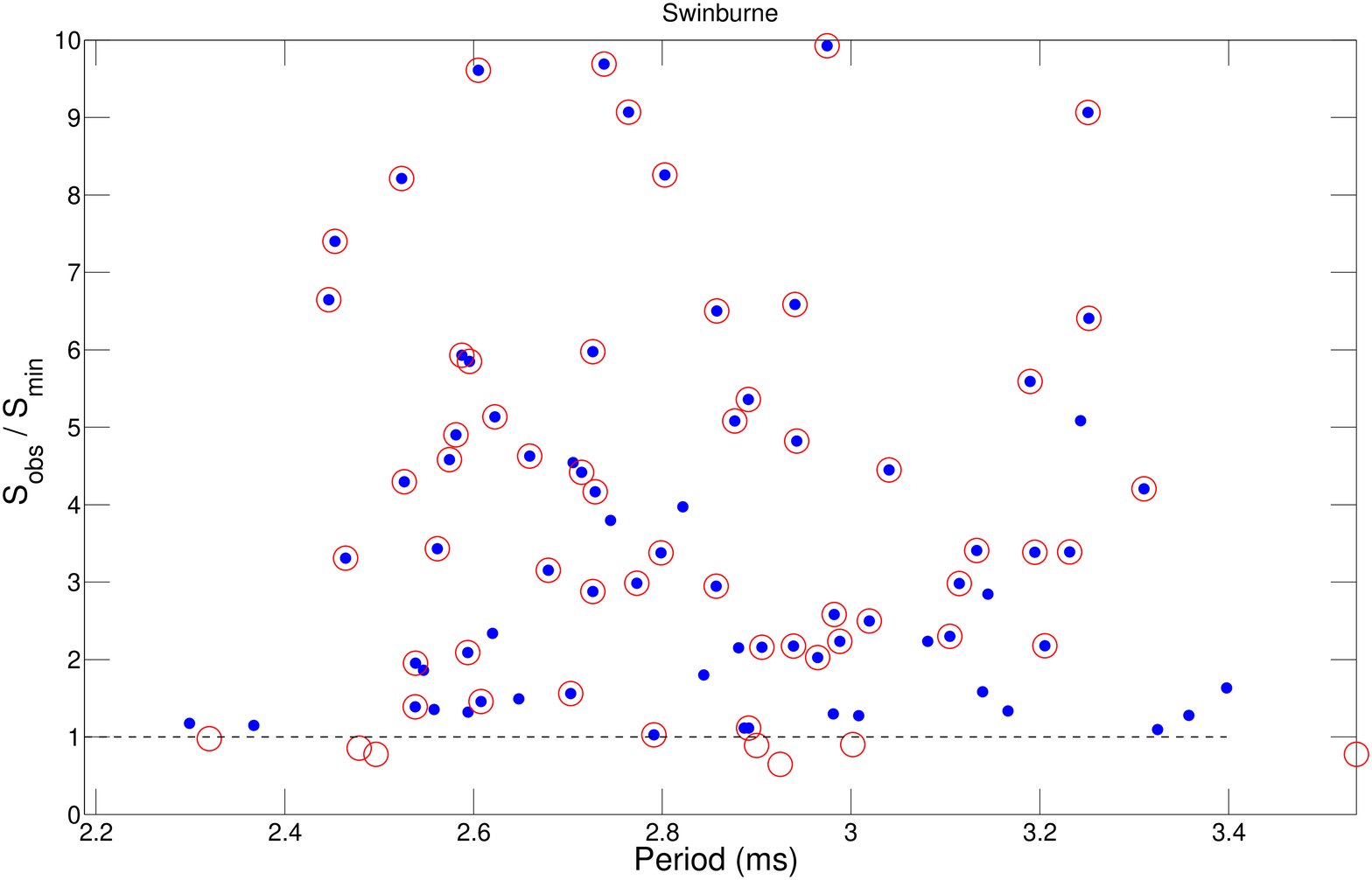} 
\caption{Definition of the radio visibility criteria for the surveys Parkes Multibeam and Swinburne. It is shown the pulsar selection,
             classification, and counting in the redefined survey sky region (table \ref{Surveys1}), and the ratio between the pulsar fluxes and the 
             threshold ones re-evaluated  by taking into account the fudge factor defined in table \ref{Surveys1}. The object categories listed in the legend are:
             \emph{ATNF} (black cross): pulsars inside the new survey coordinates region defined in Table \ref{Surveys1} that are listed in the ATNF catalogue;           
             \emph{ATNF vis (blue dot)}: pulsars that are visible according to the new survey parameters listed in Tables \ref{Surveys2} and \ref{Surveys1} and/or listed in the ATNF catalogue;
             \emph{ATNF NaN (green dot)}: pulsars listed in the ATNF catalogue for which it is not possible to evaluate the threshold flux;
             \emph{Detected (red circle)}: pulsars that have been detected according to the new survey parameters listed in Tables \ref{Surveys2} and \ref{Surveys1}.}
\label{SurveysFig5}
\end{center}
\end{figure*}

\end{document}